\documentclass[
 amsmath,amssymb,
prx,aps,preprintnumbers,floats,floatfix,superscriptaddress,eqsecnum,twocolumn,longbibliography
]{revtex4-2}

\usepackage{amssymb,amsmath,verbatim,mathtools,needspace,enumitem,etoolbox,graphicx,physics,microtype,ulem}
\usepackage{graphicx}
\usepackage{dcolumn}
\usepackage{bm}
\usepackage{hyperref}
\usepackage[mathlines]{lineno}
\usepackage{xcolor}
\usepackage{babel}
\usepackage{xspace}
\definecolor{linkcolor}{rgb}{0.0,0.3,0.5}
\definecolor{urlcolor}{rgb}{0.27,0.55,0.}
\definecolor{funcolor}{rgb}{0.65, 0.16, 0.16}
\usepackage[all]{hypcap}
\usepackage{slashbox,booktabs,amsmath}

\def\apj{\rm{ApJ}}
\def\apjl{\rm{ApJ}}

\def\aap{\rm{A\&A}}

\def\mnras{\rm{MNRAS}}

\def\araa{\rm{ARA\&A}}             

\newcommand{\beqn}{\begin{eqnarray}}
\newcommand{\enqn}{\end{eqnarray}}

\newcommand{\quotes}[1]{{``#1''}}

\newcommand{\alphanorm}{\left(\frac{\alpha}{0.1}\right)}
\newcommand{\feddnorm}{\left(\frac{f_\edd}{0.1}\frac{0.1}{\epsilon}\right)}
\newcommand{\rnorm}{\left(\frac{r}{10M_1}\right)}
\newcommand{\Monenorm}{\left(\frac{M_1}{10^6 M_{\odot}}\right)}

\newcommand{\edd}{\text{Edd}}

\newcommand{\new}[1]{{ #1}}
\newcommand{\newest}[1]{{ #1}}

\newcommand{\AEI}{\affiliation{Max Planck Institute for Gravitational Physics (Albert Einstein Institute) Am M\"{u}hlenberg 1, 14476 Potsdam, Germany}}
\newcommand{\JHU}{\affiliation{Department of Physics and Astronomy, Johns Hopkins University, 3400 N. Charles Street, Baltimore, Maryland, 21218, USA}}
\newcommand{\SISSA}{\affiliation{SISSA, Via Bonomea 265, 34136 Trieste, Italy and INFN Sezione di Trieste}}
\newcommand{\APC}{\affiliation{AstroParticule et Cosmologie (APC), Universit\'e Paris Cit\'e/CNRS 75013 Paris, France}}
\newcommand{\IFPU}{\affiliation{IFPU - Institute for Fundamental Physics of the Universe, Via Beirut 2, 34014 Trieste, Italy}}

\begin{document}

\title{Probing Accretion Physics with Gravitational Waves}

\author{Lorenzo Speri}\email{lorenzo.speri@aei.mpg.de} \AEI

\author{Andrea Antonelli} \JHU
\author{Laura Sberna} \AEI
\author{Stanislav Babak}\APC
\author{Enrico Barausse}\SISSA \IFPU
\author{Jonathan R. Gair}\AEI
\author{Michael L. Katz} \AEI

\date{\today}

\begin{abstract}
Gravitational-wave observations of extreme mass ratio inspirals (EMRIs) offer the opportunity to probe the environments of active galactic nuclei (AGN) through the torques that accretion disks induce on the binary. Within a Bayesian framework, we study how well such environmental effects can be measured using gravitational wave observations from the Laser Interferometer Space Antenna (LISA). We focus on the torque induced by planetary-type migration on quasicircular inspirals, and use different prescriptions for geometrically thin and radiatively efficient disks. We find that LISA could detect migration for a wide range of disk viscosities and accretion rates, for both $\alpha$ and $\beta$ disk prescriptions. For a typical EMRI with masses $50M_\odot+10^6M_\odot$, we find that LISA could distinguish between migration in $\alpha$ and $\beta$ disks and measure the torque amplitude with $\sim 20\%$ relative precision. Provided an accurate torque model, we also show how to turn gravitational-wave measurements of the torque into constraints on the disk properties. Furthermore, we show that, if an electromagnetic counterpart is identified, the multimessenger observations of the AGN EMRI system will yield direct measurements of the disk viscosity. 
Finally, we investigate the impact of neglecting environmental effects in the analysis of the gravitational-wave signal, finding 3$\sigma$ biases in the primary mass and spin, and showing that ignoring such effects can lead to false detection of a deviation from general relativity. 
This work demonstrates the scientific potential of gravitational observations as probes of accretion-disk physics, accessible so far through electromagnetic observations only.
\end{abstract}

\maketitle

\section{Introduction}

Extreme mass ratio inspirals (EMRIs) are a primary target of the future space-borne gravitational-wave (GW) detector LISA \citep{Audley:2017drz}. A typical EMRI involves a stellar-mass compact object with mass $\mathcal{O}(10)M_\odot$ orbiting a massive black hole (MBH) with mass $\mathcal{O}(10^6)M_\odot$. While emitting in the LISA band, the compact object will complete as many as hundreds of thousands of orbits around the MBH, making EMRIs ideal sources to test the nature of black holes (BHs), general relativity (GR), and the astrophysics of galactic nuclei~\citep{Amaro-Seoane:2007osp,Gair:2012nm,Amaro-Seoane:2012lgq,Berry:2019wgg,Amaro-Seoane:2022rxf}. 

Mergers involving MBHs commonly occur within gas-rich environments. 
Interactions with the gas can help binaries -- including MBH binaries and EMRIs -- form and become more compact. Based solely on observations of actively accreting MBHs, it is expected that $1\%$--$10\%$ of the EMRIs observed by LISA will reside in the accretion disk of active galactic nuclei (AGNs) \citep{Dittmann:2019sbm}. {However, the formation of EMRIs involves a wide range of dynamical processes and timescales, so that predictions for LISA in the literature still vary by orders of magnitude, even in the absence of gas~\cite{Amaro-Seoane:2012lgq,Berry:2019wgg,Arca-Sedda2019,Amaro-Seoane:2022rxf}.}
Recently, Refs.~\cite{Pan:2021oob,Pan:2021ksp} argued that there could be even more EMRIs taking place in AGNs than previously estimated, and that LISA could detect $10-10^4$ EMRIs per year from dense accretion disks, compared to only $1-10^2$ per year from relatively gas-free environments (see also Ref. \cite{Derdzinski:2022ltb}). 

The presence of an accretion disk can considerably modify the orbital trajectory of an EMRI emitting in the LISA band. The modification comes in the form of torques, originating either from hydrodynamical effects such as ``accretion winds''~\citep{Kocsis:2011dr,Barausse:2007dy}, or through purely gravitational torques \footnote{The direct pull from the accretion disk is negligible unless unrealistically large densities are considered~\citep{Barausse:2006vt}.} such as dynamical friction \citep{Barausse:2006vt,Barausse:2007dy,Barausse:2014tra,Barausse:2014pra}, which can take the form of planetary-style migration if the density wakes produce torques through shocks \citep{Goodman:2000jf,Kocsis:2011dr,Yunes:2011ws}. Through these processes, the disk exchanges energy and angular momentum with the system.

It is expected that the parameters characterizing EMRI signals will be measured with extreme precision by LISA, thanks to the large number of orbits that can be observed in the LISA band. For this reason, it is reasonable to believe that accretion-disk effects which are estimated to be detectable \citep{Barausse:2006vt,Barausse:2007dy,Kocsis:2011dr,Yunes:2011ws,Barausse:2014tra,Barausse:2014pra,Derdzinski:2020wlw, Zwick:2021dlg} can also be used to \emph{measure} accretion-disk properties, such as the accretion rate and disk viscosity.
This would open up the possibility of studying the properties of accretion disks through purely gravitational means.

In this work, we perform the first quantitative study of the measurability of accretion-disk effects using EMRI observations. In particular, we model \emph{migration} in the disk with two distinct analytical prescriptions -- the $\alpha$ \cite{Shakura:1972te} and $\beta$ \cite{Sakimoto:1981} -- to account for expected uncertainties in the underlying disk physics. Our analysis uses state-of-the-art waveforms \citep{Chua:2018woh,Chua:2020stf,michael_l_katz_2020_4005001,Katz:2021yft} and is performed within a realistic framework for EMRI parameter estimation. For the first time, {we show which accretion-disk quantities can be effectively constrained via gravitational waves. To achieve this, we perform a full Bayesian inference study over the parameter space of a circular-equatorial EMRI system. 
This setup is the most general in the scenario of interest: while generic EMRI trajectories can be eccentric and inclined, EMRIs formed in accretion disks are likely to have circularized and aligned with (or be born in) the disk~\citep{Cresswell2007,Bitsch2010,McKernan2012,MacLeodLin2020,Pan:2021oob,Derdzinski:2022ltb}.
} 
We show that, with an agnostic torque model employed for parameter estimation, LISA could detect and characterize migration in the EMRI GW signal.

We find that migration could be detected in both $\alpha$ and $\beta$ disks.
Our results validate earlier studies~\cite{Kocsis:2011dr,Yunes:2011ws} within a fully Bayesian setting, and extend the discussion to the measurability of the disk parameters. 
Compared to Refs.~\cite{Kocsis:2011dr,Yunes:2011ws}, we find that migration can be detected and also characterized when assuming the $\alpha$ disk model with realistic disk parameters.
Furthermore, for our reference EMRI system, we find that the GW observation alone can distinguish between disk prescriptions and constrain regions of the disk parameter space.
Assuming a torque model, we show that the GW measurement can be combined with electromagnetic observations to infer the viscosity of the host disk.
Finally, we investigate the degree to which we expect key EMRI parameters to be biased if one ignores torques in EMRI parameter inference, finding that the primary's mass and spin may be biased. {We also demonstrate that not accounting for such torques can lead to the false detection of a GR
deviation.}

While our work focuses on migration in thin accretion disks, our torque parametrization could be used to study other effects from the environment or modifications of GR, provided that the induced loss of angular momentum or energy can be written as a simple power law of the orbital separation. 

This paper is organized as follows. In Sec.~\ref{sec:accretion_effects} we present the models for the accretion-disk structure and the migration torque used throughout our work, including a very general phenomenological torque model suited for GW parameter estimation. In Sec.~\ref{sec:waveforms} we describe the vacuum waveform model and how we modify it to account for an environmental torque. We also summarize our methods for parameter estimation and the properties of our reference EMRI. Finally, in Sec.~\ref{sec:measuring_accretion_properties} we present the results: the projected constraints on the amplitude of environmental torques (Sec.~\ref{sec:constraints}), a detailed study of an EMRI signal with a detectable disk torque (including a discussion of multimessenger implications, Sec.~\ref{sec:detection}), and a study of the bias induced on the EMRI parameters {and beyond GR deviations} when environmental effects are present in the signal but ignored in the GW analysis (Sec.~\ref{sec:bias}). We discuss future prospects in Sec.~\ref{sec:discussion}. We work in units in which $G=c=1$, {unless explicitly stated}.  

\section{Accretion-disk effects}\label{sec:accretion_effects}
We begin by describing the accretion-disk prescriptions, torque model and assumptions used in this work. 
We denote the primary and secondary masses with $M_1$ and $M_2$, respectively.
We assume that the MBH is accreting at a ratio $ f_\edd\equiv \dot M_1/\dot M_{\edd, 1} = \epsilon \dot M_1/\dot L_{\edd, 1}$ of the Eddington rate $\dot M_{\edd,i} = 2.536 \times 10^{-8} M_i (\epsilon/0.1) \, {\rm yr}^{-1}$. Here $\epsilon$ denotes the efficiency of conversion of mass energy into luminosity in the disk. The accretion efficiency and the Eddington ratio enter all disk quantities in the combination $\epsilon^{-1} f_{\rm Edd}$.

The astrophysics of accretion disks is notoriously uncertain. Sophisticated numerical simulations are often required to produce realistic disks \citep{Jiang:2019bxn}, and to describe the rich phenomenology of the torques that disks generate \citep{Derdzinski:2020wlw}.
However, a fully numerical approach is intractable for this work's scope, which requires models that are fast to generate. Thus, we adopt analytical models for the disk and its torques. 
We expect future analyses of real data to be conducted in conjunction with numerical simulations \citep{Derdzinski:2018qzv,Derdzinski:2020wlw}.

We employ radiatively efficient, Newtonian, {stationary}, thin accretion-disk models, considering both the $\alpha$ and $\beta$ disk prescriptions for the standard Shakura-Sunyaev solutions for the inner (radiation-pressure dominated) region of the disk~\citep{Shakura:1972te,Abramowicz:2011xu}. For $\alpha$ disks the viscous stress is proportional to the total pressure $t_{r\phi}=\alpha (p_\text{gas}+p_\text{rad})$~\citep{Shakura:1972te}, while for $\beta$ disks $t_{r\phi}=\alpha p_\text{gas}$~\citep{Sakimoto:1981}. There is a long-standing question regarding the stability and realism of these analytic solutions \citep{Lightman:1974sm,Shakura:1976xk,Bisnovatyi1977,1978ApJ...221..652P,Jiang:2019bxn}. In particular, $\beta$ disks have raised concerns for being only loosely motivated by the absence of thermal instabilities, which appear in analytical solutions of $\alpha$ disks. Despite these instabilities, the $\alpha$ disk model is still considered a good approximation for realistic, turbulent accretion flows in the radiation-dominated regime ({although still far from reproducing all the features of radiation magnetohydrodynamic simulations, see Ref.} \citep{Jiang:2019bxn}). 
In both cases $\alpha$ is the viscosity, which parametrizes the complex (and uncertain) magnetohydrodynamic features of accretion disks. Viscosity in AGN disks is typically believed to be around $\alpha=0.1$, and as low as $\alpha=0.01$ \citep{Davis:2009sc,Jiang:2019bxn}.
However, smaller values of $\alpha$ have not been excluded yet. 

{The Shakura-Sunyaev disk with $\alpha$ and $\beta$ viscosity prescriptions predicts the following density profiles and scale height}~\citep{Shakura:1972te,Sakimoto:1981}
\begin{align}\label{eq:Sigma_H}
 \Sigma_{\alpha}\bigg[\frac{\rm kg}{\rm m^2}\bigg] \approx& 
\, 5.4 \times 10^{3}\alphanorm^{-1} \feddnorm^{-1}\rnorm^{3/2},\nonumber\\
\Sigma_{\beta}\bigg[\frac{\rm kg}{\rm m^2}\bigg] \approx&
\, 2.1\times 10^{{7}}\alphanorm^{-4/5}\feddnorm^{3/5}
\nonumber\\
&\times\Monenorm^{1/5}\rnorm^{-3/5}, \nonumber
\\
    H[\rm m]
    \approx& \, 2.3 \times 10^9 \feddnorm \left( \frac{M_1}{10^6 M_{\odot}}\right).
\end{align}
{A derivation of these quantities from the properties of the disk can be found in our Appendix~\ref{app:analytical_models}, and with more details in \cite{Frank92} and \cite{Kocsis:2011dr}.}
The corresponding disk densities are obtained as $\rho = \Sigma/2 H$, and are inversely proportional to the viscosity and the accretion rate.
The disk models that we employ are valid for geometrically thin disks only \footnote{When the thin disk condition is violated (e.g.~for super-Eddington accretion), the disk is better described by a slim-disk solution \citep{Abramowicz:1988sp,Abramowicz:2011xu}.}, where $H\ll r$. { This condition is satisfied for the entire EMRI evolution and all disk parameters explored in this work.}
{Observations of the inner regions of AGN accretion disks around 10$^6 M_{\odot}$ black holes have found electron densities $\log n_e [{\rm cm}^{-3}] \simeq 16 - 18.6$~\cite{Jiang:2019ztr}. In our disk models, these correspond to $\Sigma = n_e m_{\rm p} \times 2 H \simeq 2 \times 10^5 \, {\rm kg}/{\rm m}^2 - 3\times 10^7 \, {\rm kg}/{\rm m}^2$. }

The presence of an accretion disk induces environmental torques that modify the trajectory of the compact object orbiting the MBH. For EMRI sources in the LISA band, such torques are suppressed with respect to the effect of GW emission, but are potentially observable~\citep{Yunes:2011ws,Kocsis:2011dr,Barausse:2014tra,Barausse:2014pra,Derdzinski:2018qzv,Derdzinski:2020wlw}. {Because of this suppression, we can treat environmental torques as a perturbative effect, with the total torque acting on the secondary given by the standard, gravitational torque (see Sec.~\ref{sec:waveforms}), plus the environmental one: $\dot L = \dot L_{ \rm GW} + \dot L_{\rm disk}$.}

{Environmental torques come in a variety of forms, with different dependencies on the disk parameters and the location of the EMRI secondary in the disk. Moreover, analytic models for the torques (and the underlying disks, as discussed above) still suffer from systematic uncertainties. In order to search for environmental effects in the GW signal, we need a flexible, agnostic model. Motivated by torque models in the literature~\cite{Yunes:2011ws,Kocsis:2011dr,Barausse:2014tra}, we propose the following parametrization:}
\begin{align}\label{eq:L_GW}
    \dot L_{\rm disk} = \, &A \, \rnorm^{ n_{\small r}}\, \dot L_{ \rm GW}^{(0)},\\
    \text{with} \quad &A = \mathcal{C}\alphanorm^{ n_{\small \alpha}}  \feddnorm^{ n_{\small f_{\rm Edd}}} \Monenorm^{ n_{\small M_1}},\nonumber
\end{align}
{where we scale the torque by} $\dot L^{(0)}_{\rm GW} = -\frac{32}{5} M_2/M_1(r/M_1)^{-7/2}$, the leading order circular-orbit GW torque, and where $r$ is the distance between the compact object and the MBH.
In this model, the torque is simply characterized by an amplitude $A$ and a radial slope $n_{r}$, which corresponds to a post-Newtonian (PN) order $-n_r$PN. {We further justify the torque model~\eqref{eq:L_GW} in the following.}

{The approximation Eq.~\eqref{eq:L_GW} breaks down within the inner edge of the disk, where accretion-disk torques should wane due to the low gas density.} Modeling this phase is however not necessary: our waveform model also ignores the plunge of the EMRI system, because its duration is only a small fraction of the full inspiral \citep{Kesden:2011ma,PhysRevD.62.124022,Burke:2019yek}.

\begin{table}
\label{tab:torque_params}
\centering
\caption{
Parameters for the torque model of Eq.~\eqref{eq:L_GW} for the migration torques described in Sec.~\ref{sec:accretion_effects}. $\mathcal{C}$ is the overall torque amplitude, $n_r$ sets its dependence on the orbital separation, and $n_{\alpha,f_{\rm Edd},M_1}$ that on the disk viscosity, accretion rate and primary mass.
}
\centering
\begin{tabular}{ l| c|c }
		 & Migration ($\alpha$) 
		 & Migration ($\beta$) \\
		 \hline
		$\mathcal{C}$ & $7.2 \times 10^{-10}$  
		& $8.1 \times 10^{-6}$ \\
		$n_{\small r}$ & 8  
		& $59/10$  \\
		$n_{\small \alpha}$ & -1  
		& $-4/5$ \\
		$n_{\small f_{\rm Edd}}$ & -3   
		& $-7/5$\\
		$n_{\small M_1}$& 1 
		& $6/5$\\
\end{tabular}
\end{table}
Environmental effects in AGN disks may include the disk gravitational potential, the formation of density waves (dynamical friction or planetlike migration, depending on the scales involved), winds, and mass accretion on the primary or the secondary \citep{Barausse:2006vt,Barausse:2007dy,Yunes:2011ws,Kocsis:2011dr,Barausse:2014tra,Barausse:2014pra}. 
In this work, we
focus on migration {because, in preliminary explorations of the dephasing of the EMRI inspiral induced by the aforementioned effects,} we find migration to be the dominant one, in agreement with previous works \cite{Yunes:2011ws,Kocsis:2011dr,Barausse:2014tra,Barausse:2014pra}.

In analogy with planet-planetary disk interactions~\citep{Armitage2011,Paardekooper2022},
{a compact object spiralling through a disk can produce spiral density waves in the surrounding gas, which can, in turn, excite resonances in the disk. Some waves move inward of the secondary's orbit, and some outward. The latter extract angular momentum from the orbit, which shrinks in size. The secondary is said to have undergone \quotes{type I} migration in this case. The EMRI secondary can also undergo} \quotes{type II} migration, if a gap opens in the disk at the radius of the secondary's orbit.

{
In type-I migration, it is sufficient to approximate the disk response with linear perturbation theory~\cite{Goldreich1980}. We use the analytic formula for type-I migration for a Newtonian orbiter in a three-dimensional, isothermal disk from Ref.~\cite{Tanaka2002}, 
\begin{equation}\label{eq:Ldot_mig}
    \dot{L}_{\rm disk} = - c_1 M_2 \eta \Sigma \frac{r^3}{H^2} \quad {\rm with} \quad c_1 =1.4 + 0.5 \gamma,
\end{equation}
where $\eta = M_2/M_1$ is the mass ratio and $\gamma=3/2$ ($\alpha$ disk) or $\gamma=-3/5$ ($\beta$ disk). 
This takes precisely the form Eq.~\eqref{eq:L_GW}, with amplitude and powers given in Table \ref{tab:torque_params}. This torque is negative (i.e., the migration is inward) and decreasing along the inspiral, as can be seen substituting the expressions for the surface density and disk height Eq.~\eqref{eq:Sigma_H}. Being proportional to the disk density and inversely proportional to the disk height Eq.~\eqref{eq:Sigma_H}, the migration torque Eq.~\eqref{eq:Ldot_mig} scales with inverse powers of 
the disk parameters $\alpha$ and $f_{\rm Edd}$.

This formula was derived in the context of planetary disks, where it successfully reproduces numerical studies~\cite{Tanaka2002,Casoli2009,Paardekooper2009} {(see also the review \cite{Baruteau2014})}. Unfortunately, numerical simulations have not yet reproduced the realistic conditions of an EMRI migrating in a nuclear accretion disk, such as relativistic inspirals and turbulent, radiating, nonisothermal disks. However, hydrodynamical simulation for higher mass-ratio systems (intermediate mass ratio inspirals) on a post-Newtonian trajectory~\cite{Derdzinski:2018qzv,Derdzinski:2020wlw} found that the analytic formula of Ref.~\cite{Tanaka2002} successfully reproduces the torque amplitude (within an order of magnitude) and its dependence on the radius.}

{The type-I torque described above does not take into account the effect of gas on horseshoe or corotation orbits near the inspiraling body. This is sometimes referred to as ``type-III" migration, and has been shown to lead to rapid runaway migration in the context of protoplanetary dynamics \cite{Malik_2015}. This corotation torque should only be significant when the mass in the disk near the EMRI orbit is comparable or greater than the secondary mass. In our system, the local disk mass is orders of magnitude smaller than the secondary mass while the system emits in the LISA band~\cite{Kocsis:2011dr}. If our EMRI enters the type-III regime, our type-I model is likely be a conservative choice for the amplitude of the torque.}

Because of the turbulent nature of AGN disks {and the rapid evolution of gas around the inspiraling body}, migration can oscillate over time and average to both negative and positive values~\citep{Derdzinski:2020wlw}. Here we use the positive, constant overall torque amplitude predicted in Ref.~\cite{Tanaka2002}. {Oscillations around the average migration torque were found to be suppressed for smaller mass ratios~\citep{Derdzinski:2020wlw}} and could be modeled separately; see e.g., Ref.~\cite{Zwick:2021dlg}. 

{Type-II migration takes place when the orbiter's tidal torques remove gas from the orbit faster 
than viscous forces refill it, creating a lower-density annular gap in the disk. The conditions for gap opening in the two disk models we consider are given in Ref.~\cite{Kocsis:2011dr}, Eqs.~(42)-(46). While migration torques are of type-I in most of the parameter space relevant for EMRI GW observations, the type-II regime may also occur for some disk-parameter and EMRI radii.
In this regime, previous work on EMRIs}~\citep{Kocsis:2011dr,Barausse:2014tra} used the quasistationary approximation of Syer and Clarke~\cite{SyerClarke1995}. {However, contrary to the standard paradigm, recent simulations have shown that gas can flow through the gap and that long timescales might be required for the torque to reach the quasistationary estimate \citep{Duffell_2014,Durmann2017,Kanagawa2018,Scardoni2020}.} To avoid overestimating the torque {and introducing additional model dependence in our analysis} when entering the type-II regime, we use the type-I formula whenever migration is active, as suggested by recent numerical simulations of migration in intermediate mass-ratio inspirals~\citep{Derdzinski:2018qzv,Derdzinski:2020wlw}.

Migration can also be quenched at certain radii in the disk \new{(see condition (113) in~\citep{Kocsis:2011dr})}, a transition which could not be described by Eq.~\eqref{eq:L_GW}. We verify that the EMRI does not cross the quenching radius in the LISA band across the parameter space explored in this work.

{To summarize, in the following we use the type-I migration torque of Table~\ref{tab:torque_params} to simulate the impact of the disk on the compact object, and we recover it in terms of the agnostic parametrization of Eq.~\eqref{eq:L_GW}. This formulation is crucial for three reasons. Firstly, it reduces the parameter space dimension and allows us to perform parameter estimation efficiently. It compresses the information encoded in $f_{\rm Edd},\alpha,\epsilon, \mathcal{C}, n_r, n_\alpha, n_{f_{\rm Edd}}, n_{M_1}$ into the amplitude $A$ and slope $n_r$, which characterize the size of the effect and the disk type, respectively. Secondly, this parametrization allows us to extract the astrophysical parameters \emph{a posteriori}. Once the disk type is determined through the slope $n_r$, we can infer the astrophysical parameters from the measured amplitude $A\rightarrow (f_{\rm Edd},\alpha,\epsilon)$ and Table~\ref{tab:torque_params}. Furthermore, if a more accurate model for the mapping $A\rightarrow (f_{\rm Edd},\alpha,\epsilon)$ is available, we can just update the mapping without repeating the slow parameter estimation procedure. Lastly, the constraints obtained with this parametrization can also be used to test general relativity. In fact, for some fixed value of $n_r$ (or $-n_r$PN order) the amplitude $A$ can represent the size of a GR deviation through the parametrized
post-Einsteinian expansion \citep{Cornish:2011ys} commonly used to test general relativity with gravitational waves \citep{LIGOScientific:2021sio}.
}

\section{EMRI Waveforms \label{sec:waveforms}}
The GW emission from EMRI systems will only provide measurements of their parameters with unprecedented precision~\citep{Barack:2003fp,Glampedakis:2005cf,Barack:2006pq,Babak:2017tow} if their waveforms are accurately modeled.
EMRI waveform models rely on perturbative solutions in which the Einstein equations are expanded about the limit of small mass ratio $\eta = M_2/M_1\ll 1$ \citep{Barack:2018yvs}. 
In this limit, the orbital evolution of the compact object in vacuum is governed by the Kerr geodesic equations with a forcing term, called the gravitational self-force \citep{PhysRevD.55.3457,PhysRevD.56.3381}. The self-force takes into account the finite size and mass of the body and its backreaction on the background Kerr spacetime. As a result, the orbit of the compact object slowly decays into the MBH due to the emission of GWs. 
The presence of an accretion disk can further enhance the decay of the compact object's orbit.

In this work we employ the \texttt{FastEMRIWaveform} (\texttt{few}) waveform model package \citep{Chua:2018woh,Chua:2020stf,michael_l_katz_2020_4005001,Katz:2021yft} and modify the trajectory evolution to take into account the presence of environmental effects.
We model the evolution of Kerr circular-equatorial orbits at adiabatic order \footnote{Since the environmental effects considered in this work appear at negative PN orders, we do not expect post-adiabatic corrections to significantly affect our results.} by interpolating the Teukolsky fluxes $\dot{L}_\text{GW}$ using the \texttt{Teukolsky} package in the \texttt{Black Hole Perturbation Toolkit} \cite{BHPToolkit}. We model the disk-induced effects by writing the rate of change of angular momentum as $\dot L = \dot{L}_\text{GW} + \dot{L}_\text{disk} $, with the environmental contribution from Eq.~\eqref{eq:L_GW}. This implementation provides a fast and accurate adiabatic trajectory that can be fed into a waveform-generation formalism through the augmented analytic kludges \citep{Chua:2017ujo}. 

{We assume that the disk rotation axis is aligned with the spin of the primary BH. While this might not be the case for the outer regions of a disk, we expect the inner regions to be aligned by the Bardeen-Petterson effect \cite{Bardeen:1975zz, Bogdanovic:2007hp, Perego:2009cw}.} We also restrict our analysis to circular and planar orbits (i.e., in the plane of the disk), as disk-induced density waves are expected to damp the EMRI inclination and eccentricity long before it enters the LISA band~\citep{Cresswell2007,Bitsch2010,McKernan2012,MacLeodLin2020,Pan:2021oob}. {Compact bodies are also likely to form preferentially within the disk, giving rise to planar EMRIs~\cite{Derdzinski:2022ltb}.}
We conservatively assume prograde orbits, as retrograde orbits can suffer from even larger disk-induced torques~\citep{Ostriker:1998fa,Barausse:2007ph,Barausse:2007dy}. Compact objects formed in the disk are expected to be on prograde orbits~\citep{Syer1991,Fabj2020}, and prograde EMRIs can be seen to much greater distances, so we expect these to dominate detected LISA events \citep{Babak:2017tow}.
However, our implementation is flexible enough to model generic orbits once prescriptions for disk-induced effects become available for this scenario.

A quasicircular equatorial EMRI waveform is described by the masses of the two bodies $M_1, M_2$; the dimensionless spin parameter $a$ of the primary; the initial phase and radius, $\Phi_0$ and $r_0$; the azimuthal and polar angles, $\theta_K$ and $\phi_K$, describing the orientation of the spin-angular momentum vector of the MBH; the polar and azimuthal sky location angles, $\theta_S$ and $\phi_S$, given in the Solar System barycenter frame; and the luminosity distance $d_L$. 
The presence of accretion effects introduces two additional parameters, $A$ and $n_r$. We will refer to $(\theta_S,\phi_S, \theta_K, \phi_K, d_L)$ as extrinsic parameters, and to $(M_1,M_2, a, r_0, \Phi_0)$ as intrinsic parameters.

The speed of generation of \texttt{few} waveforms allows us to perform Bayesian analyses with Markov chain Monte Carlo (MCMC) methods. In this work, we use {our version of parallel-tempered MCMC, based on both \texttt{emcee} \citep{ForemanMackey2013Emcee} and \texttt{ptemcee} \citep{Vousden_2015} with five adaptive temperatures and 16 walkers per temperature. The sampler was run until chains were longer than $50\hat{\tau}$, where $\hat{\tau}$ is the average autocorrelation time, determined across chains~\cite{autocorrelation}. This ensures the convergence of the samples. We sample using a} standard Gaussian likelihood $\mathcal{L} = \exp{-1/2 \braket{s - h(\vb*{\lambda})}{s- h(\vb*{\lambda})}}$. Here we assume stationary Gaussian noise, we define the data stream $s$, GW signal $h(\vb*{\lambda})$ with parameters $\vb*{\lambda}$, and the inner product 
\begin{equation}
    \braket{a (t)}{b (t)} =4 \Re \int _{0} ^\infty \frac{\tilde{a} ^* (f) \tilde{b} (f) }{S_n (f)} \, \dd f 
\end{equation}
weighted by the LISA power spectral density (PSD), $S_n (f)$ \cite{Babak:2021mhe,robsonConstructionUseLISA2019} and tilde means Fourier transform. In all the studies reported, we inject noise-free data streams since this yields results averaged over noise realizations. {A noise injection would not affect the shape of the posteriors, but only shift the recovered parameters away from the injected ones. Since our conclusions are based on the measurement precision obtained by the width of the posteriors, this choice does not affect our results.}
We assume uniform priors in all parameters, restricted to a narrow range around the true parameters. All posteriors shown in this work have a support that is much tighter than the prior ranges.

{Because of the high computational cost of the analysis} we focus on a reference EMRI with masses $M_1=10^6 M_\odot$ and $M_2=50 M_\odot$, and primary spin $a=0.9$. {Black holes of 50$M_{\odot}$ or more have already been observed in binaries by LIGO and Virgo~\cite{LIGOScientific:2021psn} Collaborations, although with a lower rate compared to lighter BHs.} 
Our choice of a relatively heavy secondary is motivated by the fact that black holes are expected to grow via accretion when originating in an AGN disk~\citep{Tagawa2022,Derdzinski:2022ltb} (for possible formation scenarios see Ref.~\citep{Levin:2003ej,Levin:2006uc}). 
{Depending on the accretion rate and the time available for growth, a much larger secondary mass could be reached, in principle. However, we decided conservatively to set $M_2=50 M_\odot$ to be within the range $M_2\in [10, 60]M_\odot$ reported in the LISA Science Requirements Document \citep{LISASciRD}, and because a larger secondary mass would only improve our results. Since the SNR grows linearly with $M_2$, a larger secondary mass would provide tighter constraints on environmental effects. Furthermore, due to the negative-PN nature of the disk torques, for a fixed four years inspiral the compact object could explore farther regions from the central MBH, yielding tighter constraints on environmental effects.}

We set the {EMRI initial radius at the beginning of the LISA observation to $r_0 = 15.482 \, M_1$ ($= 7.45 \times 10^{-7} \rm{pc}$), such that the compact object plunges into the MBH after 4 years}. {We assume that LISA will observe the signal in its entirety, consistently with the nominal science mission duration~\cite{LISASciRD} and neglecting gaps in the detector's data.}
We fix the other parameters to randomly chosen values of $(\Phi_0, \theta_S,\phi_S, \theta_K, \phi_K) = (3.0, 0.54, 5.36, 1.73, 3.2)$ and a luminosity distance $d_L = 1.456$ Gpc \footnote{This luminosity distance corresponds to the redshift $z=0.276$ for a flat $\Lambda$CDM cosmology with Hubble constant $H_0 = 67.74 \, {\rm km/s/Mpc}$ and matter density $\Omega_M=0.3075$.}, chosen to give a signal-to-noise ratio SNR$=\sqrt{\braket{h}{h}}$=50.
A detailed investigation of how different configurations affect the detectability of environmental effects with EMRIs is beyond the scope of this work. {We stress that this choice of parameters is typical for observable EMRIs and well within the range provided in the LISA Science Requirements Document\citep{LISASciRD}, where a ``Golden'' EMRI system has SNR$ > 50$, spin $a>0.9$, and a prograde orbit.
}

\section{EMRI parameter estimation with environmental torques \label{sec:measuring_accretion_properties}}

There are three ways in which environmental effects could be relevant to GW observations of EMRIs:
\begin{enumerate}
    \item If environmental effects are absent or too weak to affect the waveform, EMRI observations could be used to set an upper limit on the torque amplitudes. The same analysis can be used to forecast the detectability of a given effect, and it only requires knowledge of the torque's radial dependence (Sec.~\ref{sec:constraints});
    \item If environmental effects are strong enough and can be modeled with a simple power of the radius, their presence can be detected (Sec.~\ref{sec:detection}).
    Provided reliable torque models, we can use such detections to constrain some of the properties of the disk, especially in coordination with electromagnetic observations;
    \item If we analyze GW data ignoring environmental effects, and these are relatively strong, EMRI parameter estimation will be biased (Sec.~\ref{sec:bias}).
\end{enumerate}
We discuss each of these scenarios in the following sections.

\subsection{Constraints on environmental torques}\label{sec:constraints}
\begin{figure}[ht]
\centering
\includegraphics[width=\linewidth]{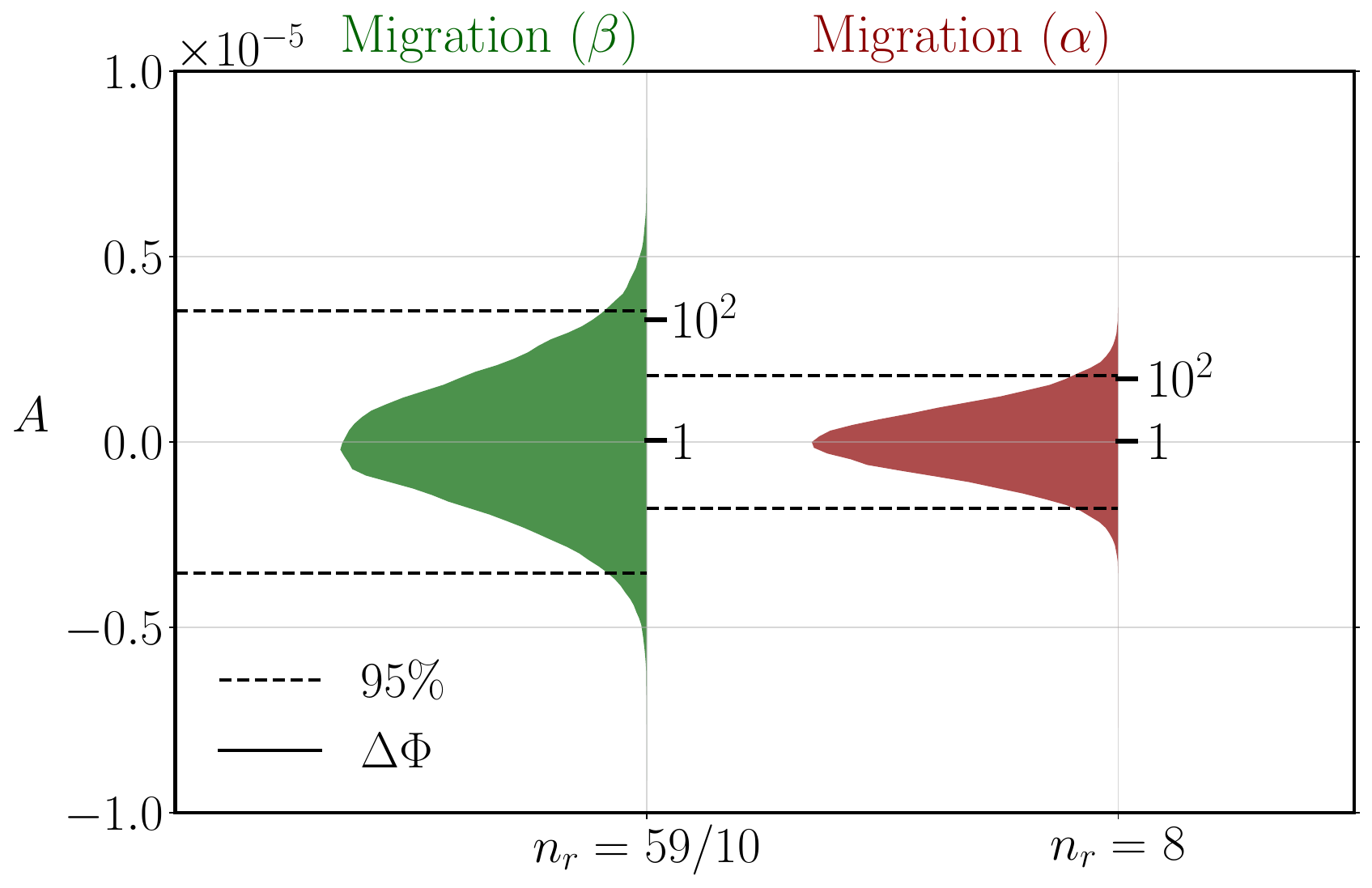}
\caption{Marginalized posterior of the amplitude of the migration torque in {$\beta$ (left-hand side green) and $\alpha$ (right-hand side red) disks} for null injections ($A=0$). {The two disk models predict different radial dependences for the torque, $n_r=59/10$ for $\beta$ and $n_r=8$ for $\alpha$ disks (see Table~\ref{tab:torque_params} for the dependence on accretion rate $f_\textrm{Edd}$ and viscosity $\alpha$). We consider a torque detectable} if its amplitude is above the $95\%$ percentile line (dashed black line). Assuming an accretion-disk model, this implies a bound on combinations of $\alpha$ and $f_{\rm Edd}$ (for $\epsilon=0.1$).
{We also tick the amplitudes at which the phase difference between a disk-embedded ($A\neq0$) and vacuum system ($A=0$) reaches $\Delta\Phi=1$, 100 at the end of the inspiral.} The EMRI is observed by LISA 4 years before plunge with SNR$=50$, and has central black hole spin $a=0.9$ and masses $M_1=10^6 M_{\odot}$, $M_2=50  M_{\odot}$.
}
\label{fig:violinplot}
\end{figure}
Assuming the disk models and migration torque from Sec.~\ref{sec:accretion_effects}, we ask, when are environmental effects large enough to be detectable?

The dephasing $\Delta\Phi$ between vacuum and matter-influenced waveforms is a commonly used metric for detectability, with $\Delta\Phi\gtrsim 1$ a reference threshold. However, a significant dephasing is a necessary, but not sufficient condition for detectability. Using this threshold as a proxy can lead to overestimating detectability, implying that it must be used as a qualitative, but not quantitative metric.
To better assess the detectability of environmental effects, we perform a Bayesian parameter estimation over the EMRI parameters and the torque amplitude for a vacuum signal. This approach is similar to tests of GR with parametrized post-Einsteinian expansions of the phase~\citep{Yunes:2009ke}. We sample over the intrinsic EMRI parameters as well as the torque amplitude, fixing the slope $n_r$ and the extrinsic parameters for computational efficiency. Sampling over the extrinsic parameters does not affect our results, because these are not strongly correlated with the amplitude $A$ (see Sec.~\ref{sec:detection}).
 
The resulting posterior for $A$ has a variance that carries information about the detectability of the effect (slope) selected. {In this work, we assume that any effect with an amplitude falling inside the 95\% (symmetric) bound is $2\sigma$-consistent with noise, under our assumed (flat) prior. Therefore any effect with $A<A_{95\%}$ is considered indistinguishable from an EMRI in vacuum, and viceversa. In reality, the indistinguishability of the two hypotheses should be assessed by calculating the Bayes factors between the disk and vacuum (possibly eccentric and inclined) hypotheses when injecting different amplitudes. This is a computationally expensive analysis which is beyond the scope of this work.
} 

We show the results obtained for our reference EMRI in Fig.~\ref{fig:violinplot} for the two slopes predicted for the migration torque by the two disk models (Table \ref{tab:torque_params}). We find that the symmetric 95\% bounds for $\beta$ and $\alpha$ disks are $A_{95\%}=\{3.5, 1.8\}\times 10^{-6}$.
The constraint on $A$ becomes tighter as the slope increases, consistent with the fact that larger slopes correspond to more negative PN orders probed by the long inspiral.

In Fig.~\ref{fig:violinplot}, we also indicate at what amplitude the torque induces a dephasing of $1$ and $10^2$ rad. This shows torques may not be detected even for dephasings much larger than a radian, confirming that the previously adopted requirement $\Delta\Phi \gtrsim 1$ overestimates the detectability of environmental effects.

A more accurate estimate of detectability consists in requiring $A>A_{95\%}$. For migration in $\beta$ disks, we find that this implies
\begin{equation}\label{eq:detectability_beta}
    \left(\frac{\alpha}{0.1}\right)^{-4/5}\feddnorm^{-7/5} > 0.4 \quad \text{($\beta$ disks).}
\end{equation}
For typical parameters $\alpha=f_{\rm Edd}=\epsilon=0.1$ the effect is detectable, as found in \cite{Yunes:2011ws}. The dependence on the disk accretion rate and viscosity implies that lower values would lead to more observable effects.

Referring to the same typical parameters, Ref.~\cite{Kocsis:2011dr} found that migration in the more realistic $\alpha$ disks could not be observed. 
In this case, we find the condition
\begin{equation}
\label{eq:detectability_alpha}
    \left(\frac{\alpha}{0.1}\right)^{-1}\feddnorm^{-3} > 2 \times 10^3 \quad \text{($\alpha$ disks).}
\end{equation}
This implies that, while $\alpha=f_{\rm Edd}=\epsilon=0.1$ are still excluded, the effect could be detected in disks with lower accretion rates and/or viscosity. For instance, for values $\alpha \lesssim 0.05$ the effect is measurable for $f_{\rm Edd}\lesssim 0.01$, $\epsilon=0.1$. These values are within what is expected from global simulations \citep{Jiang:2019bxn} and x-ray observations of AGNs \citep{Aird:2017cbs}. 

{The conditions \eqref{eq:detectability_alpha} and \eqref{eq:detectability_beta} are based on the ``typical" parameters for EMRI in accretion disks used in our signal injection. These detection thresholds would degrade, for instance, if LISA observed a shorter portion of the inspiral.}

While here we present the bounds for the radial slopes predicted for the migration torque in two disk models, this analysis can be easily generalized to other effects at different slopes, or PN orders. 
In Fig.~\ref{fig:bounds_A_nr} of Appendix~\ref{app:upper_limit}, we present the symmetric $95 \%$ bound on the torque amplitude as a function of the slope for $-4\leq n_r \leq 10$. We find that for $n_r\gtrsim3$ the upper limit follows the relation
\begin{equation}
\log_{10} A_{95\%} = -4.63 -0.14 \, n_{r} .
\end{equation}
This can be readily used as an approximate bound for other effects.

\subsection{Detection of environmental torques}\label{sec:detection}

\begin{figure}
\centering
\includegraphics[width=\linewidth]{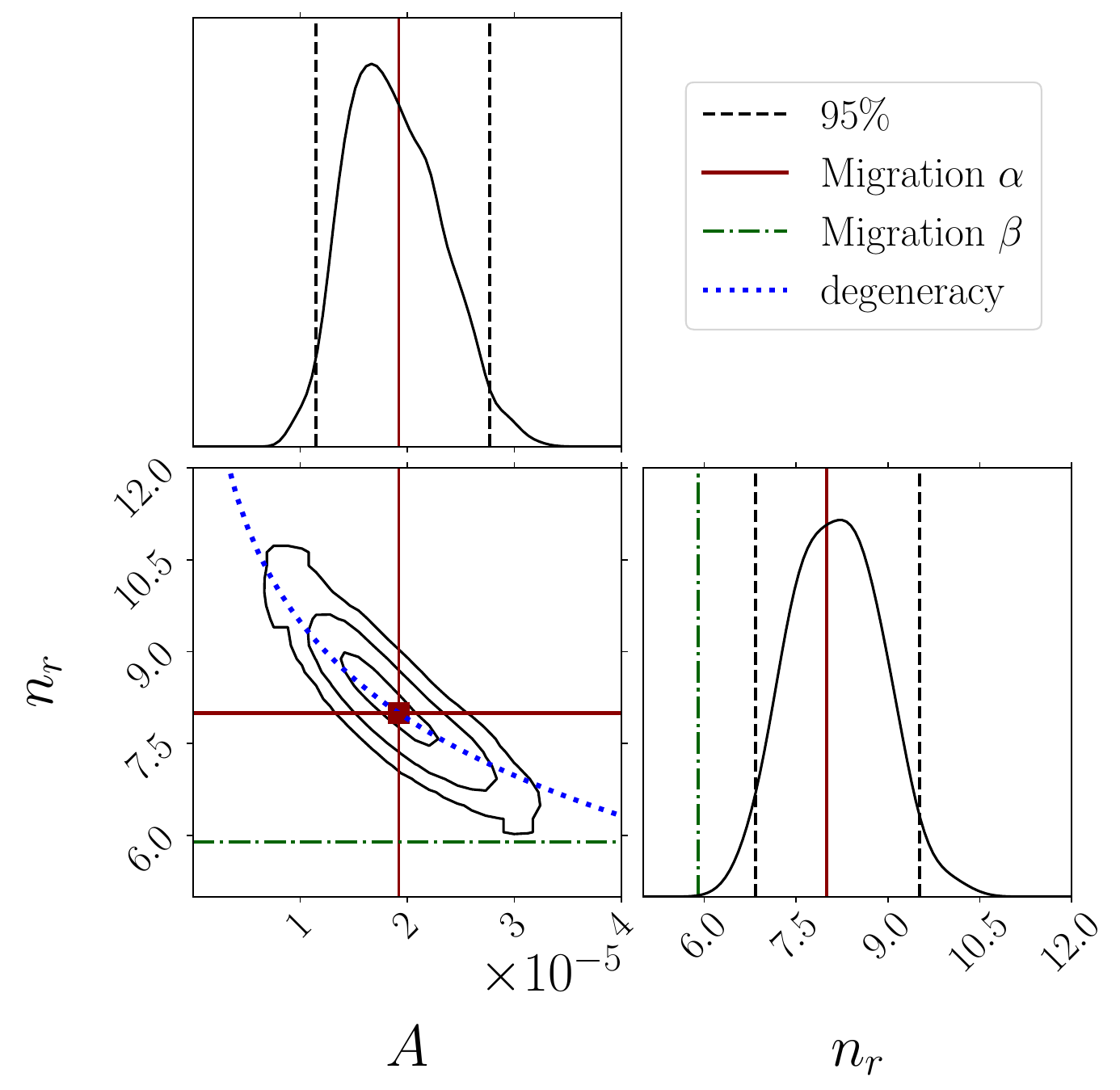}
\caption{Posterior distribution of the torque parameters obtained from the GW observation of an EMRI system affected by migration in an $\alpha$ disk. We inject a torque amplitude $A=1.92\times 10^{-5}$, consistent with disk parameters e.g.~$f_{\rm Edd}=0.005,\alpha=0.03$, and a torque slope $n_r=8$ as predicted for migration in $\alpha$ disks. 
The medians and the 95\% credible intervals of the inferred amplitude and slope are $A=1.84^{+0.9}_{-0.7} \times 10^{-5}$ and $n_r = 8.15 ^{+1.37} _{-1.34} $.
Migration in $\beta$ disks predicts a slope $n_r=59/10$ (green dash-dotted line), which is excluded at more than $95\%$ (black dashed lines). 
The expected degeneracy between the amplitude and slope parameters is shown as a blue dotted line.
The EMRI configuration is the same as in Fig.~\ref{fig:violinplot} and the complete corner plot can be found in Appendix~\ref{app:full_posterior} in Fig.~\ref{fig:fullcorner_plot_alpha}. {The contours show the 1$\sigma$,2$\sigma$,3$\sigma$ Gaussian credible regions.}
}

\label{fig:corner_plot}
\end{figure}

We now consider the more optimistic scenario in which disk effects are above the detection threshold. Can we measure and characterize environmental torques? Provided a reliable model, can we infer the parameters of the disk hosting the EMRI?

We again investigate this scenario by performing a Bayesian analysis for our reference EMRI. We limit our study to the more realistic $\alpha$ disks, injecting a signal with slope $n_r=8$ and amplitude $A=1.92\times 10^{-5}$.
Choosing $\alpha=0.03$ as in Ref.~\cite{Derdzinski:2018qzv}, this amplitude corresponds to a value $f_{\rm Edd}=0.005$ (with $\epsilon=0.1$){, and to a surface density $\Sigma_\alpha \approx 3.6 \times 10^5 \,$ kg/m$^2$ at $r=10\, M_1$, consistent with observations~\cite{Jiang:2019ztr}}. 
However, our agnostic procedure means the constraint applies to all combinations of $\alpha$, $f_{\rm Edd}$ and $\epsilon$ resulting in the same amplitude through Eq.~\eqref{eq:L_GW}.

The full posterior can be found in Appendix~\ref{app:full_posterior} in Fig.~\ref{fig:fullcorner_plot_alpha}.
As expected for typical EMRI observations, the intrinsic parameters are measured with $\sim 10^{-5}$ relative precision. The sky localization error is $\Delta\Omega = 1.8$ deg$^2$ \citep{Cutler:1997ta,Lang:2006bsg} and the relative luminosity distance error is $6\%$ \footnote{We ignore errors on the luminosity distance due to lensing and peculiar velocity because they are an order of magnitude smaller \citep{Speri:2020hwc}.}. The comoving volume error for this source is $\approx 5\times 10^{-5} \,{\rm Gpc}^{3}$, which means that this source would be promising for follow-up electromagnetic campaigns.

The marginalized posteriors of the environmental parameters $A$ and $n_r$ are shown in Fig.~\ref{fig:corner_plot}. 
The posterior of $A$ is inconsistent with $A=0$ at more than $3 \sigma$, as expected from our previous discussion.
Moreover, the measurement of $n_r$ can be used as a discriminator between disk (or torque) models. 
In our example, we can distinguish that the injection is due to migration in $\alpha$ and not $\beta$ disks, since the latter predicts a slope $n_r=59/10$ (dash-dotted green line) that is excluded by the posterior at more than $95\%$.

\begin{figure*}
\centering
\includegraphics[trim={1.0cm 0.cm 3.5cm 0.cm},width=0.44\textwidth]{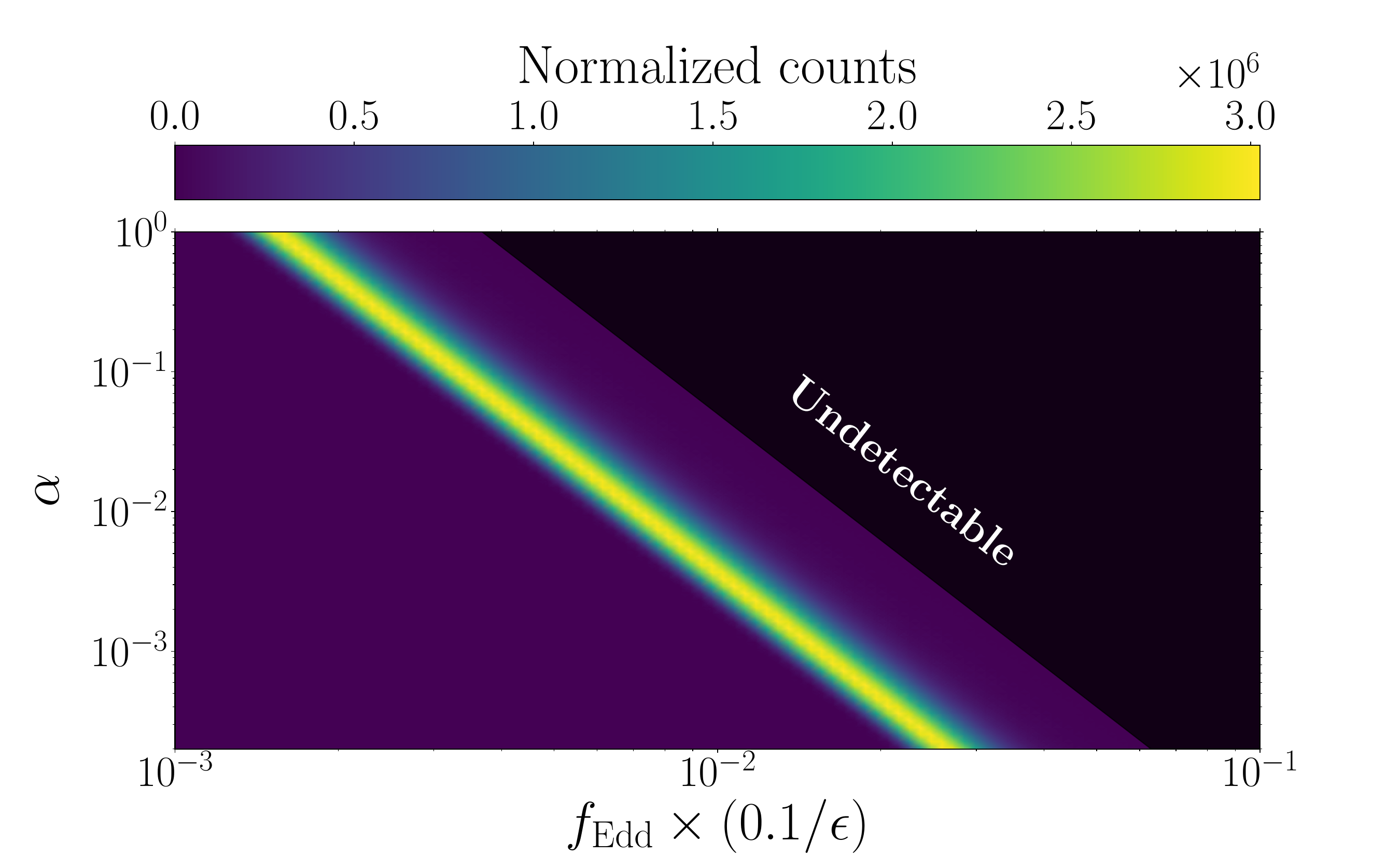}
\hspace{0.05\textwidth}
\includegraphics[trim={1.0cm 0.cm 3.5cm 0.cm},width=0.44\textwidth]{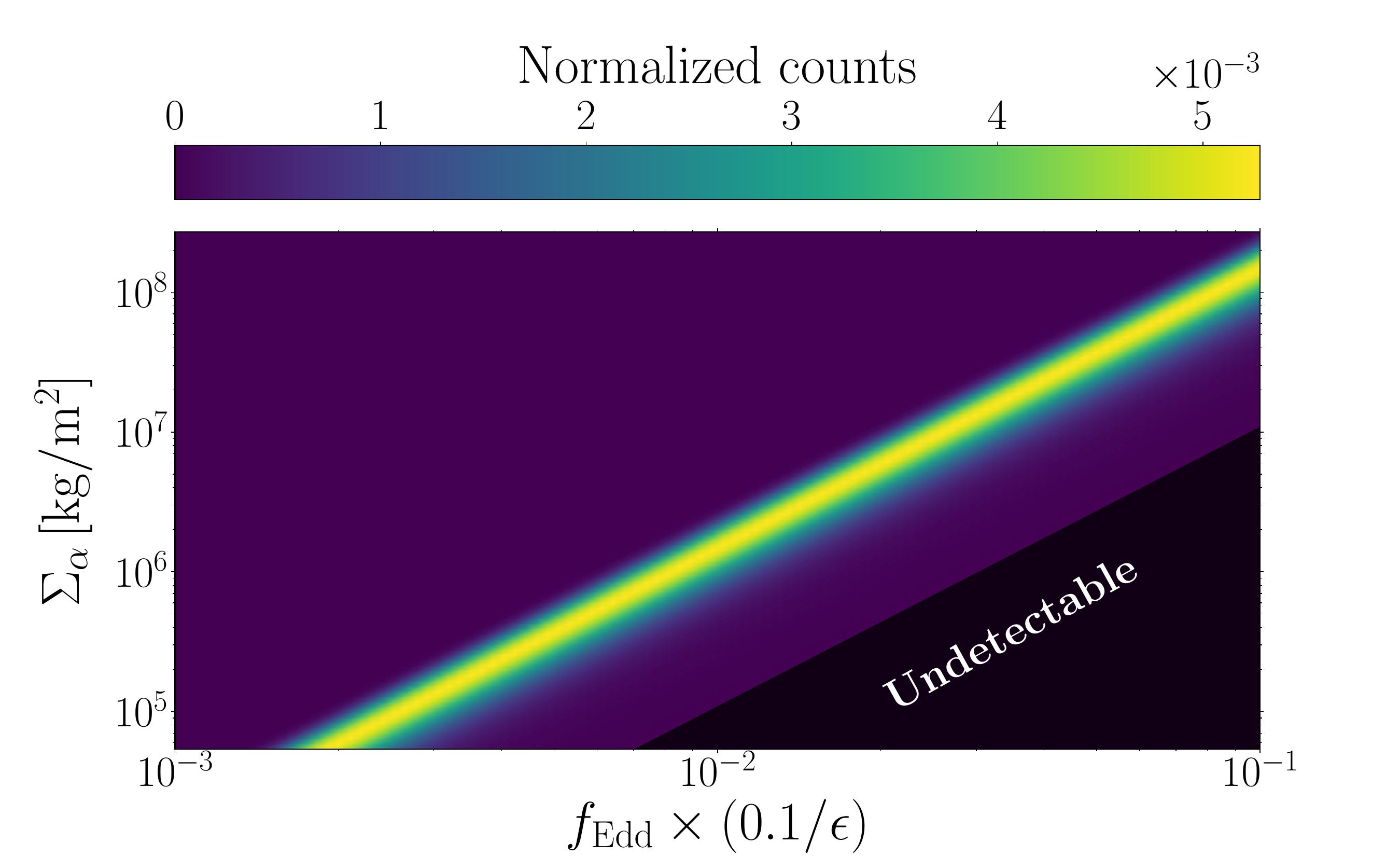}
\includegraphics[trim={1.0cm 0.cm 2.5cm 0.cm},width=0.47\textwidth]{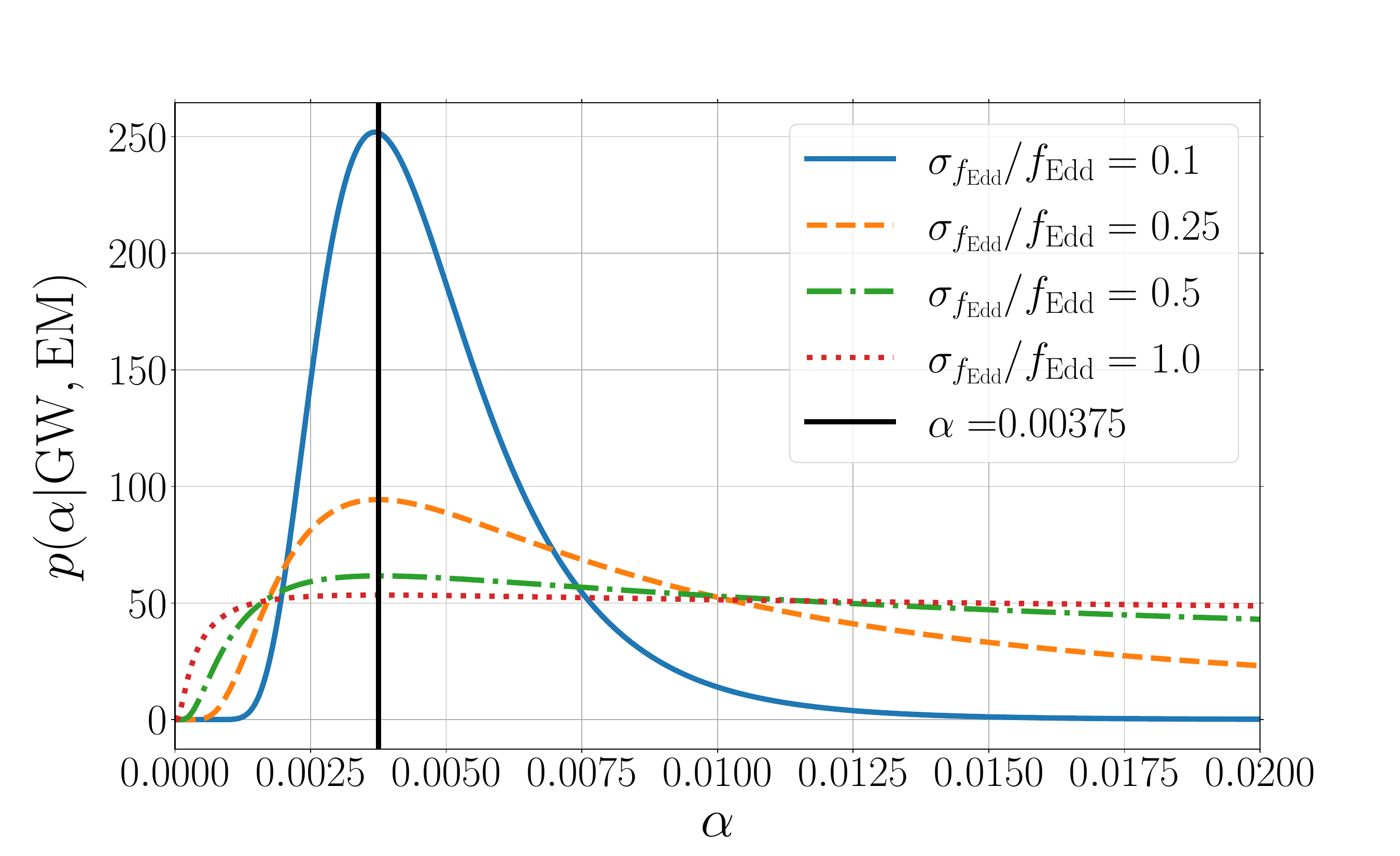}
\hspace{0.015\textwidth}
\includegraphics[trim={1.0cm 0.cm 2.5cm 0.cm},width=0.47\textwidth]{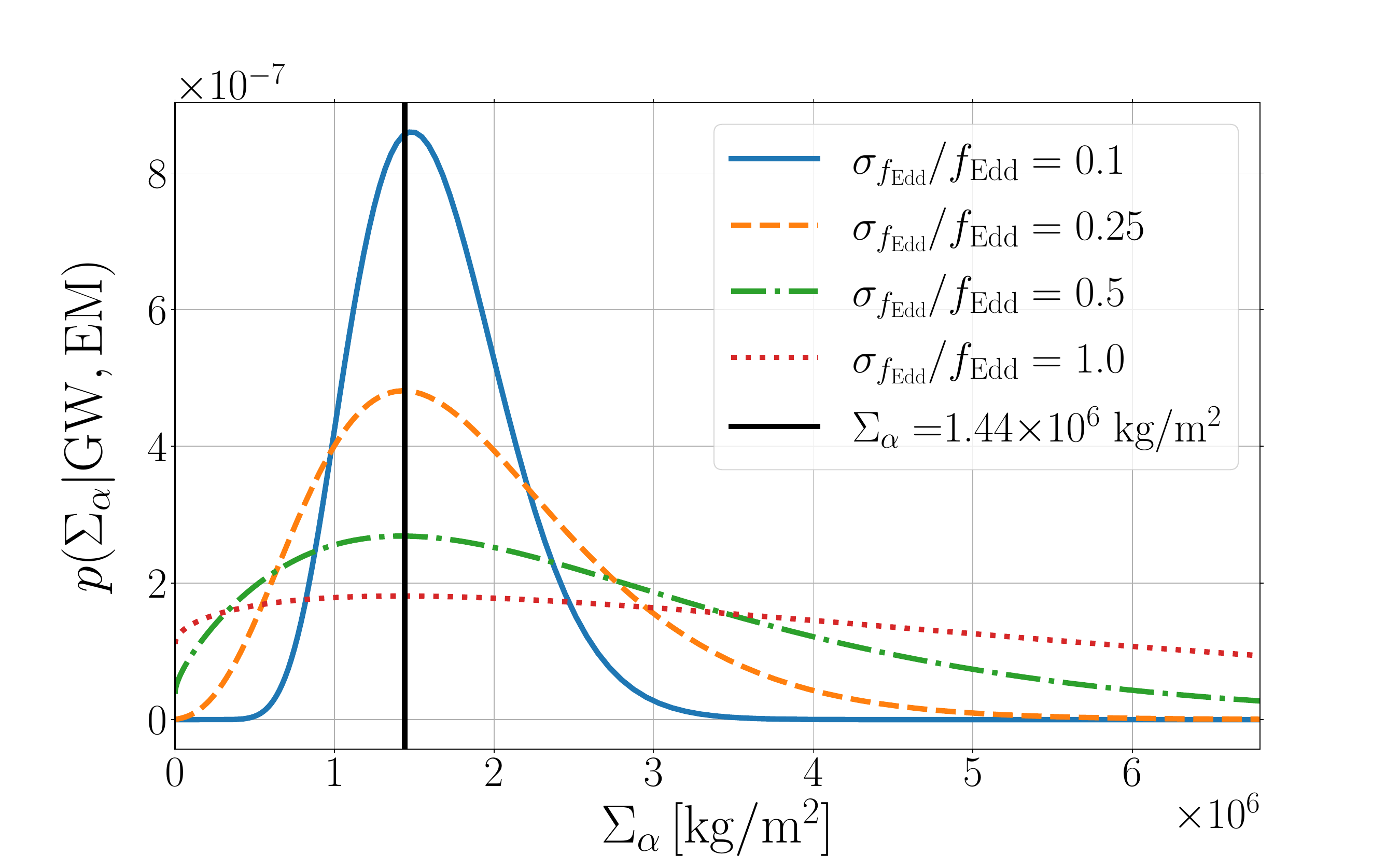}
\caption{\emph{Upper panels}: derived posterior distribution of the disk accretion rate $f_{\rm Edd}$ and the disk viscosity $\alpha$ and (left) \new{or the disk surface density (right)} obtained from an EMRI GW observation. The samples of primary mass $M_1$ and torque amplitude $A$ from Figs.~\ref{fig:corner_plot} and \ref{fig:fullcorner_plot_alpha} are converted into a constraints on viscosity $\alpha$ and accretion rate $f_{\rm Edd}$ through the relation \eqref{eq:from_A_M_to_f_a} \new{and to $\Sigma_\alpha$ through Eq.~\eqref{eq:Sigma_H} at $r=10M_1$}. The black shaded area shows the undetectable region according to the criterion in Eq.~\eqref{eq:detectability_alpha}.
\emph{Lower panels}: posterior distribution of the disk viscosity \new{(left) or disk surface density (right)}, assuming that the disk accretion rate is determined with uncertainty $\sigma_{f_{\rm Edd}}$ through an EM follow-up observation. Here we assume $A=1.92\times 10^{-5}$, consistent with $f_{\rm Edd}=0.01$, $\alpha=0.00375$ or $\Sigma_\alpha=1.44 \times 10^6 $~kg/m$^2$.
}
\label{fig:alpha_fEdd}
\end{figure*}

We find a strong degeneracy between the amplitude $A$ and slope $n_r$, with correlation coefficient $-0.97$. This means that the waveform template does not change significantly if we vary the amplitude $A$ and slope $n_r$ at the same time along such a correlation. The reason for the degeneracy is that the environmental torque has the largest impact on the waveform at the beginning of the inspiral, namely when the radius roughly coincides with the initial separation $r\approx r_0$. As a result, the torque is well approximated by a constant $ \dot L_\text{disk} (r_0) \approx A \, r_0^{n_r} $.
This gives rise to the observed correlation, as shown in Fig~\ref{fig:corner_plot} (blue dotted line). This degeneracy is characteristic of this model and, therefore, could be used as a model consistency check.

The correlation coefficients of the amplitude and slope with the intrinsic parameters are of the order $\sim 0.6-0.7$, whereas with the sky localization and distance are of order $\sim 0.2$ and $0.08$, respectively. As expected, extrinsic parameters are not strongly correlated with the accretion parameters.

Our analysis shows that a sufficiently strong disk torque can be detected with a model-independent template, as long as the torque can be described as a power law of the radius.
Provided that we have a trusted model for the torque slope and amplitude, the latter carries further information about the effect, which can be used in conjunction with measurements of the mass to extract the values of $\alpha$ and $f_{\rm Edd}$ consistent with the observation.
This is represented in the top left panel of Fig.~\ref{fig:alpha_fEdd} as a derived distribution from the samples for $A$ and $M_1$ in Figs~\ref{fig:corner_plot} and \ref{fig:fullcorner_plot_alpha}. These are related to $\alpha$ and $f_{\rm Edd}$ through
the second line of Eq.~\eqref{eq:L_GW}:
\begin{equation}
\label{eq:from_A_M_to_f_a}
\frac{A}{\mathcal{C}} \Monenorm^{-n_{\small M_1}}= \alphanorm^{ n_{\small \alpha}}  \feddnorm^{ n_{\small f_{\rm Edd}}}\, .\nonumber
\end{equation}
We also show the undetectable region according to the criterion in Eq.~\eqref{eq:detectability_alpha} (black-shaded region).
We find that purely GW observations can single out a narrow region of the parameter space $(\alpha,f_{\rm Edd})$, or equivalently $(\Sigma_\alpha,f_{\rm Edd})$. \new{Note that this parameter space region is determined through the aforementioned relation. Therefore, if there was an analytic prescription defining the validity region of Eq.~\eqref{eq:L_GW}, it would be possible to further constrain the parameter space of $(\alpha,f_{\rm Edd})$. Extremely low (large) values of $\alpha$ ($\Sigma_\alpha$) should not be misinterpreted as actual possible constraints, but only as a consequence of the degenerate measurement of $\alpha,f_{\rm Edd}$ ($\Sigma_\alpha,f_{\rm Edd}$) consistent with the amplitude torque $A$.}

In order to break the degeneracy and fully characterize the disk, we need electromagnetic observations.
If the host AGN for this EMRI is identified in a follow-up campaign, electromagnetic observations across the spectrum could be used to determine the bolometric luminosity of the central engine. The bolometric luminosity, together with the GW measurement of the central BH mass and spin, could be used to determine the underlying efficiency and accretion rate of the AGN. Finally, this multimessenger determination of the accretion rate $f_{\rm Edd}$ could be used to extract the disk viscosity $\alpha$ from the joint posterior provided by the GW analysis.
We give a concrete example of this procedure for $f_{\rm Edd}=0.01$ and $\alpha=0.00375$ consistent with amplitude $A=1.92\times 10^{-5}$ \new{and a surface density $\Sigma_\alpha \approx 1.44 \times 10^6 $~kg/m$^2$ at $r=10\, M_1$, well within observational limits~\cite{Jiang:2019ztr}}. We use the numerical fits provided in \cite{Madau:2014pta} to relate the (observable) bolometric luminosity to the intrinsic accretion rate and radiative efficiency \footnote{Note that we use different definitions for the accretion rate compared to Ref. \cite{Madau:2014pta}.}. In the lower panels of Fig.~\ref{fig:alpha_fEdd}, we show the constraints obtained on the viscosity (or on the disk surface density) when the AGN accretion rate is inferred with error $\sigma_{f_{\rm Edd}}$ \footnote{At the relatively low accretion rates considered here, the bolometric luminosity is proportional to the Eddington ratio $f_{\rm Edd}$. The BH mass and spin determined through the GW signal have negligible uncertainty, see Fig.~\ref{fig:bias}. Therefore, the relative precision of a luminosity measurement translates directly into a relative precision on the Eddington ratio.}. This example showcases the potential for multimessenger observations of EMRIs in accretion disks. 

Electromagnetic detection and host association will be somewhat challenging for our reference EMRI. \new{In our error volume, there could be up to $\mathcal{O}(10-100)$ black holes with a mass measurement from EM observations consistent with $M_1$ within $0.5$ dex (as estimated from the black hole mass function~\cite{Shankar2013})},
too many for unique identification of the source. However, other properties of the source (e.g., the inclination of the MBH spin) could be used to simplify the identification problem. 
Accounting for band-dependent bolometric corrections~\citep{Hopkins:2006fq}, we can also estimate whether electromagnetic missions contemporary to LISA will be able to detect the source.
In the X-rays, the future Athena wide field imager, with field of view $\Delta \Omega=0.4 \, {\rm deg}^2$ \citep{Meidinger2018}, will require for this source a total integration time of $\sim 2.8$ days.
The near-infrared or optical emission of this system is also within the sensitivity limit of the near-infrared camera instrument on the James Webb Space Telescope~\citep{Gardner:2006ky}.
\new{As for the radio emission, the low-mass MBHs ($\lesssim 10^7 M_\odot$) typical of EMRI systems detectable by LISA are not expected to host powerful radio jets. However, radio emissions may be observed from, e.g., synchroton emission or even relativistic proto-jets~\cite{Jarvela2021}.}

\subsection{Biased parameter estimation from environmental effects}\label{sec:bias}

Finally, we consider the case in which the EMRI GW signal is analyzed ignoring environmental effects and ask: how would ignoring environmental torques affect the inference of the EMRI parameters and \new{tests of general relativity}?

We investigate this aspect by analyzing the same GW signal from a migrating EMRI injected in the previous section. This time, we analyze the data using two waveform templates that do not allow for environmental torques (\quotes{vacuum template} and \quotes{GR deviation template}) \footnote{Since this procedure is computationally expensive, we run this analysis only over the intrinsic EMRI parameters. We do not expect this choice to affect the conclusions. \new{For more information on EMRI searches we refer the reader to \cite{Babak:2017tow}} }.

\new{Firstly, we search for a signal using the vacuum template}. We perform several runs where the MCMC walkers are allowed to explore a parameter space with priors extending up to $5\%$ around the true value \footnote{Even though we use a naive, brute-force search method and explore only a portion of the parameter space, our results qualitatively suggest how challenging EMRI search and inference studies could be when ignoring environmental effects.}. When using an incorrect template, we are not guaranteed to find \new{a maximum likelihood point when trying to match the full signal present in four years of LISA data}. In fact, the migration torque we consider here is strong enough that we cannot find any match\new{, i.e.~any maximum likelihood point.}

\new{In the last part of the inspiral the orbital decay due to GW emission is stronger than the disk-torque dissipation. Therefore, we expect to match the signal with a vacuum template when considering only the portion of the data closer to the plunge. We refine our search by considering the last $3.5, 3, 2.5, \textrm{and}, 2$ years of data.
We find a maximum likelihood only when we analyze the last two years of data. This maximum likelihood $\mathcal{L}\approx0.07$, is approximately 14 times smaller than the one obtained with the correct template (``migration template'').
}
\begin{figure}
\centering
\includegraphics[width=\linewidth]{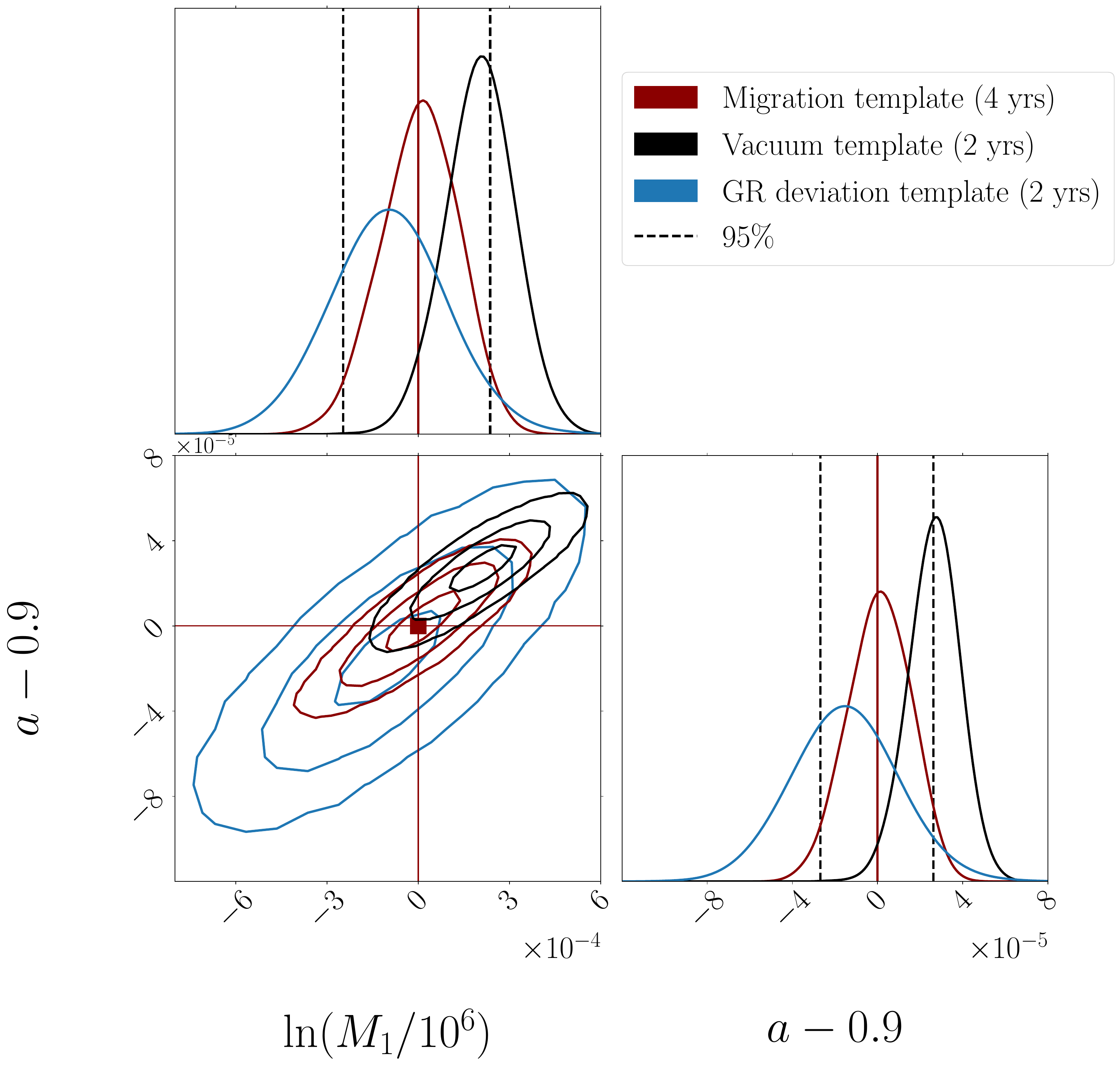}
\caption{Posterior distributions of the primary mass $M_1$ and spin $a$ for the analysis of the same GW signal affected by environmental effects injected in Fig.~\ref{fig:corner_plot}\new{ and analyzed with three different waveform models (templates)}. \new{The contours show the 1,2,3-sigma Gaussian credible regions and the black dashed lines show the 95\% credible intervals.} The migration template \new{(dark red)} takes into account the presence of environmental effects and is able to match the full four-year inspiral, whereas the vacuum \new{(black)} and \new{the GR deviation (blue) templates} can match only the last two years of inspiral and recovers a biased value of primary mass and spin. The vacuum template is 3-sigma biased, whereas the 1$\sigma$ contours from the GR deviation posteriors are consistent with the injected parameters.}
\label{fig:bias}
\end{figure}

Figure~\ref{fig:bias} shows the posterior for primary mass and spin recovered by the analysis of the last 2 years of data with the vacuum template.
For reference, we also show the posterior distribution using the template matching the injection (\quotes{migration template}). We find that the vacuum-template posteriors are significantly biased, i.e.~they are shifted 3$\sigma$ away from the true values. In particular, unaccounted (inward) migration leads to an overestimation of the mass and spin of the primary, as it increases the rate of inspiral.

The absolute size of these biases is small and would not adversely impact any conclusions about the astrophysics of the sources. However, if a similar bias occurred on a parameter that characterizes a deviation from GR, it could shift the inferred value of that parameter away from zero and possibly lead to false detection of a GR deviation. 

\begin{figure}
\centering
\includegraphics[width=\linewidth]{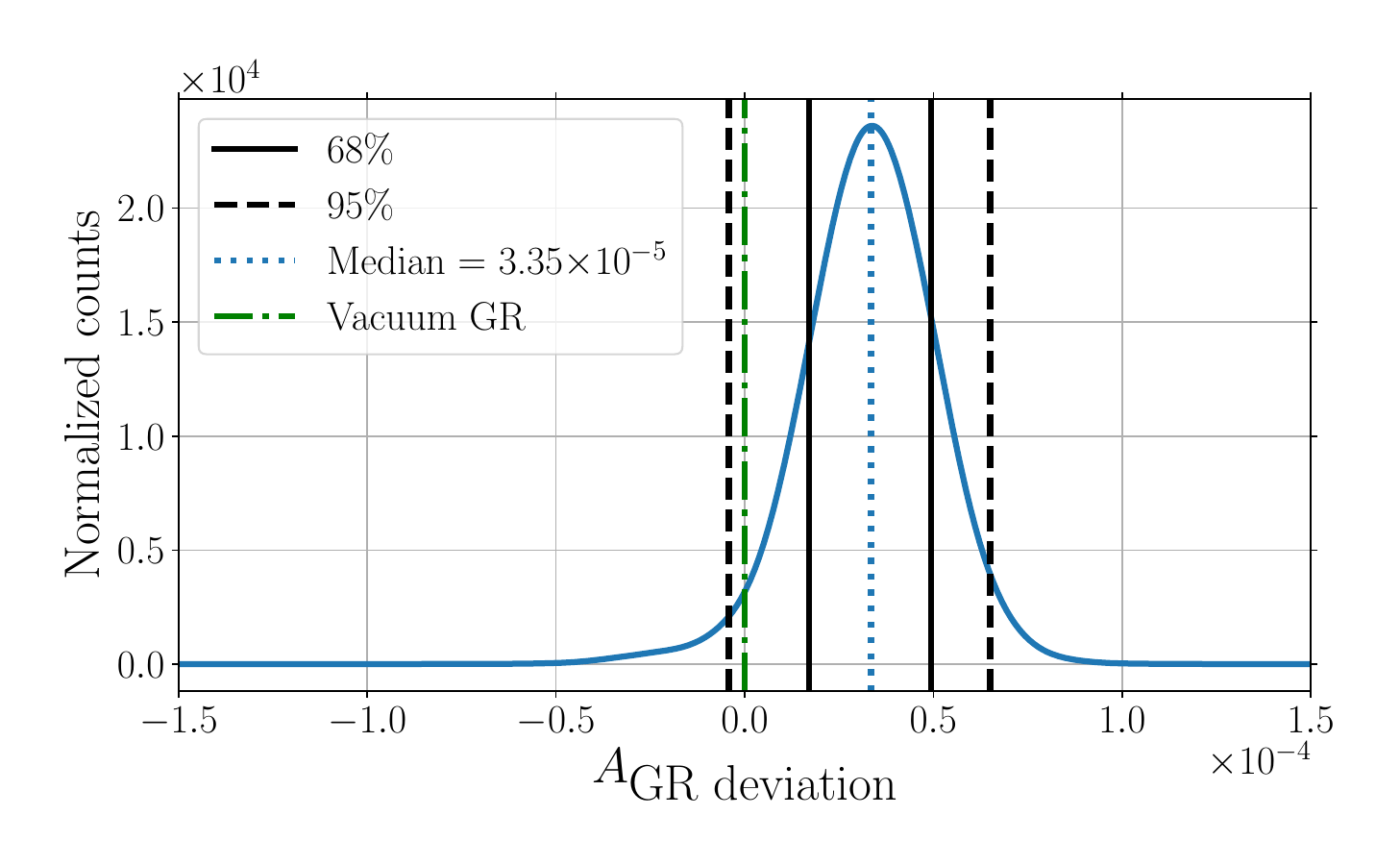}
\caption{Demonstration of false detection of a GR deviation when analyzing GW signals affected by environmental effects. We show the posterior distributions of the amplitude of a GR deviation at $-4$PN for the analysis of the GW signal affected by migration $\alpha$ as in Fig.~\ref{fig:corner_plot}. The black solid and dashed lines show the $68\%$ and $95\%$ credible intervals, respectively, whereas the blue dotted line shows the median of the posterior. The expected amplitude for vacuum GR is shown as a dash-dotted green line.}
\label{fig:beyondGR}
\end{figure}

\new{We verify this by searching for a deviation from GR in the two years of data with a standard parametrized-PN model. In particular, we allow for a deviation from GR coming from a time-varying gravitational constant \citep{Yunes:2009bv, Chamberlain:2017fjl,Barbieri:2022zge, Wang:2022yxb}. This deviation will manifest itself in the waveform as a $-4$PN term and, therefore, can be accounted for in our agnostic model of Eq.~(\ref{eq:L_GW}) by fixing $n_r = 4$. The amplitude of Eq.~(\ref{eq:L_GW}) encodes the information about the size of the GR deviation. After analyzing the last two years of data, we show the posterior distribution for primary mass and spin in Fig.~\ref{fig:bias}, and the marginalized posterior of the amplitude in Fig.~\ref{fig:beyondGR}. 
The posterior for the primary mass and spin shown Fig.~\ref{fig:bias} is slightly shifted from the injected parameters and broader than the two posteriors obtained with the migration and vacuum templates. However, this broadening makes it consistent at 1-sigma with the injected parameter values.
The posterior for the amplitude of the GR deviation shown in Fig.~\ref{fig:beyondGR} is centered around $A_\textrm{GR deviation}=3.35^{+1.6}_{-1.6} \times 10^{-5}$ and it is $1.87 \sigma$ inconsistent with GR, where the amplitude is expected to be zero. \newest{This demonstrates the degeneracy between disk effects and modifications of gravity in EMRI signals, and should motivate further studies into how to test GR with systems potentially affected by the environment. }}

We expect to perform exquisitely sensitive tests of GR with EMRI observations \citep{Maselli:2021men,Barsanti:2021ydd,Barausse:2014tra,Babak:2017tow, Chamberlain:2017fjl}, but our analysis suggests it will be important to allow for additional environmental perturbations when carrying out these tests.
Additionally, if environmental effects are ignored, the residuals between the template and the signal might affect parameter estimation of other sources \cite{Antonelli:2021vwg}. 

\section{Discussion}
\label{sec:discussion}
In this paper, for the first time, we quantitatively study how to measure accretion-disk-induced torques with GW observations from EMRIs. The analysis we carry out assumes the binary is affected by migration in radiatively efficient and geometrically thin disks \citep{Yunes:2011ws,Kocsis:2011dr}, and it is based upon a realistic waveform-generation formalism for EMRI parameter estimation within a fully Bayesian framework.

We investigate three different scenarios in which torques from accretion disks could affect EMRI parameter inference. In the first scenario, we measure how well accretion-disk effects are bounded to zero if they are absent, and we find that the migration torque amplitude can be constrained with a precision of $\sim 5\times 10^{-6}$. This allows us to infer when environmental effects are strong enough to be detectable. Interpreting the constraints as coming from migration torques, we confirm previous estimates in Ref.~\cite{Yunes:2011ws} that migration is observable assuming $\beta$ disk prescriptions, but we also point out that the same is true for a wide range of accretion-disk parameters with $\alpha$ disks.
Our work is the first realistic assessment of the detectability of an environmental effect in an EMRI, identifying the region of parameter space that GW observations can realistically probe.

In the second scenario, we analyze the GW signal of a typical migrating EMRI in an $\alpha$ disk. 
We find that we can distinguish between different disk prescriptions and constrain the amplitude of the environmental effects with $\sim20\%$ relative error.
Using the (agnostic) measurements of torque amplitude and mass, we infer 2D marginalized posteriors for the disk viscosity and accretion rate. 
Moreover, assuming a multimessenger measurement of the bolometric luminosity with $10\%$ precision, we show that the viscosity can be measured with $50\%$ precision.

In the third and final scenario, we investigate the size of biases in EMRI parameter estimates caused by ignoring a strong migration torque. Our proof-of-principle analysis shows that the size of the bias that one should expect from reasonable migration torques will not significantly affect the inference of the astrophysics of galactic nuclei. \new{However, we demonstrated that these biases adversely affect tests of general-relativity with EMRIs \cite{Barausse:2014tra}, and that unmodeled environmental effects can lead to a false ``detection" of a deviation from GR. \newest{This emphasizes the importance of including environmental effects when performing tests of GR.}
Should a population of EMRIs be detected, we {also} expect beyond-GR effects to be universal across the population, unlike environmental effects. Bayesian model selection pipelines should therefore be able to tell the two models apart.}

Our work highlights the science potential of EMRIs embedded in accretion disks and the need for accurate torque models.
In this work, we use prescriptions for EMRI migration that are designed for planetary (type-I) migration in a 3D isothermal disk~\citep{Tanaka2002}. This model has several limitations: for instance, it does not account for the fact that migration torques can change significantly close to the inner edge of the disk~\citep{Tsang2011}, where migration can halt altogether~\citep{Miranda2018}. In general, EMRI migration differs from planetary migration in that the secondary object inspirals rapidly due to GW emission~\citep{Derdzinski:2018qzv}. Targeted numerical simulations, although limited by the wide range of scales and timescales involved, will therefore be crucial to accurately model migration torques for GW observations. The first such simulations (in 2D) have shown promising results \citep{Derdzinski:2018qzv,Derdzinski:2020wlw}. Similarly to what happened in planetary science, we could see in the next decade a progression from 2D to 3D simulations and the inclusion of more and more physics (radiation, magnetic fields, temperature and entropy gradients, thinner disks, etc.).

Another important takeaway point from this work concerns EMRI search and parameter inference strategies. Phenomenological models capable of capturing a host of environmental effects are likely to be needed in future analyses. Our work discusses a possible way of doing this. Our analysis should be considered as a proof-of-principle study of the impact of environmental effects on inference on EMRIs. The code used in this work will be available in the near future as an extension of the \texttt{few} package~\citep{michael_l_katz_2020_4005001}.

While this work relies on a reference EMRI configuration, we expect the detectability of environmental effects to improve when the small compact object explores farther regions around the MBH. This is due to the negative PN order of the effects considered in this work, which become dominant over gravitational emission at low frequencies. For fixed inspiral length and primary mass, a larger secondary mass {(due to accretion~\citep{Tagawa2022,Derdzinski:2022ltb} and/or hierarchical mergers~\citep{McKernan2012,Bellovary2016} in the disk)} would not only have higher SNR, but also be observable at larger radii. \new{Therefore, intermediate massive black hole binary systems might be the best sources for detecting environmental effects. However, accurate waveforms for such systems are not available yet, and the environmental effects might require different modeling from the one presented here \citep{Derdzinski:2020wlw}.}

\new{ In our work, we have assumed circular equatorial orbits. A realistic, agnostic analysis of an EMRI signal with LISA would allow for non-zero eccentricity and inclination, even when looking for accretion-disk torques (or at least perform Bayesian comparisons between the two hypotheses). 
There is also a possibility that eccentricity may evolve as a result of environmental torques, as seen in binaries with a circumbinary disk~\citep{Zrake:2020zkw}. Since EMRI observations are able to measure eccentricity with $10^{-5}$ relative precision, the study of such a scenario would require accurate eccentric vacuum trajectories, and a reliable model for the environment-driven evolution. We also did not consider the possibility of searching for the migrating signal with an eccentric or inclined vacuum waveform. In this case, we expect to be able to tell if the EMRI was truly eccentric by studying the harmonic content of the signal.
We plan to investigate the effects of eccentricity and inclination in future work, when accurate generic orbits around Kerr BHs become available.}

While our study shows that accretion-disk properties can be resolved with EMRIs \new{if observed for 4 or possibly more years (up to 10)}, it remains to be seen if the same holds true when {considering different source classes detectable by LISA~\cite{Zwick:2022dih}, or when taking into account} competing torques, such as from dark-matter spikes \citep{Hannuksela:2019vip,Annulli:2020ilw,Kavanagh:2020cfn,Coogan:2021uqv,Cardoso:2021wlq,Vicente:2022ivh,Speeney:2022ryg,Cole:2022ucw}, hierarchical triples \citep{Cardoso:2021vjq}, or modifications to GR \citep{Barausse:2014tra,Maselli:2020zgv,Maselli:2021men,Barsanti:2022ana}. A detailed study of the distinguishability between these different effects will be the subject of future work.
\newline

\textit{Acknowledgments.} 
The authors wish to thank V. Cardoso, F. Duque, Z. Haiman, S. Hughes, J. Krolik and N. Stone, for very useful discussions. Special thanks to A. Derdzinski for insightful discussions and feedback on our results. A.A. is supported by NSF Grants No. PHY-1912550, AST-2006538, PHY-090003 and PHY-20043, and NASA Grants No. 17-ATP17-0225, 19-ATP19-0051 and 20-LPS20-0011. {S. B. acknowledges support form the French space agency CNES in the framework of LISA.} E.B. acknowledges financial support provided under the European Union's H2020 ERC Consolidator Grant ``GRavity from Astrophysical to Microscopic Scales'' grant agreement no. GRAMS-815673, and  from the MISTI Global Seed Funds MIT-Friuli Venezia-Giulia.
This work was supported by the EU Horizon 2020 Research and
Innovation Programme under the Marie Sklodowska-Curie Grant Agreement
No. 101007855.
This work makes use of the Black Hole Perturbation Toolkit \cite{BHPToolkit}, \texttt{numpy} \cite{harris2020array}, \texttt{matplotlib} \cite{Hunter:2007} and \texttt{scipy} \cite{2020SciPy-NMeth}.

\bibliography{EMRI_accretion}

\begin{thebibliography}{133}%
\makeatletter
\providecommand \@ifxundefined [1]{%
 \@ifx{#1\undefined}
}%
\providecommand \@ifnum [1]{%
 \ifnum #1\expandafter \@firstoftwo
 \else \expandafter \@secondoftwo
 \fi
}%
\providecommand \@ifx [1]{%
 \ifx #1\expandafter \@firstoftwo
 \else \expandafter \@secondoftwo
 \fi
}%
\providecommand \natexlab [1]{#1}%
\providecommand \enquote  [1]{``#1''}%
\providecommand \bibnamefont  [1]{#1}%
\providecommand \bibfnamefont [1]{#1}%
\providecommand \citenamefont [1]{#1}%
\providecommand \href@noop [0]{\@secondoftwo}%
\providecommand \href [0]{\begingroup \@sanitize@url \@href}%
\providecommand \@href[1]{\@@startlink{#1}\@@href}%
\providecommand \@@href[1]{\endgroup#1\@@endlink}%
\providecommand \@sanitize@url [0]{\catcode `\\12\catcode `\$12\catcode
  `\&12\catcode `\#12\catcode `\^12\catcode `\_12\catcode `\%12\relax}%
\providecommand \@@startlink[1]{}%
\providecommand \@@endlink[0]{}%
\providecommand \url  [0]{\begingroup\@sanitize@url \@url }%
\providecommand \@url [1]{\endgroup\@href {#1}{\urlprefix }}%
\providecommand \urlprefix  [0]{URL }%
\providecommand \Eprint [0]{\href }%
\providecommand \doibase [0]{https://doi.org/}%
\providecommand \selectlanguage [0]{\@gobble}%
\providecommand \bibinfo  [0]{\@secondoftwo}%
\providecommand \bibfield  [0]{\@secondoftwo}%
\providecommand \translation [1]{[#1]}%
\providecommand \BibitemOpen [0]{}%
\providecommand \bibitemStop [0]{}%
\providecommand \bibitemNoStop [0]{.\EOS\space}%
\providecommand \EOS [0]{\spacefactor3000\relax}%
\providecommand \BibitemShut  [1]{\csname bibitem#1\endcsname}%
\let\auto@bib@innerbib\@empty
\bibitem [{\citenamefont {Amaro-Seoane}\ \emph {et~al.}(2017)\citenamefont
  {Amaro-Seoane} \emph {et~al.}}]{Audley:2017drz}%
  \BibitemOpen
  \bibfield  {author} {\bibinfo {author} {\bibfnamefont {P.}~\bibnamefont
  {Amaro-Seoane}} \emph {et~al.} (\bibinfo {collaboration} {LISA}),\
  }\href@noop {} {\bibinfo {title} {{Laser Interferometer Space Antenna}}}
  (\bibinfo {year} {2017}),\ \Eprint {https://arxiv.org/abs/1702.00786}
  {arXiv:1702.00786 [astro-ph.IM]} \BibitemShut {NoStop}%
\bibitem [{\citenamefont {Amaro-Seoane}\ \emph {et~al.}(2007)\citenamefont
  {Amaro-Seoane}, \citenamefont {Gair}, \citenamefont {Freitag}, \citenamefont
  {Coleman~Miller}, \citenamefont {Mandel}, \citenamefont {Cutler},\ and\
  \citenamefont {Babak}}]{Amaro-Seoane:2007osp}%
  \BibitemOpen
  \bibfield  {author} {\bibinfo {author} {\bibfnamefont {P.}~\bibnamefont
  {Amaro-Seoane}}, \bibinfo {author} {\bibfnamefont {J.~R.}\ \bibnamefont
  {Gair}}, \bibinfo {author} {\bibfnamefont {M.}~\bibnamefont {Freitag}},
  \bibinfo {author} {\bibfnamefont {M.}~\bibnamefont {Coleman~Miller}},
  \bibinfo {author} {\bibfnamefont {I.}~\bibnamefont {Mandel}}, \bibinfo
  {author} {\bibfnamefont {C.~J.}\ \bibnamefont {Cutler}},\ and\ \bibinfo
  {author} {\bibfnamefont {S.}~\bibnamefont {Babak}},\ }\bibfield  {title}
  {\bibinfo {title} {{Astrophysics, detection and science applications of
  intermediate- and extreme mass-ratio inspirals}},\ }\href
  {https://doi.org/10.1088/0264-9381/24/17/R01} {\bibfield  {journal} {\bibinfo
   {journal} {Class. Quant. Grav.}\ }\textbf {\bibinfo {volume} {24}},\
  \bibinfo {pages} {R113} (\bibinfo {year} {2007})},\ \Eprint
  {https://arxiv.org/abs/astro-ph/0703495} {arXiv:astro-ph/0703495}
  \BibitemShut {NoStop}%
\bibitem [{\citenamefont {Gair}\ \emph {et~al.}(2013)\citenamefont {Gair},
  \citenamefont {Vallisneri}, \citenamefont {Larson},\ and\ \citenamefont
  {Baker}}]{Gair:2012nm}%
  \BibitemOpen
  \bibfield  {author} {\bibinfo {author} {\bibfnamefont {J.~R.}\ \bibnamefont
  {Gair}}, \bibinfo {author} {\bibfnamefont {M.}~\bibnamefont {Vallisneri}},
  \bibinfo {author} {\bibfnamefont {S.~L.}\ \bibnamefont {Larson}},\ and\
  \bibinfo {author} {\bibfnamefont {J.~G.}\ \bibnamefont {Baker}},\ }\bibfield
  {title} {\bibinfo {title} {{Testing General Relativity with Low-Frequency,
  Space-Based Gravitational-Wave Detectors}},\ }\href
  {https://doi.org/10.12942/lrr-2013-7} {\bibfield  {journal} {\bibinfo
  {journal} {Living Rev. Rel.}\ }\textbf {\bibinfo {volume} {16}},\ \bibinfo
  {pages} {7} (\bibinfo {year} {2013})},\ \Eprint
  {https://arxiv.org/abs/1212.5575} {arXiv:1212.5575 [gr-qc]} \BibitemShut
  {NoStop}%
\bibitem [{\citenamefont {Amaro-Seoane}(2018)}]{Amaro-Seoane:2012lgq}%
  \BibitemOpen
  \bibfield  {author} {\bibinfo {author} {\bibfnamefont {P.}~\bibnamefont
  {Amaro-Seoane}},\ }\bibfield  {title} {\bibinfo {title} {{Relativistic
  dynamics and extreme mass ratio inspirals}},\ }\href
  {https://doi.org/10.1007/s41114-018-0013-8} {\bibfield  {journal} {\bibinfo
  {journal} {Living Rev. Rel.}\ }\textbf {\bibinfo {volume} {21}},\ \bibinfo
  {pages} {4} (\bibinfo {year} {2018})},\ \Eprint
  {https://arxiv.org/abs/1205.5240} {arXiv:1205.5240 [astro-ph.CO]}
  \BibitemShut {NoStop}%
\bibitem [{\citenamefont {Berry}\ \emph {et~al.}(2019)\citenamefont {Berry},
  \citenamefont {Hughes}, \citenamefont {Sopuerta}, \citenamefont {Chua},
  \citenamefont {Heffernan}, \citenamefont {Holley-Bockelmann}, \citenamefont
  {Mihaylov}, \citenamefont {Miller},\ and\ \citenamefont
  {Sesana}}]{Berry:2019wgg}%
  \BibitemOpen
  \bibfield  {author} {\bibinfo {author} {\bibfnamefont {C.~P.~L.}\
  \bibnamefont {Berry}}, \bibinfo {author} {\bibfnamefont {S.~A.}\ \bibnamefont
  {Hughes}}, \bibinfo {author} {\bibfnamefont {C.~F.}\ \bibnamefont
  {Sopuerta}}, \bibinfo {author} {\bibfnamefont {A.~J.~K.}\ \bibnamefont
  {Chua}}, \bibinfo {author} {\bibfnamefont {A.}~\bibnamefont {Heffernan}},
  \bibinfo {author} {\bibfnamefont {K.}~\bibnamefont {Holley-Bockelmann}},
  \bibinfo {author} {\bibfnamefont {D.~P.}\ \bibnamefont {Mihaylov}}, \bibinfo
  {author} {\bibfnamefont {M.~C.}\ \bibnamefont {Miller}},\ and\ \bibinfo
  {author} {\bibfnamefont {A.}~\bibnamefont {Sesana}},\ }\href@noop {}
  {\bibinfo {title} {{The unique potential of extreme mass-ratio inspirals for
  gravitational-wave astronomy}}} (\bibinfo {year} {2019}),\ \Eprint
  {https://arxiv.org/abs/1903.03686} {arXiv:1903.03686 [astro-ph.HE]}
  \BibitemShut {NoStop}%
\bibitem [{\citenamefont {Amaro-Seoane}\ \emph {et~al.}(2022)\citenamefont
  {Amaro-Seoane} \emph {et~al.}}]{Amaro-Seoane:2022rxf}%
  \BibitemOpen
  \bibfield  {author} {\bibinfo {author} {\bibfnamefont {P.}~\bibnamefont
  {Amaro-Seoane}} \emph {et~al.},\ }\href@noop {} {\bibinfo {title}
  {{Astrophysics with the Laser Interferometer Space Antenna}}} (\bibinfo
  {year} {2022}),\ \Eprint {https://arxiv.org/abs/2203.06016} {arXiv:2203.06016
  [gr-qc]} \BibitemShut {NoStop}%
\bibitem [{\citenamefont {Dittmann}\ and\ \citenamefont
  {Miller}(2020)}]{Dittmann:2019sbm}%
  \BibitemOpen
  \bibfield  {author} {\bibinfo {author} {\bibfnamefont {A.~J.}\ \bibnamefont
  {Dittmann}}\ and\ \bibinfo {author} {\bibfnamefont {M.~C.}\ \bibnamefont
  {Miller}},\ }\bibfield  {title} {\bibinfo {title} {{Star formation in
  accretion discs and SMBH growth}},\ }\href
  {https://doi.org/10.1093/mnras/staa463} {\bibfield  {journal} {\bibinfo
  {journal} {Mon. Not. Roy. Astron. Soc.}\ }\textbf {\bibinfo {volume} {493}},\
  \bibinfo {pages} {3732} (\bibinfo {year} {2020})},\ \Eprint
  {https://arxiv.org/abs/1911.08685} {arXiv:1911.08685 [astro-ph.HE]}
  \BibitemShut {NoStop}%
\bibitem [{\citenamefont {{Arca-Sedda}}\ and\ \citenamefont
  {{Capuzzo-Dolcetta}}(2019)}]{Arca-Sedda2019}%
  \BibitemOpen
  \bibfield  {author} {\bibinfo {author} {\bibfnamefont {M.}~\bibnamefont
  {{Arca-Sedda}}}\ and\ \bibinfo {author} {\bibfnamefont {R.}~\bibnamefont
  {{Capuzzo-Dolcetta}}},\ }\bibfield  {title} {\bibinfo {title} {{The MEGaN
  project II. Gravitational waves from intermediate-mass and binary black holes
  around a supermassive black hole}},\ }\href
  {https://doi.org/10.1093/mnras/sty3096} {\bibfield  {journal} {\bibinfo
  {journal} {\mnras}\ }\textbf {\bibinfo {volume} {483}},\ \bibinfo {pages}
  {152} (\bibinfo {year} {2019})},\ \Eprint {https://arxiv.org/abs/1709.05567}
  {arXiv:1709.05567 [astro-ph.GA]} \BibitemShut {NoStop}%
\bibitem [{\citenamefont {Pan}\ \emph {et~al.}(2021)\citenamefont {Pan},
  \citenamefont {Lyu},\ and\ \citenamefont {Yang}}]{Pan:2021oob}%
  \BibitemOpen
  \bibfield  {author} {\bibinfo {author} {\bibfnamefont {Z.}~\bibnamefont
  {Pan}}, \bibinfo {author} {\bibfnamefont {Z.}~\bibnamefont {Lyu}},\ and\
  \bibinfo {author} {\bibfnamefont {H.}~\bibnamefont {Yang}},\ }\bibfield
  {title} {\bibinfo {title} {{Wet extreme mass ratio inspirals may be more
  common for spaceborne gravitational wave detection}},\ }\href
  {https://doi.org/10.1103/PhysRevD.104.063007} {\bibfield  {journal} {\bibinfo
   {journal} {Phys. Rev. D}\ }\textbf {\bibinfo {volume} {104}},\ \bibinfo
  {pages} {063007} (\bibinfo {year} {2021})},\ \Eprint
  {https://arxiv.org/abs/2104.01208} {arXiv:2104.01208 [astro-ph.HE]}
  \BibitemShut {NoStop}%
\bibitem [{\citenamefont {Pan}\ and\ \citenamefont {Yang}(2021)}]{Pan:2021ksp}%
  \BibitemOpen
  \bibfield  {author} {\bibinfo {author} {\bibfnamefont {Z.}~\bibnamefont
  {Pan}}\ and\ \bibinfo {author} {\bibfnamefont {H.}~\bibnamefont {Yang}},\
  }\bibfield  {title} {\bibinfo {title} {{Formation Rate of Extreme Mass Ratio
  Inspirals in Active Galactic Nuclei}},\ }\href
  {https://doi.org/10.1103/PhysRevD.103.103018} {\bibfield  {journal} {\bibinfo
   {journal} {Phys. Rev. D}\ }\textbf {\bibinfo {volume} {103}},\ \bibinfo
  {pages} {103018} (\bibinfo {year} {2021})},\ \Eprint
  {https://arxiv.org/abs/2101.09146} {arXiv:2101.09146 [astro-ph.HE]}
  \BibitemShut {NoStop}%
\bibitem [{\citenamefont {Derdzinski}\ and\ \citenamefont
  {Mayer}(2022)}]{Derdzinski:2022ltb}%
  \BibitemOpen
  \bibfield  {author} {\bibinfo {author} {\bibfnamefont {A.}~\bibnamefont
  {Derdzinski}}\ and\ \bibinfo {author} {\bibfnamefont {L.}~\bibnamefont
  {Mayer}},\ }\href@noop {} {\bibinfo {title} {{In-situ extreme mass ratio
  inspirals via sub-parsec formation and migration of stars in thin,
  gravitationally unstable AGN discs}}} (\bibinfo {year} {2022}),\ \Eprint
  {https://arxiv.org/abs/2205.10382} {arXiv:2205.10382 [astro-ph.GA]}
  \BibitemShut {NoStop}%
\bibitem [{\citenamefont {Kocsis}\ \emph {et~al.}(2011)\citenamefont {Kocsis},
  \citenamefont {Yunes},\ and\ \citenamefont {Loeb}}]{Kocsis:2011dr}%
  \BibitemOpen
  \bibfield  {author} {\bibinfo {author} {\bibfnamefont {B.}~\bibnamefont
  {Kocsis}}, \bibinfo {author} {\bibfnamefont {N.}~\bibnamefont {Yunes}},\ and\
  \bibinfo {author} {\bibfnamefont {A.}~\bibnamefont {Loeb}},\ }\bibfield
  {title} {\bibinfo {title} {{Observable Signatures of EMRI Black Hole Binaries
  Embedded in Thin Accretion Disks}},\ }\href
  {https://doi.org/10.1103/PhysRevD.86.049907} {\bibfield  {journal} {\bibinfo
  {journal} {Phys. Rev. D}\ }\textbf {\bibinfo {volume} {84}},\ \bibinfo
  {pages} {024032} (\bibinfo {year} {2011})},\ \Eprint
  {https://arxiv.org/abs/1104.2322} {arXiv:1104.2322 [astro-ph.GA]}
  \BibitemShut {NoStop}%
\bibitem [{\citenamefont {Barausse}\ and\ \citenamefont
  {Rezzolla}(2008)}]{Barausse:2007dy}%
  \BibitemOpen
  \bibfield  {author} {\bibinfo {author} {\bibfnamefont {E.}~\bibnamefont
  {Barausse}}\ and\ \bibinfo {author} {\bibfnamefont {L.}~\bibnamefont
  {Rezzolla}},\ }\bibfield  {title} {\bibinfo {title} {{The Influence of the
  hydrodynamic drag from an accretion torus on extreme mass-ratio inspirals}},\
  }\href {https://doi.org/10.1103/PhysRevD.77.104027} {\bibfield  {journal}
  {\bibinfo  {journal} {Phys. Rev. D}\ }\textbf {\bibinfo {volume} {77}},\
  \bibinfo {pages} {104027} (\bibinfo {year} {2008})},\ \Eprint
  {https://arxiv.org/abs/0711.4558} {arXiv:0711.4558 [gr-qc]} \BibitemShut
  {NoStop}%
\bibitem [{Note1()}]{Note1}%
  \BibitemOpen
  \bibinfo {note} {The direct pull from the accretion disk is negligible unless
  unrealistically large densities are considered~\protect \citep
  {Barausse:2006vt}.}\BibitemShut {Stop}%
\bibitem [{\citenamefont {Barausse}\ \emph {et~al.}(2007)\citenamefont
  {Barausse}, \citenamefont {Rezzolla}, \citenamefont {Petroff},\ and\
  \citenamefont {Ansorg}}]{Barausse:2006vt}%
  \BibitemOpen
  \bibfield  {author} {\bibinfo {author} {\bibfnamefont {E.}~\bibnamefont
  {Barausse}}, \bibinfo {author} {\bibfnamefont {L.}~\bibnamefont {Rezzolla}},
  \bibinfo {author} {\bibfnamefont {D.}~\bibnamefont {Petroff}},\ and\ \bibinfo
  {author} {\bibfnamefont {M.}~\bibnamefont {Ansorg}},\ }\bibfield  {title}
  {\bibinfo {title} {{Gravitational waves from Extreme Mass Ratio Inspirals in
  non-pure Kerr spacetimes}},\ }\href
  {https://doi.org/10.1103/PhysRevD.75.064026} {\bibfield  {journal} {\bibinfo
  {journal} {Phys. Rev. D}\ }\textbf {\bibinfo {volume} {75}},\ \bibinfo
  {pages} {064026} (\bibinfo {year} {2007})},\ \Eprint
  {https://arxiv.org/abs/gr-qc/0612123} {arXiv:gr-qc/0612123} \BibitemShut
  {NoStop}%
\bibitem [{\citenamefont {Barausse}\ \emph {et~al.}(2014)\citenamefont
  {Barausse}, \citenamefont {Cardoso},\ and\ \citenamefont
  {Pani}}]{Barausse:2014tra}%
  \BibitemOpen
  \bibfield  {author} {\bibinfo {author} {\bibfnamefont {E.}~\bibnamefont
  {Barausse}}, \bibinfo {author} {\bibfnamefont {V.}~\bibnamefont {Cardoso}},\
  and\ \bibinfo {author} {\bibfnamefont {P.}~\bibnamefont {Pani}},\ }\bibfield
  {title} {\bibinfo {title} {{Can environmental effects spoil precision
  gravitational-wave astrophysics?}},\ }\href
  {https://doi.org/10.1103/PhysRevD.89.104059} {\bibfield  {journal} {\bibinfo
  {journal} {Phys. Rev. D}\ }\textbf {\bibinfo {volume} {89}},\ \bibinfo
  {pages} {104059} (\bibinfo {year} {2014})},\ \Eprint
  {https://arxiv.org/abs/1404.7149} {arXiv:1404.7149 [gr-qc]} \BibitemShut
  {NoStop}%
\bibitem [{\citenamefont {Barausse}\ \emph {et~al.}(2015)\citenamefont
  {Barausse}, \citenamefont {Cardoso},\ and\ \citenamefont
  {Pani}}]{Barausse:2014pra}%
  \BibitemOpen
  \bibfield  {author} {\bibinfo {author} {\bibfnamefont {E.}~\bibnamefont
  {Barausse}}, \bibinfo {author} {\bibfnamefont {V.}~\bibnamefont {Cardoso}},\
  and\ \bibinfo {author} {\bibfnamefont {P.}~\bibnamefont {Pani}},\ }\bibfield
  {title} {\bibinfo {title} {{Environmental Effects for Gravitational-wave
  Astrophysics}},\ }\href {https://doi.org/10.1088/1742-6596/610/1/012044}
  {\bibfield  {journal} {\bibinfo  {journal} {J. Phys. Conf. Ser.}\ }\textbf
  {\bibinfo {volume} {610}},\ \bibinfo {pages} {012044} (\bibinfo {year}
  {2015})},\ \Eprint {https://arxiv.org/abs/1404.7140} {arXiv:1404.7140
  [astro-ph.CO]} \BibitemShut {NoStop}%
\bibitem [{\citenamefont {Goodman}\ and\ \citenamefont
  {Rafikov}(2001)}]{Goodman:2000jf}%
  \BibitemOpen
  \bibfield  {author} {\bibinfo {author} {\bibfnamefont {J.}~\bibnamefont
  {Goodman}}\ and\ \bibinfo {author} {\bibfnamefont {R.~R.}\ \bibnamefont
  {Rafikov}},\ }\bibfield  {title} {\bibinfo {title} {{Planetary torques as the
  viscosity of protoplanetary disks}},\ }\href {https://doi.org/10.1086/320572}
  {\bibfield  {journal} {\bibinfo  {journal} {Astrophys. J.}\ }\textbf
  {\bibinfo {volume} {552}},\ \bibinfo {pages} {793} (\bibinfo {year}
  {2001})},\ \Eprint {https://arxiv.org/abs/astro-ph/0010576}
  {arXiv:astro-ph/0010576} \BibitemShut {NoStop}%
\bibitem [{\citenamefont {Yunes}\ \emph {et~al.}(2011)\citenamefont {Yunes},
  \citenamefont {Kocsis}, \citenamefont {Loeb},\ and\ \citenamefont
  {Haiman}}]{Yunes:2011ws}%
  \BibitemOpen
  \bibfield  {author} {\bibinfo {author} {\bibfnamefont {N.}~\bibnamefont
  {Yunes}}, \bibinfo {author} {\bibfnamefont {B.}~\bibnamefont {Kocsis}},
  \bibinfo {author} {\bibfnamefont {A.}~\bibnamefont {Loeb}},\ and\ \bibinfo
  {author} {\bibfnamefont {Z.}~\bibnamefont {Haiman}},\ }\bibfield  {title}
  {\bibinfo {title} {{Imprint of Accretion Disk-Induced Migration on
  Gravitational Waves from Extreme Mass Ratio Inspirals}},\ }\href
  {https://doi.org/10.1103/PhysRevLett.107.171103} {\bibfield  {journal}
  {\bibinfo  {journal} {Phys. Rev. Lett.}\ }\textbf {\bibinfo {volume} {107}},\
  \bibinfo {pages} {171103} (\bibinfo {year} {2011})},\ \Eprint
  {https://arxiv.org/abs/1103.4609} {arXiv:1103.4609 [astro-ph.CO]}
  \BibitemShut {NoStop}%
\bibitem [{\citenamefont {Derdzinski}\ \emph {et~al.}(2021)\citenamefont
  {Derdzinski}, \citenamefont {D'Orazio}, \citenamefont {Duffell},
  \citenamefont {Haiman},\ and\ \citenamefont
  {MacFadyen}}]{Derdzinski:2020wlw}%
  \BibitemOpen
  \bibfield  {author} {\bibinfo {author} {\bibfnamefont {A.}~\bibnamefont
  {Derdzinski}}, \bibinfo {author} {\bibfnamefont {D.}~\bibnamefont
  {D'Orazio}}, \bibinfo {author} {\bibfnamefont {P.}~\bibnamefont {Duffell}},
  \bibinfo {author} {\bibfnamefont {Z.}~\bibnamefont {Haiman}},\ and\ \bibinfo
  {author} {\bibfnamefont {A.}~\bibnamefont {MacFadyen}},\ }\bibfield  {title}
  {\bibinfo {title} {{Evolution of gas disc\textendash{}embedded intermediate
  mass ratio inspirals in the $LISA$ band}},\ }\href
  {https://doi.org/10.1093/mnras/staa3976} {\bibfield  {journal} {\bibinfo
  {journal} {Mon. Not. Roy. Astron. Soc.}\ }\textbf {\bibinfo {volume} {501}},\
  \bibinfo {pages} {3540} (\bibinfo {year} {2021})},\ \Eprint
  {https://arxiv.org/abs/2005.11333} {arXiv:2005.11333 [astro-ph.HE]}
  \BibitemShut {NoStop}%
\bibitem [{\citenamefont {Zwick}\ \emph {et~al.}(2021)\citenamefont {Zwick},
  \citenamefont {Derdzinski}, \citenamefont {Garg}, \citenamefont {Capelo},\
  and\ \citenamefont {Mayer}}]{Zwick:2021dlg}%
  \BibitemOpen
  \bibfield  {author} {\bibinfo {author} {\bibfnamefont {L.}~\bibnamefont
  {Zwick}}, \bibinfo {author} {\bibfnamefont {A.}~\bibnamefont {Derdzinski}},
  \bibinfo {author} {\bibfnamefont {M.}~\bibnamefont {Garg}}, \bibinfo {author}
  {\bibfnamefont {P.~R.}\ \bibnamefont {Capelo}},\ and\ \bibinfo {author}
  {\bibfnamefont {L.}~\bibnamefont {Mayer}},\ }\href
  {https://doi.org/10.1093/mnras/stac299} {\bibinfo {title} {{Dirty waveforms:
  multiband harmonic content of gas-embedded gravitational wave sources}}}
  (\bibinfo {year} {2021}),\ \Eprint {https://arxiv.org/abs/2110.09097}
  {arXiv:2110.09097 [astro-ph.HE]} \BibitemShut {NoStop}%
\bibitem [{\citenamefont {Shakura}\ and\ \citenamefont
  {Sunyaev}(1973)}]{Shakura:1972te}%
  \BibitemOpen
  \bibfield  {author} {\bibinfo {author} {\bibfnamefont {N.~I.}\ \bibnamefont
  {Shakura}}\ and\ \bibinfo {author} {\bibfnamefont {R.~A.}\ \bibnamefont
  {Sunyaev}},\ }\bibfield  {title} {\bibinfo {title} {{Black holes in binary
  systems. Observational appearance}},\ }\href@noop {} {\bibfield  {journal}
  {\bibinfo  {journal} {Astron. Astrophys.}\ }\textbf {\bibinfo {volume}
  {24}},\ \bibinfo {pages} {337} (\bibinfo {year} {1973})}\BibitemShut
  {NoStop}%
\bibitem [{\citenamefont {Sakimoto}\ and\ \citenamefont
  {Coroniti}(1981)}]{Sakimoto:1981}%
  \BibitemOpen
  \bibfield  {author} {\bibinfo {author} {\bibfnamefont {P.}~\bibnamefont
  {Sakimoto}}\ and\ \bibinfo {author} {\bibfnamefont {F.}~\bibnamefont
  {Coroniti}},\ }\bibfield  {title} {\bibinfo {title} {Accretion disk models
  for qsos and active galactic nuclei - the role of magnetic viscosity},\
  }\href {https://doi.org/10.1086/159005} {\bibfield  {journal} {\bibinfo
  {journal} {Astrophys. J.}\ }\textbf {\bibinfo {volume} {247}} (\bibinfo
  {year} {1981})}\BibitemShut {NoStop}%
\bibitem [{\citenamefont {Chua}\ \emph {et~al.}(2019)\citenamefont {Chua},
  \citenamefont {Galley},\ and\ \citenamefont {Vallisneri}}]{Chua:2018woh}%
  \BibitemOpen
  \bibfield  {author} {\bibinfo {author} {\bibfnamefont {A.~J.~K.}\
  \bibnamefont {Chua}}, \bibinfo {author} {\bibfnamefont {C.~R.}\ \bibnamefont
  {Galley}},\ and\ \bibinfo {author} {\bibfnamefont {M.}~\bibnamefont
  {Vallisneri}},\ }\bibfield  {title} {\bibinfo {title} {{Reduced-order
  modeling with artificial neurons for gravitational-wave inference}},\ }\href
  {https://doi.org/10.1103/PhysRevLett.122.211101} {\bibfield  {journal}
  {\bibinfo  {journal} {Phys. Rev. Lett.}\ }\textbf {\bibinfo {volume} {122}},\
  \bibinfo {pages} {211101} (\bibinfo {year} {2019})},\ \Eprint
  {https://arxiv.org/abs/1811.05491} {arXiv:1811.05491 [astro-ph.IM]}
  \BibitemShut {NoStop}%
\bibitem [{\citenamefont {Chua}\ \emph {et~al.}(2021)\citenamefont {Chua},
  \citenamefont {Katz}, \citenamefont {Warburton},\ and\ \citenamefont
  {Hughes}}]{Chua:2020stf}%
  \BibitemOpen
  \bibfield  {author} {\bibinfo {author} {\bibfnamefont {A.~J.~K.}\
  \bibnamefont {Chua}}, \bibinfo {author} {\bibfnamefont {M.~L.}\ \bibnamefont
  {Katz}}, \bibinfo {author} {\bibfnamefont {N.}~\bibnamefont {Warburton}},\
  and\ \bibinfo {author} {\bibfnamefont {S.~A.}\ \bibnamefont {Hughes}},\
  }\bibfield  {title} {\bibinfo {title} {{Rapid generation of fully
  relativistic extreme-mass-ratio-inspiral waveform templates for LISA data
  analysis}},\ }\href {https://doi.org/10.1103/PhysRevLett.126.051102}
  {\bibfield  {journal} {\bibinfo  {journal} {Phys. Rev. Lett.}\ }\textbf
  {\bibinfo {volume} {126}},\ \bibinfo {pages} {051102} (\bibinfo {year}
  {2021})},\ \Eprint {https://arxiv.org/abs/2008.06071} {arXiv:2008.06071
  [gr-qc]} \BibitemShut {NoStop}%
\bibitem [{\citenamefont {Katz}\ \emph {et~al.}(2020)\citenamefont {Katz},
  \citenamefont {Chua}, \citenamefont {Warburton},\ and\ \citenamefont
  {Hughes.}}]{michael_l_katz_2020_4005001}%
  \BibitemOpen
  \bibfield  {author} {\bibinfo {author} {\bibfnamefont {M.~L.}\ \bibnamefont
  {Katz}}, \bibinfo {author} {\bibfnamefont {A.~J.~K.}\ \bibnamefont {Chua}},
  \bibinfo {author} {\bibfnamefont {N.}~\bibnamefont {Warburton}},\ and\
  \bibinfo {author} {\bibfnamefont {S.~A.}\ \bibnamefont {Hughes.}},\ }\href
  {https://doi.org/10.5281/zenodo.4005001} {\bibinfo {title}
  {{BlackHolePerturbationToolkit/FastEMRIWaveforms: Official Release}}}
  (\bibinfo {year} {2020})\BibitemShut {NoStop}%
\bibitem [{\citenamefont {Katz}\ \emph {et~al.}(2021)\citenamefont {Katz},
  \citenamefont {Chua}, \citenamefont {Speri}, \citenamefont {Warburton},\ and\
  \citenamefont {Hughes}}]{Katz:2021yft}%
  \BibitemOpen
  \bibfield  {author} {\bibinfo {author} {\bibfnamefont {M.~L.}\ \bibnamefont
  {Katz}}, \bibinfo {author} {\bibfnamefont {A.~J.~K.}\ \bibnamefont {Chua}},
  \bibinfo {author} {\bibfnamefont {L.}~\bibnamefont {Speri}}, \bibinfo
  {author} {\bibfnamefont {N.}~\bibnamefont {Warburton}},\ and\ \bibinfo
  {author} {\bibfnamefont {S.~A.}\ \bibnamefont {Hughes}},\ }\bibfield  {title}
  {\bibinfo {title} {{Fast extreme-mass-ratio-inspiral waveforms: New tools for
  millihertz gravitational-wave data analysis}},\ }\href
  {https://doi.org/10.1103/PhysRevD.104.064047} {\bibfield  {journal} {\bibinfo
   {journal} {Phys. Rev. D}\ }\textbf {\bibinfo {volume} {104}},\ \bibinfo
  {pages} {064047} (\bibinfo {year} {2021})},\ \Eprint
  {https://arxiv.org/abs/2104.04582} {arXiv:2104.04582 [gr-qc]} \BibitemShut
  {NoStop}%
\bibitem [{\citenamefont {{Cresswell}}\ \emph {et~al.}(2007)\citenamefont
  {{Cresswell}}, \citenamefont {{Dirksen}}, \citenamefont {{Kley}},\ and\
  \citenamefont {{Nelson}}}]{Cresswell2007}%
  \BibitemOpen
  \bibfield  {author} {\bibinfo {author} {\bibfnamefont {P.}~\bibnamefont
  {{Cresswell}}}, \bibinfo {author} {\bibfnamefont {G.}~\bibnamefont
  {{Dirksen}}}, \bibinfo {author} {\bibfnamefont {W.}~\bibnamefont {{Kley}}},\
  and\ \bibinfo {author} {\bibfnamefont {R.~P.}\ \bibnamefont {{Nelson}}},\
  }\bibfield  {title} {\bibinfo {title} {{On the evolution of eccentric and
  inclined protoplanets embedded in protoplanetary disks}},\ }\href
  {https://doi.org/10.1051/0004-6361:20077666} {\bibfield  {journal} {\bibinfo
  {journal} {\aap}\ }\textbf {\bibinfo {volume} {473}},\ \bibinfo {pages} {329}
  (\bibinfo {year} {2007})},\ \Eprint {https://arxiv.org/abs/0707.2225}
  {arXiv:0707.2225 [astro-ph]} \BibitemShut {NoStop}%
\bibitem [{\citenamefont {{Bitsch}}\ and\ \citenamefont
  {{Kley}}(2010)}]{Bitsch2010}%
  \BibitemOpen
  \bibfield  {author} {\bibinfo {author} {\bibfnamefont {B.}~\bibnamefont
  {{Bitsch}}}\ and\ \bibinfo {author} {\bibfnamefont {W.}~\bibnamefont
  {{Kley}}},\ }\bibfield  {title} {\bibinfo {title} {{Orbital evolution of
  eccentric planets in radiative discs}},\ }\href
  {https://doi.org/10.1051/0004-6361/201014414} {\bibfield  {journal} {\bibinfo
   {journal} {\aap}\ }\textbf {\bibinfo {volume} {523}},\ \bibinfo {eid} {A30}
  (\bibinfo {year} {2010})},\ \Eprint {https://arxiv.org/abs/1008.2656}
  {arXiv:1008.2656 [astro-ph.EP]} \BibitemShut {NoStop}%
\bibitem [{\citenamefont {{McKernan}}\ \emph {et~al.}(2012)\citenamefont
  {{McKernan}}, \citenamefont {{Ford}}, \citenamefont {{Lyra}},\ and\
  \citenamefont {{Perets}}}]{McKernan2012}%
  \BibitemOpen
  \bibfield  {author} {\bibinfo {author} {\bibfnamefont {B.}~\bibnamefont
  {{McKernan}}}, \bibinfo {author} {\bibfnamefont {K.~E.~S.}\ \bibnamefont
  {{Ford}}}, \bibinfo {author} {\bibfnamefont {W.}~\bibnamefont {{Lyra}}},\
  and\ \bibinfo {author} {\bibfnamefont {H.~B.}\ \bibnamefont {{Perets}}},\
  }\bibfield  {title} {\bibinfo {title} {{Intermediate mass black holes in AGN
  discs - I. Production and growth}},\ }\href
  {https://doi.org/10.1111/j.1365-2966.2012.21486.x} {\bibfield  {journal}
  {\bibinfo  {journal} {\mnras}\ }\textbf {\bibinfo {volume} {425}},\ \bibinfo
  {pages} {460} (\bibinfo {year} {2012})},\ \Eprint
  {https://arxiv.org/abs/1206.2309} {arXiv:1206.2309 [astro-ph.GA]}
  \BibitemShut {NoStop}%
\bibitem [{\citenamefont {{MacLeod}}\ and\ \citenamefont
  {{Lin}}(2020)}]{MacLeodLin2020}%
  \BibitemOpen
  \bibfield  {author} {\bibinfo {author} {\bibfnamefont {M.}~\bibnamefont
  {{MacLeod}}}\ and\ \bibinfo {author} {\bibfnamefont {D.~N.~C.}\ \bibnamefont
  {{Lin}}},\ }\bibfield  {title} {\bibinfo {title} {{The Effect of Star-Disk
  Interactions on Highly Eccentric Stellar Orbits in Active Galactic Nuclei: A
  Disk Loss Cone and Implications for Stellar Tidal Disruption Events}},\
  }\href {https://doi.org/10.3847/1538-4357/ab64db} {\bibfield  {journal}
  {\bibinfo  {journal} {\apj}\ }\textbf {\bibinfo {volume} {889}},\ \bibinfo
  {eid} {94} (\bibinfo {year} {2020})},\ \Eprint
  {https://arxiv.org/abs/1909.09645} {arXiv:1909.09645 [astro-ph.SR]}
  \BibitemShut {NoStop}%
\bibitem [{\citenamefont {Jiang}\ \emph
  {et~al.}(2019{\natexlab{a}})\citenamefont {Jiang}, \citenamefont {Blaes},
  \citenamefont {Stone},\ and\ \citenamefont {Davis}}]{Jiang:2019bxn}%
  \BibitemOpen
  \bibfield  {author} {\bibinfo {author} {\bibfnamefont {Y.-F.}\ \bibnamefont
  {Jiang}}, \bibinfo {author} {\bibfnamefont {O.}~\bibnamefont {Blaes}},
  \bibinfo {author} {\bibfnamefont {J.}~\bibnamefont {Stone}},\ and\ \bibinfo
  {author} {\bibfnamefont {S.~W.}\ \bibnamefont {Davis}},\ }\href
  {https://doi.org/10.3847/1538-4357/ab4a00} {\bibinfo {title} {{Global
  Radiation Magneto-hydrodynamic Simulations of Sub-Eddington Accretion Disks
  around Supermassive Black Holes}}} (\bibinfo {year} {2019}{\natexlab{a}}),\
  \Eprint {https://arxiv.org/abs/1904.01674} {arXiv:1904.01674 [astro-ph.HE]}
  \BibitemShut {NoStop}%
\bibitem [{\citenamefont {Derdzinski}\ \emph {et~al.}(2019)\citenamefont
  {Derdzinski}, \citenamefont {D'Orazio}, \citenamefont {Duffell},
  \citenamefont {Haiman},\ and\ \citenamefont
  {MacFadyen}}]{Derdzinski:2018qzv}%
  \BibitemOpen
  \bibfield  {author} {\bibinfo {author} {\bibfnamefont {A.~M.}\ \bibnamefont
  {Derdzinski}}, \bibinfo {author} {\bibfnamefont {D.}~\bibnamefont
  {D'Orazio}}, \bibinfo {author} {\bibfnamefont {P.}~\bibnamefont {Duffell}},
  \bibinfo {author} {\bibfnamefont {Z.}~\bibnamefont {Haiman}},\ and\ \bibinfo
  {author} {\bibfnamefont {A.}~\bibnamefont {MacFadyen}},\ }\bibfield  {title}
  {\bibinfo {title} {{Probing gas disc physics with LISA: simulations of an
  intermediate mass ratio inspiral in an accretion disc}},\ }\href
  {https://doi.org/10.1093/mnras/stz1026} {\bibfield  {journal} {\bibinfo
  {journal} {Mon. Not. Roy. Astron. Soc.}\ }\textbf {\bibinfo {volume} {486}},\
  \bibinfo {pages} {2754} (\bibinfo {year} {2019})},\ \bibinfo {note}
  {[Erratum: Mon.Not.Roy.Astron.Soc. 489, 4860--4861 (2019)]},\ \Eprint
  {https://arxiv.org/abs/1810.03623} {arXiv:1810.03623 [astro-ph.HE]}
  \BibitemShut {NoStop}%
\bibitem [{\citenamefont {Abramowicz}\ and\ \citenamefont
  {Fragile}(2013)}]{Abramowicz:2011xu}%
  \BibitemOpen
  \bibfield  {author} {\bibinfo {author} {\bibfnamefont {M.~A.}\ \bibnamefont
  {Abramowicz}}\ and\ \bibinfo {author} {\bibfnamefont {P.~C.}\ \bibnamefont
  {Fragile}},\ }\bibfield  {title} {\bibinfo {title} {{Foundations of Black
  Hole Accretion Disk Theory}},\ }\href {https://doi.org/10.12942/lrr-2013-1}
  {\bibfield  {journal} {\bibinfo  {journal} {Living Rev. Rel.}\ }\textbf
  {\bibinfo {volume} {16}},\ \bibinfo {pages} {1} (\bibinfo {year} {2013})},\
  \Eprint {https://arxiv.org/abs/1104.5499} {arXiv:1104.5499 [astro-ph.HE]}
  \BibitemShut {NoStop}%
\bibitem [{\citenamefont {Lightman}\ and\ \citenamefont
  {Eardley}(1974)}]{Lightman:1974sm}%
  \BibitemOpen
  \bibfield  {author} {\bibinfo {author} {\bibfnamefont {A.~P.}\ \bibnamefont
  {Lightman}}\ and\ \bibinfo {author} {\bibfnamefont {D.~M.}\ \bibnamefont
  {Eardley}},\ }\bibfield  {title} {\bibinfo {title} {{Black Holes in Binary
  Systems: Instability of Disk Accretion}},\ }\href
  {https://doi.org/10.1086/181377} {\bibfield  {journal} {\bibinfo  {journal}
  {Astrophys. J. Lett.}\ }\textbf {\bibinfo {volume} {187}},\ \bibinfo {pages}
  {L1} (\bibinfo {year} {1974})}\BibitemShut {NoStop}%
\bibitem [{\citenamefont {Shakura}\ and\ \citenamefont
  {Sunyaev}(1976)}]{Shakura:1976xk}%
  \BibitemOpen
  \bibfield  {author} {\bibinfo {author} {\bibfnamefont {N.~I.}\ \bibnamefont
  {Shakura}}\ and\ \bibinfo {author} {\bibfnamefont {R.~A.}\ \bibnamefont
  {Sunyaev}},\ }\bibfield  {title} {\bibinfo {title} {{A Theory of the
  instability of disk accretion on to black holes and the variability of binary
  X-ray sources, galactic nuclei and quasars}},\ }\href@noop {} {\bibfield
  {journal} {\bibinfo  {journal} {Mon. Not. Roy. Astron. Soc.}\ }\textbf
  {\bibinfo {volume} {175}},\ \bibinfo {pages} {613} (\bibinfo {year}
  {1976})}\BibitemShut {NoStop}%
\bibitem [{\citenamefont {{Bisnovatyi-Kogan}}\ and\ \citenamefont
  {{Blinnikov}}(1977)}]{Bisnovatyi1977}%
  \BibitemOpen
  \bibfield  {author} {\bibinfo {author} {\bibfnamefont {G.~S.}\ \bibnamefont
  {{Bisnovatyi-Kogan}}}\ and\ \bibinfo {author} {\bibfnamefont {S.~I.}\
  \bibnamefont {{Blinnikov}}},\ }\bibfield  {title} {\bibinfo {title} {{Disk
  accretion onto a black hole at subcritical luminosity.}},\ }\href@noop {}
  {\bibfield  {journal} {\bibinfo  {journal} {\aap}\ }\textbf {\bibinfo
  {volume} {59}},\ \bibinfo {pages} {111} (\bibinfo {year} {1977})}\BibitemShut
  {NoStop}%
\bibitem [{\citenamefont {{Piran}}(1978)}]{1978ApJ...221..652P}%
  \BibitemOpen
  \bibfield  {author} {\bibinfo {author} {\bibfnamefont {T.}~\bibnamefont
  {{Piran}}},\ }\bibfield  {title} {\bibinfo {title} {{The role of viscosity
  and cooling mechanisms in the stability of accretion disks.}},\ }\href
  {https://doi.org/10.1086/156069} {\bibfield  {journal} {\bibinfo  {journal}
  {\apj}\ }\textbf {\bibinfo {volume} {221}},\ \bibinfo {pages} {652} (\bibinfo
  {year} {1978})}\BibitemShut {NoStop}%
\bibitem [{\citenamefont {Davis}\ \emph {et~al.}(2010)\citenamefont {Davis},
  \citenamefont {Stone},\ and\ \citenamefont {Pessah}}]{Davis:2009sc}%
  \BibitemOpen
  \bibfield  {author} {\bibinfo {author} {\bibfnamefont {S.~W.}\ \bibnamefont
  {Davis}}, \bibinfo {author} {\bibfnamefont {J.~M.}\ \bibnamefont {Stone}},\
  and\ \bibinfo {author} {\bibfnamefont {M.~E.}\ \bibnamefont {Pessah}},\
  }\bibfield  {title} {\bibinfo {title} {{Sustained Magnetorotational
  Turbulence in Local Simulations of Stratified Disks with Zero Net Magnetic
  Flux}},\ }\href {https://doi.org/10.1088/0004-637X/713/1/52} {\bibfield
  {journal} {\bibinfo  {journal} {Astrophys. J.}\ }\textbf {\bibinfo {volume}
  {713}},\ \bibinfo {pages} {52} (\bibinfo {year} {2010})},\ \Eprint
  {https://arxiv.org/abs/0909.1570} {arXiv:0909.1570 [astro-ph.HE]}
  \BibitemShut {NoStop}%
\bibitem [{\citenamefont {Frank}\ \emph {et~al.}(2002)\citenamefont {Frank},
  \citenamefont {King},\ and\ \citenamefont {Raine}}]{Frank92}%
  \BibitemOpen
  \bibfield  {author} {\bibinfo {author} {\bibfnamefont {J.}~\bibnamefont
  {Frank}}, \bibinfo {author} {\bibfnamefont {A.}~\bibnamefont {King}},\ and\
  \bibinfo {author} {\bibfnamefont {D.}~\bibnamefont {Raine}},\ }\href
  {https://doi.org/10.1017/CBO9781139164245} {\emph {\bibinfo {title}
  {Accretion Power in Astrophysics}}},\ \bibinfo {edition} {3rd}\ ed.\
  (\bibinfo  {publisher} {Cambridge University Press},\ \bibinfo {year}
  {2002})\BibitemShut {NoStop}%
\bibitem [{Note2()}]{Note2}%
  \BibitemOpen
  \bibinfo {note} {When the thin disk condition is violated (e.g.~for
  super-Eddington accretion), the disk is better described by a slim-disk
  solution \protect \citep {Abramowicz:1988sp,Abramowicz:2011xu}.}\BibitemShut
  {Stop}%
\bibitem [{\citenamefont {Jiang}\ \emph
  {et~al.}(2019{\natexlab{b}})\citenamefont {Jiang}, \citenamefont {Fabian},
  \citenamefont {Dauser}, \citenamefont {Gallo}, \citenamefont {Garcia},
  \citenamefont {Kara}, \citenamefont {Parker}, \citenamefont {Tomsick},
  \citenamefont {Walton},\ and\ \citenamefont {Reynolds}}]{Jiang:2019ztr}%
  \BibitemOpen
  \bibfield  {author} {\bibinfo {author} {\bibfnamefont {J.}~\bibnamefont
  {Jiang}}, \bibinfo {author} {\bibfnamefont {A.~C.}\ \bibnamefont {Fabian}},
  \bibinfo {author} {\bibfnamefont {T.}~\bibnamefont {Dauser}}, \bibinfo
  {author} {\bibfnamefont {L.}~\bibnamefont {Gallo}}, \bibinfo {author}
  {\bibfnamefont {J.~A.}\ \bibnamefont {Garcia}}, \bibinfo {author}
  {\bibfnamefont {E.}~\bibnamefont {Kara}}, \bibinfo {author} {\bibfnamefont
  {M.~L.}\ \bibnamefont {Parker}}, \bibinfo {author} {\bibfnamefont {J.~A.}\
  \bibnamefont {Tomsick}}, \bibinfo {author} {\bibfnamefont {D.~J.}\
  \bibnamefont {Walton}},\ and\ \bibinfo {author} {\bibfnamefont {C.~S.}\
  \bibnamefont {Reynolds}},\ }\bibfield  {title} {\bibinfo {title} {{High
  Density Reflection Spectroscopy \textendash{} II. The density of the inner
  black hole accretion disc in AGN}},\ }\href
  {https://doi.org/10.1093/mnras/stz2326} {\bibfield  {journal} {\bibinfo
  {journal} {Mon. Not. Roy. Astron. Soc.}\ }\textbf {\bibinfo {volume} {489}},\
  \bibinfo {pages} {3436} (\bibinfo {year} {2019}{\natexlab{b}})},\ \Eprint
  {https://arxiv.org/abs/1908.07272} {arXiv:1908.07272 [astro-ph.HE]}
  \BibitemShut {NoStop}%
\bibitem [{\citenamefont {Kesden}(2011)}]{Kesden:2011ma}%
  \BibitemOpen
  \bibfield  {author} {\bibinfo {author} {\bibfnamefont {M.}~\bibnamefont
  {Kesden}},\ }\bibfield  {title} {\bibinfo {title} {{Transition from adiabatic
  inspiral to plunge into a spinning black hole}},\ }\href
  {https://doi.org/10.1103/PhysRevD.83.104011} {\bibfield  {journal} {\bibinfo
  {journal} {Phys. Rev. D}\ }\textbf {\bibinfo {volume} {83}},\ \bibinfo
  {pages} {104011} (\bibinfo {year} {2011})},\ \Eprint
  {https://arxiv.org/abs/1101.3749} {arXiv:1101.3749 [gr-qc]} \BibitemShut
  {NoStop}%
\bibitem [{\citenamefont {Ori}\ and\ \citenamefont
  {Thorne}(2000)}]{PhysRevD.62.124022}%
  \BibitemOpen
  \bibfield  {author} {\bibinfo {author} {\bibfnamefont {A.}~\bibnamefont
  {Ori}}\ and\ \bibinfo {author} {\bibfnamefont {K.~S.}\ \bibnamefont
  {Thorne}},\ }\bibfield  {title} {\bibinfo {title} {Transition from inspiral
  to plunge for a compact body in a circular equatorial orbit around a massive,
  spinning black hole},\ }\href {https://doi.org/10.1103/PhysRevD.62.124022}
  {\bibfield  {journal} {\bibinfo  {journal} {Phys. Rev. D}\ }\textbf {\bibinfo
  {volume} {62}},\ \bibinfo {pages} {124022} (\bibinfo {year}
  {2000})}\BibitemShut {NoStop}%
\bibitem [{\citenamefont {Burke}\ \emph {et~al.}(2020)\citenamefont {Burke},
  \citenamefont {Gair},\ and\ \citenamefont {Sim\'on}}]{Burke:2019yek}%
  \BibitemOpen
  \bibfield  {author} {\bibinfo {author} {\bibfnamefont {O.}~\bibnamefont
  {Burke}}, \bibinfo {author} {\bibfnamefont {J.~R.}\ \bibnamefont {Gair}},\
  and\ \bibinfo {author} {\bibfnamefont {J.}~\bibnamefont {Sim\'on}},\
  }\bibfield  {title} {\bibinfo {title} {{Transition from Inspiral to Plunge: A
  Complete Near-Extremal Trajectory and Associated Waveform}},\ }\href
  {https://doi.org/10.1103/PhysRevD.101.064026} {\bibfield  {journal} {\bibinfo
   {journal} {Phys. Rev. D}\ }\textbf {\bibinfo {volume} {101}},\ \bibinfo
  {pages} {064026} (\bibinfo {year} {2020})},\ \Eprint
  {https://arxiv.org/abs/1909.12846} {arXiv:1909.12846 [gr-qc]} \BibitemShut
  {NoStop}%
\bibitem [{\citenamefont {{Armitage}}(2011)}]{Armitage2011}%
  \BibitemOpen
  \bibfield  {author} {\bibinfo {author} {\bibfnamefont {P.~J.}\ \bibnamefont
  {{Armitage}}},\ }\bibfield  {title} {\bibinfo {title} {{Dynamics of
  Protoplanetary Disks}},\ }\href
  {https://doi.org/10.1146/annurev-astro-081710-102521} {\bibfield  {journal}
  {\bibinfo  {journal} {\araa}\ }\textbf {\bibinfo {volume} {49}},\ \bibinfo
  {pages} {195} (\bibinfo {year} {2011})},\ \Eprint
  {https://arxiv.org/abs/1011.1496} {arXiv:1011.1496 [astro-ph.SR]}
  \BibitemShut {NoStop}%
\bibitem [{\citenamefont {{Paardekooper}}\ \emph {et~al.}(2022)\citenamefont
  {{Paardekooper}}, \citenamefont {{Dong}}, \citenamefont {{Duffell}},
  \citenamefont {{Fung}}, \citenamefont {{Masset}}, \citenamefont {{Ogilvie}},\
  and\ \citenamefont {{Tanaka}}}]{Paardekooper2022}%
  \BibitemOpen
  \bibfield  {author} {\bibinfo {author} {\bibfnamefont {S.-J.}\ \bibnamefont
  {{Paardekooper}}}, \bibinfo {author} {\bibfnamefont {R.}~\bibnamefont
  {{Dong}}}, \bibinfo {author} {\bibfnamefont {P.}~\bibnamefont {{Duffell}}},
  \bibinfo {author} {\bibfnamefont {J.}~\bibnamefont {{Fung}}}, \bibinfo
  {author} {\bibfnamefont {F.~S.}\ \bibnamefont {{Masset}}}, \bibinfo {author}
  {\bibfnamefont {G.}~\bibnamefont {{Ogilvie}}},\ and\ \bibinfo {author}
  {\bibfnamefont {H.}~\bibnamefont {{Tanaka}}},\ }\bibfield  {title} {\bibinfo
  {title} {{Planet-Disk Interactions}},\ }\href@noop {} {\bibfield  {journal}
  {\bibinfo  {journal} {arXiv e-prints}\ ,\ \bibinfo {eid} {arXiv:2203.09595}}
  (\bibinfo {year} {2022})},\ \Eprint {https://arxiv.org/abs/2203.09595}
  {arXiv:2203.09595 [astro-ph.EP]} \BibitemShut {NoStop}%
\bibitem [{\citenamefont {{Goldreich}}\ and\ \citenamefont
  {{Tremaine}}(1980)}]{Goldreich1980}%
  \BibitemOpen
  \bibfield  {author} {\bibinfo {author} {\bibfnamefont {P.}~\bibnamefont
  {{Goldreich}}}\ and\ \bibinfo {author} {\bibfnamefont {S.}~\bibnamefont
  {{Tremaine}}},\ }\bibfield  {title} {\bibinfo {title} {{Disk-satellite
  interactions.}},\ }\href {https://doi.org/10.1086/158356} {\bibfield
  {journal} {\bibinfo  {journal} {\apj}\ }\textbf {\bibinfo {volume} {241}},\
  \bibinfo {pages} {425} (\bibinfo {year} {1980})}\BibitemShut {NoStop}%
\bibitem [{\citenamefont {{Tanaka}}\ \emph {et~al.}(2002)\citenamefont
  {{Tanaka}}, \citenamefont {{Takeuchi}},\ and\ \citenamefont
  {{Ward}}}]{Tanaka2002}%
  \BibitemOpen
  \bibfield  {author} {\bibinfo {author} {\bibfnamefont {H.}~\bibnamefont
  {{Tanaka}}}, \bibinfo {author} {\bibfnamefont {T.}~\bibnamefont
  {{Takeuchi}}},\ and\ \bibinfo {author} {\bibfnamefont {W.~R.}\ \bibnamefont
  {{Ward}}},\ }\bibfield  {title} {\bibinfo {title} {{Three-Dimensional
  Interaction between a Planet and an Isothermal Gaseous Disk. I. Corotation
  and Lindblad Torques and Planet Migration}},\ }\href
  {https://doi.org/10.1086/324713} {\bibfield  {journal} {\bibinfo  {journal}
  {\apj}\ }\textbf {\bibinfo {volume} {565}},\ \bibinfo {pages} {1257}
  (\bibinfo {year} {2002})}\BibitemShut {NoStop}%
\bibitem [{\citenamefont {{Casoli}}\ and\ \citenamefont
  {{Masset}}(2009)}]{Casoli2009}%
  \BibitemOpen
  \bibfield  {author} {\bibinfo {author} {\bibfnamefont {J.}~\bibnamefont
  {{Casoli}}}\ and\ \bibinfo {author} {\bibfnamefont {F.~S.}\ \bibnamefont
  {{Masset}}},\ }\bibfield  {title} {\bibinfo {title} {{On the Horseshoe Drag
  of a Low-Mass Planet. I. Migration in Isothermal Disks}},\ }\href
  {https://doi.org/10.1088/0004-637X/703/1/845} {\bibfield  {journal} {\bibinfo
   {journal} {\apj}\ }\textbf {\bibinfo {volume} {703}},\ \bibinfo {pages}
  {845} (\bibinfo {year} {2009})},\ \Eprint {https://arxiv.org/abs/0907.4677}
  {arXiv:0907.4677 [astro-ph.EP]} \BibitemShut {NoStop}%
\bibitem [{\citenamefont {{Paardekooper}}\ and\ \citenamefont
  {{Papaloizou}}(2009)}]{Paardekooper2009}%
  \BibitemOpen
  \bibfield  {author} {\bibinfo {author} {\bibfnamefont {S.~J.}\ \bibnamefont
  {{Paardekooper}}}\ and\ \bibinfo {author} {\bibfnamefont {J.~C.~B.}\
  \bibnamefont {{Papaloizou}}},\ }\bibfield  {title} {\bibinfo {title} {{On
  corotation torques, horseshoe drag and the possibility of sustained stalled
  or outward protoplanetary migration}},\ }\href
  {https://doi.org/10.1111/j.1365-2966.2009.14511.x} {\bibfield  {journal}
  {\bibinfo  {journal} {\mnras}\ }\textbf {\bibinfo {volume} {394}},\ \bibinfo
  {pages} {2283} (\bibinfo {year} {2009})},\ \Eprint
  {https://arxiv.org/abs/0901.2265} {arXiv:0901.2265 [astro-ph.EP]}
  \BibitemShut {NoStop}%
\bibitem [{\citenamefont {{Baruteau}}\ \emph {et~al.}(2014)\citenamefont
  {{Baruteau}}, \citenamefont {{Crida}}, \citenamefont {{Paardekooper}},
  \citenamefont {{Masset}}, \citenamefont {{Guilet}}, \citenamefont {{Bitsch}},
  \citenamefont {{Nelson}}, \citenamefont {{Kley}},\ and\ \citenamefont
  {{Papaloizou}}}]{Baruteau2014}%
  \BibitemOpen
  \bibfield  {author} {\bibinfo {author} {\bibfnamefont {C.}~\bibnamefont
  {{Baruteau}}}, \bibinfo {author} {\bibfnamefont {A.}~\bibnamefont {{Crida}}},
  \bibinfo {author} {\bibfnamefont {S.~J.}\ \bibnamefont {{Paardekooper}}},
  \bibinfo {author} {\bibfnamefont {F.}~\bibnamefont {{Masset}}}, \bibinfo
  {author} {\bibfnamefont {J.}~\bibnamefont {{Guilet}}}, \bibinfo {author}
  {\bibfnamefont {B.}~\bibnamefont {{Bitsch}}}, \bibinfo {author}
  {\bibfnamefont {R.}~\bibnamefont {{Nelson}}}, \bibinfo {author}
  {\bibfnamefont {W.}~\bibnamefont {{Kley}}},\ and\ \bibinfo {author}
  {\bibfnamefont {J.}~\bibnamefont {{Papaloizou}}},\ }\bibfield  {title}
  {\bibinfo {title} {{Planet-Disk Interactions and Early Evolution of Planetary
  Systems}},\ }in\ \href
  {https://doi.org/10.2458/azu_uapress_9780816531240-ch029} {\emph {\bibinfo
  {booktitle} {Protostars and Planets VI}}},\ \bibinfo {editor} {edited by\
  \bibinfo {editor} {\bibfnamefont {H.}~\bibnamefont {{Beuther}}}, \bibinfo
  {editor} {\bibfnamefont {R.~S.}\ \bibnamefont {{Klessen}}}, \bibinfo {editor}
  {\bibfnamefont {C.~P.}\ \bibnamefont {{Dullemond}}},\ and\ \bibinfo {editor}
  {\bibfnamefont {T.}~\bibnamefont {{Henning}}}}\ (\bibinfo {year} {2014})\ p.\
  \bibinfo {pages} {667},\ \Eprint {https://arxiv.org/abs/1312.4293}
  {arXiv:1312.4293 [astro-ph.EP]} \BibitemShut {NoStop}%
\bibitem [{\citenamefont {Malik}\ \emph {et~al.}(2015)\citenamefont {Malik},
  \citenamefont {Meru}, \citenamefont {Mayer},\ and\ \citenamefont
  {Meyer}}]{Malik_2015}%
  \BibitemOpen
  \bibfield  {author} {\bibinfo {author} {\bibfnamefont {M.}~\bibnamefont
  {Malik}}, \bibinfo {author} {\bibfnamefont {F.}~\bibnamefont {Meru}},
  \bibinfo {author} {\bibfnamefont {L.}~\bibnamefont {Mayer}},\ and\ \bibinfo
  {author} {\bibfnamefont {M.}~\bibnamefont {Meyer}},\ }\bibfield  {title}
  {\bibinfo {title} {On the gap-opening criterion of migrating planets in
  protoplanetary disks},\ }\href {https://doi.org/10.1088/0004-637X/802/1/56}
  {\bibfield  {journal} {\bibinfo  {journal} {The Astrophysical Journal}\
  }\textbf {\bibinfo {volume} {802}},\ \bibinfo {pages} {56} (\bibinfo {year}
  {2015})}\BibitemShut {NoStop}%
\bibitem [{\citenamefont {{Syer}}\ and\ \citenamefont
  {{Clarke}}(1995)}]{SyerClarke1995}%
  \BibitemOpen
  \bibfield  {author} {\bibinfo {author} {\bibfnamefont {D.}~\bibnamefont
  {{Syer}}}\ and\ \bibinfo {author} {\bibfnamefont {C.~J.}\ \bibnamefont
  {{Clarke}}},\ }\bibfield  {title} {\bibinfo {title} {{Satellites in discs:
  regulating the accretion luminosity}},\ }\href
  {https://doi.org/10.1093/mnras/277.3.758} {\bibfield  {journal} {\bibinfo
  {journal} {\mnras}\ }\textbf {\bibinfo {volume} {277}},\ \bibinfo {pages}
  {758} (\bibinfo {year} {1995})},\ \Eprint
  {https://arxiv.org/abs/astro-ph/9505021} {arXiv:astro-ph/9505021 [astro-ph]}
  \BibitemShut {NoStop}%
\bibitem [{\citenamefont {Duffell}\ \emph {et~al.}(2014)\citenamefont
  {Duffell}, \citenamefont {Haiman}, \citenamefont {MacFadyen}, \citenamefont
  {D'Orazio},\ and\ \citenamefont {Farris}}]{Duffell_2014}%
  \BibitemOpen
  \bibfield  {author} {\bibinfo {author} {\bibfnamefont {P.~C.}\ \bibnamefont
  {Duffell}}, \bibinfo {author} {\bibfnamefont {Z.}~\bibnamefont {Haiman}},
  \bibinfo {author} {\bibfnamefont {A.~I.}\ \bibnamefont {MacFadyen}}, \bibinfo
  {author} {\bibfnamefont {D.~J.}\ \bibnamefont {D'Orazio}},\ and\ \bibinfo
  {author} {\bibfnamefont {B.~D.}\ \bibnamefont {Farris}},\ }\bibfield  {title}
  {\bibinfo {title} {{THE} {MIGRATION} {OF} {GAP}-{OPENING} {PLANETS} {IS}
  {NOT} {LOCKED} {TO} {VISCOUS} {DISK} {EVOLUTION}},\ }\href
  {https://doi.org/10.1088/2041-8205/792/1/l10} {\bibfield  {journal} {\bibinfo
   {journal} {The Astrophysical Journal}\ }\textbf {\bibinfo {volume} {792}},\
  \bibinfo {pages} {L10} (\bibinfo {year} {2014})}\BibitemShut {NoStop}%
\bibitem [{\citenamefont {{D{\"u}rmann}}\ and\ \citenamefont
  {{Kley}}(2017)}]{Durmann2017}%
  \BibitemOpen
  \bibfield  {author} {\bibinfo {author} {\bibfnamefont {C.}~\bibnamefont
  {{D{\"u}rmann}}}\ and\ \bibinfo {author} {\bibfnamefont {W.}~\bibnamefont
  {{Kley}}},\ }\bibfield  {title} {\bibinfo {title} {{The accretion of
  migrating giant planets}},\ }\href
  {https://doi.org/10.1051/0004-6361/201629074} {\bibfield  {journal} {\bibinfo
   {journal} {\aap}\ }\textbf {\bibinfo {volume} {598}},\ \bibinfo {eid} {A80}
  (\bibinfo {year} {2017})},\ \Eprint {https://arxiv.org/abs/1611.01070}
  {arXiv:1611.01070 [astro-ph.EP]} \BibitemShut {NoStop}%
\bibitem [{\citenamefont {{Kanagawa}}\ \emph {et~al.}(2018)\citenamefont
  {{Kanagawa}}, \citenamefont {{Tanaka}},\ and\ \citenamefont
  {{Szuszkiewicz}}}]{Kanagawa2018}%
  \BibitemOpen
  \bibfield  {author} {\bibinfo {author} {\bibfnamefont {K.~D.}\ \bibnamefont
  {{Kanagawa}}}, \bibinfo {author} {\bibfnamefont {H.}~\bibnamefont
  {{Tanaka}}},\ and\ \bibinfo {author} {\bibfnamefont {E.}~\bibnamefont
  {{Szuszkiewicz}}},\ }\bibfield  {title} {\bibinfo {title} {{Radial Migration
  of Gap-opening Planets in Protoplanetary Disks. I. The Case of a Single
  Planet}},\ }\href {https://doi.org/10.3847/1538-4357/aac8d9} {\bibfield
  {journal} {\bibinfo  {journal} {\apj}\ }\textbf {\bibinfo {volume} {861}},\
  \bibinfo {eid} {140} (\bibinfo {year} {2018})},\ \Eprint
  {https://arxiv.org/abs/1805.11101} {arXiv:1805.11101 [astro-ph.EP]}
  \BibitemShut {NoStop}%
\bibitem [{\citenamefont {{Scardoni}}\ \emph {et~al.}(2020)\citenamefont
  {{Scardoni}}, \citenamefont {{Rosotti}}, \citenamefont {{Lodato}},\ and\
  \citenamefont {{Clarke}}}]{Scardoni2020}%
  \BibitemOpen
  \bibfield  {author} {\bibinfo {author} {\bibfnamefont {C.~E.}\ \bibnamefont
  {{Scardoni}}}, \bibinfo {author} {\bibfnamefont {G.~P.}\ \bibnamefont
  {{Rosotti}}}, \bibinfo {author} {\bibfnamefont {G.}~\bibnamefont
  {{Lodato}}},\ and\ \bibinfo {author} {\bibfnamefont {C.~J.}\ \bibnamefont
  {{Clarke}}},\ }\bibfield  {title} {\bibinfo {title} {{Type II migration
  strikes back - an old paradigm for planet migration in discs}},\ }\href
  {https://doi.org/10.1093/mnras/stz3534} {\bibfield  {journal} {\bibinfo
  {journal} {\mnras}\ }\textbf {\bibinfo {volume} {492}},\ \bibinfo {pages}
  {1318} (\bibinfo {year} {2020})},\ \Eprint {https://arxiv.org/abs/1912.07313}
  {arXiv:1912.07313 [astro-ph.EP]} \BibitemShut {NoStop}%
\bibitem [{\citenamefont {Cornish}\ \emph {et~al.}(2011)\citenamefont
  {Cornish}, \citenamefont {Sampson}, \citenamefont {Yunes},\ and\
  \citenamefont {Pretorius}}]{Cornish:2011ys}%
  \BibitemOpen
  \bibfield  {author} {\bibinfo {author} {\bibfnamefont {N.}~\bibnamefont
  {Cornish}}, \bibinfo {author} {\bibfnamefont {L.}~\bibnamefont {Sampson}},
  \bibinfo {author} {\bibfnamefont {N.}~\bibnamefont {Yunes}},\ and\ \bibinfo
  {author} {\bibfnamefont {F.}~\bibnamefont {Pretorius}},\ }\bibfield  {title}
  {\bibinfo {title} {{Gravitational Wave Tests of General Relativity with the
  Parameterized Post-Einsteinian Framework}},\ }\href
  {https://doi.org/10.1103/PhysRevD.84.062003} {\bibfield  {journal} {\bibinfo
  {journal} {Phys. Rev. D}\ }\textbf {\bibinfo {volume} {84}},\ \bibinfo
  {pages} {062003} (\bibinfo {year} {2011})},\ \Eprint
  {https://arxiv.org/abs/1105.2088} {arXiv:1105.2088 [gr-qc]} \BibitemShut
  {NoStop}%
\bibitem [{\citenamefont {Abbott}\ \emph
  {et~al.}(2021{\natexlab{a}})\citenamefont {Abbott} \emph
  {et~al.}}]{LIGOScientific:2021sio}%
  \BibitemOpen
  \bibfield  {author} {\bibinfo {author} {\bibfnamefont {R.}~\bibnamefont
  {Abbott}} \emph {et~al.} (\bibinfo {collaboration} {LIGO Scientific, VIRGO,
  KAGRA}),\ }\href@noop {} {\bibinfo {title} {{Tests of General Relativity with
  GWTC-3}}} (\bibinfo {year} {2021}{\natexlab{a}}),\ \Eprint
  {https://arxiv.org/abs/2112.06861} {arXiv:2112.06861 [gr-qc]} \BibitemShut
  {NoStop}%
\bibitem [{\citenamefont {Barack}\ and\ \citenamefont
  {Cutler}(2004)}]{Barack:2003fp}%
  \BibitemOpen
  \bibfield  {author} {\bibinfo {author} {\bibfnamefont {L.}~\bibnamefont
  {Barack}}\ and\ \bibinfo {author} {\bibfnamefont {C.}~\bibnamefont
  {Cutler}},\ }\bibfield  {title} {\bibinfo {title} {{LISA capture sources:
  Approximate waveforms, signal-to-noise ratios, and parameter estimation
  accuracy}},\ }\href {https://doi.org/10.1103/PhysRevD.69.082005} {\bibfield
  {journal} {\bibinfo  {journal} {Phys. Rev. D}\ }\textbf {\bibinfo {volume}
  {69}},\ \bibinfo {pages} {082005} (\bibinfo {year} {2004})},\ \Eprint
  {https://arxiv.org/abs/gr-qc/0310125} {arXiv:gr-qc/0310125} \BibitemShut
  {NoStop}%
\bibitem [{\citenamefont {Glampedakis}\ and\ \citenamefont
  {Babak}(2006)}]{Glampedakis:2005cf}%
  \BibitemOpen
  \bibfield  {author} {\bibinfo {author} {\bibfnamefont {K.}~\bibnamefont
  {Glampedakis}}\ and\ \bibinfo {author} {\bibfnamefont {S.}~\bibnamefont
  {Babak}},\ }\bibfield  {title} {\bibinfo {title} {{Mapping spacetimes with
  LISA: Inspiral of a test-body in a `quasi-Kerr' field}},\ }\href
  {https://doi.org/10.1088/0264-9381/23/12/013} {\bibfield  {journal} {\bibinfo
   {journal} {Class. Quant. Grav.}\ }\textbf {\bibinfo {volume} {23}},\
  \bibinfo {pages} {4167} (\bibinfo {year} {2006})},\ \Eprint
  {https://arxiv.org/abs/gr-qc/0510057} {arXiv:gr-qc/0510057} \BibitemShut
  {NoStop}%
\bibitem [{\citenamefont {Barack}\ and\ \citenamefont
  {Cutler}(2007)}]{Barack:2006pq}%
  \BibitemOpen
  \bibfield  {author} {\bibinfo {author} {\bibfnamefont {L.}~\bibnamefont
  {Barack}}\ and\ \bibinfo {author} {\bibfnamefont {C.}~\bibnamefont
  {Cutler}},\ }\bibfield  {title} {\bibinfo {title} {{Using LISA EMRI sources
  to test off-Kerr deviations in the geometry of massive black holes}},\ }\href
  {https://doi.org/10.1103/PhysRevD.75.042003} {\bibfield  {journal} {\bibinfo
  {journal} {Phys. Rev. D}\ }\textbf {\bibinfo {volume} {75}},\ \bibinfo
  {pages} {042003} (\bibinfo {year} {2007})},\ \Eprint
  {https://arxiv.org/abs/gr-qc/0612029} {arXiv:gr-qc/0612029} \BibitemShut
  {NoStop}%
\bibitem [{\citenamefont {Babak}\ \emph {et~al.}(2017)\citenamefont {Babak},
  \citenamefont {Gair}, \citenamefont {Sesana}, \citenamefont {Barausse},
  \citenamefont {Sopuerta}, \citenamefont {Berry}, \citenamefont {Berti},
  \citenamefont {Amaro-Seoane}, \citenamefont {Petiteau},\ and\ \citenamefont
  {Klein}}]{Babak:2017tow}%
  \BibitemOpen
  \bibfield  {author} {\bibinfo {author} {\bibfnamefont {S.}~\bibnamefont
  {Babak}}, \bibinfo {author} {\bibfnamefont {J.}~\bibnamefont {Gair}},
  \bibinfo {author} {\bibfnamefont {A.}~\bibnamefont {Sesana}}, \bibinfo
  {author} {\bibfnamefont {E.}~\bibnamefont {Barausse}}, \bibinfo {author}
  {\bibfnamefont {C.~F.}\ \bibnamefont {Sopuerta}}, \bibinfo {author}
  {\bibfnamefont {C.~P.~L.}\ \bibnamefont {Berry}}, \bibinfo {author}
  {\bibfnamefont {E.}~\bibnamefont {Berti}}, \bibinfo {author} {\bibfnamefont
  {P.}~\bibnamefont {Amaro-Seoane}}, \bibinfo {author} {\bibfnamefont
  {A.}~\bibnamefont {Petiteau}},\ and\ \bibinfo {author} {\bibfnamefont
  {A.}~\bibnamefont {Klein}},\ }\bibfield  {title} {\bibinfo {title} {{Science
  with the space-based interferometer LISA. V: Extreme mass-ratio inspirals}},\
  }\href {https://doi.org/10.1103/PhysRevD.95.103012} {\bibfield  {journal}
  {\bibinfo  {journal} {Phys. Rev. D}\ }\textbf {\bibinfo {volume} {95}},\
  \bibinfo {pages} {103012} (\bibinfo {year} {2017})},\ \Eprint
  {https://arxiv.org/abs/1703.09722} {arXiv:1703.09722 [gr-qc]} \BibitemShut
  {NoStop}%
\bibitem [{\citenamefont {Barack}\ and\ \citenamefont
  {Pound}(2019)}]{Barack:2018yvs}%
  \BibitemOpen
  \bibfield  {author} {\bibinfo {author} {\bibfnamefont {L.}~\bibnamefont
  {Barack}}\ and\ \bibinfo {author} {\bibfnamefont {A.}~\bibnamefont {Pound}},\
  }\bibfield  {title} {\bibinfo {title} {{Self-force and radiation reaction in
  general relativity}},\ }\href {https://doi.org/10.1088/1361-6633/aae552}
  {\bibfield  {journal} {\bibinfo  {journal} {Rept. Prog. Phys.}\ }\textbf
  {\bibinfo {volume} {82}},\ \bibinfo {pages} {016904} (\bibinfo {year}
  {2019})},\ \Eprint {https://arxiv.org/abs/1805.10385} {arXiv:1805.10385
  [gr-qc]} \BibitemShut {NoStop}%
\bibitem [{\citenamefont {Mino}\ \emph {et~al.}(1997)\citenamefont {Mino},
  \citenamefont {Sasaki},\ and\ \citenamefont {Tanaka}}]{PhysRevD.55.3457}%
  \BibitemOpen
  \bibfield  {author} {\bibinfo {author} {\bibfnamefont {Y.}~\bibnamefont
  {Mino}}, \bibinfo {author} {\bibfnamefont {M.}~\bibnamefont {Sasaki}},\ and\
  \bibinfo {author} {\bibfnamefont {T.}~\bibnamefont {Tanaka}},\ }\bibfield
  {title} {\bibinfo {title} {Gravitational radiation reaction to a particle
  motion},\ }\href {https://doi.org/10.1103/PhysRevD.55.3457} {\bibfield
  {journal} {\bibinfo  {journal} {Phys. Rev. D}\ }\textbf {\bibinfo {volume}
  {55}},\ \bibinfo {pages} {3457} (\bibinfo {year} {1997})}\BibitemShut
  {NoStop}%
\bibitem [{\citenamefont {Quinn}\ and\ \citenamefont
  {Wald}(1997)}]{PhysRevD.56.3381}%
  \BibitemOpen
  \bibfield  {author} {\bibinfo {author} {\bibfnamefont {T.~C.}\ \bibnamefont
  {Quinn}}\ and\ \bibinfo {author} {\bibfnamefont {R.~M.}\ \bibnamefont
  {Wald}},\ }\bibfield  {title} {\bibinfo {title} {Axiomatic approach to
  electromagnetic and gravitational radiation reaction of particles in curved
  spacetime},\ }\href {https://doi.org/10.1103/PhysRevD.56.3381} {\bibfield
  {journal} {\bibinfo  {journal} {Phys. Rev. D}\ }\textbf {\bibinfo {volume}
  {56}},\ \bibinfo {pages} {3381} (\bibinfo {year} {1997})}\BibitemShut
  {NoStop}%
\bibitem [{Note3()}]{Note3}%
  \BibitemOpen
  \bibinfo {note} {Since the environmental effects considered in this work
  appear at negative PN orders, we do not expect post-adiabatic corrections to
  significantly affect our results.}\BibitemShut {Stop}%
\bibitem [{BHP()}]{BHPToolkit}%
  \BibitemOpen
  \href@noop {} {\bibinfo {title} {{Black Hole Perturbation Toolkit}}},\
  \bibinfo {howpublished}
  {(\href{http://bhptoolkit.org/}{bhptoolkit.org})}\BibitemShut {NoStop}%
\bibitem [{\citenamefont {Chua}\ \emph {et~al.}(2017)\citenamefont {Chua},
  \citenamefont {Moore},\ and\ \citenamefont {Gair}}]{Chua:2017ujo}%
  \BibitemOpen
  \bibfield  {author} {\bibinfo {author} {\bibfnamefont {A.~J.~K.}\
  \bibnamefont {Chua}}, \bibinfo {author} {\bibfnamefont {C.~J.}\ \bibnamefont
  {Moore}},\ and\ \bibinfo {author} {\bibfnamefont {J.~R.}\ \bibnamefont
  {Gair}},\ }\bibfield  {title} {\bibinfo {title} {{Augmented kludge waveforms
  for detecting extreme-mass-ratio inspirals}},\ }\href
  {https://doi.org/10.1103/PhysRevD.96.044005} {\bibfield  {journal} {\bibinfo
  {journal} {Phys. Rev. D}\ }\textbf {\bibinfo {volume} {96}},\ \bibinfo
  {pages} {044005} (\bibinfo {year} {2017})},\ \Eprint
  {https://arxiv.org/abs/1705.04259} {arXiv:1705.04259 [gr-qc]} \BibitemShut
  {NoStop}%
\bibitem [{\citenamefont {Bardeen}\ and\ \citenamefont
  {Petterson}(1975)}]{Bardeen:1975zz}%
  \BibitemOpen
  \bibfield  {author} {\bibinfo {author} {\bibfnamefont {J.~M.}\ \bibnamefont
  {Bardeen}}\ and\ \bibinfo {author} {\bibfnamefont {J.~A.}\ \bibnamefont
  {Petterson}},\ }\bibfield  {title} {\bibinfo {title} {{The Lense-Thirring
  Effect and Accretion Disks around Kerr Black Holes}},\ }\href
  {https://doi.org/10.1086/181711} {\bibfield  {journal} {\bibinfo  {journal}
  {Astrophys. J. Lett.}\ }\textbf {\bibinfo {volume} {195}},\ \bibinfo {pages}
  {L65} (\bibinfo {year} {1975})}\BibitemShut {NoStop}%
\bibitem [{\citenamefont {Bogdanovic}\ \emph {et~al.}(2007)\citenamefont
  {Bogdanovic}, \citenamefont {Reynolds},\ and\ \citenamefont
  {Miller}}]{Bogdanovic:2007hp}%
  \BibitemOpen
  \bibfield  {author} {\bibinfo {author} {\bibfnamefont {T.}~\bibnamefont
  {Bogdanovic}}, \bibinfo {author} {\bibfnamefont {C.~S.}\ \bibnamefont
  {Reynolds}},\ and\ \bibinfo {author} {\bibfnamefont {M.~C.}\ \bibnamefont
  {Miller}},\ }\bibfield  {title} {\bibinfo {title} {{Alignment of the spins of
  supermassive black holes prior to merger}},\ }\href
  {https://doi.org/10.1086/518769} {\bibfield  {journal} {\bibinfo  {journal}
  {Astrophys. J. Lett.}\ }\textbf {\bibinfo {volume} {661}},\ \bibinfo {pages}
  {L147} (\bibinfo {year} {2007})},\ \Eprint
  {https://arxiv.org/abs/astro-ph/0703054} {arXiv:astro-ph/0703054}
  \BibitemShut {NoStop}%
\bibitem [{\citenamefont {Perego}\ \emph {et~al.}(2009)\citenamefont {Perego},
  \citenamefont {Dotti}, \citenamefont {Colpi},\ and\ \citenamefont
  {Volonteri}}]{Perego:2009cw}%
  \BibitemOpen
  \bibfield  {author} {\bibinfo {author} {\bibfnamefont {A.}~\bibnamefont
  {Perego}}, \bibinfo {author} {\bibfnamefont {M.}~\bibnamefont {Dotti}},
  \bibinfo {author} {\bibfnamefont {M.}~\bibnamefont {Colpi}},\ and\ \bibinfo
  {author} {\bibfnamefont {M.}~\bibnamefont {Volonteri}},\ }\bibfield  {title}
  {\bibinfo {title} {{Mass and spin coevolution during the alignment of a black
  hole in a warped accretion disc}},\ }\href
  {https://doi.org/10.1111/j.1365-2966.2009.15427.x} {\bibfield  {journal}
  {\bibinfo  {journal} {Mon. Not. Roy. Astron. Soc.}\ }\textbf {\bibinfo
  {volume} {399}},\ \bibinfo {pages} {2249} (\bibinfo {year} {2009})},\ \Eprint
  {https://arxiv.org/abs/0907.3742} {arXiv:0907.3742 [astro-ph.CO]}
  \BibitemShut {NoStop}%
\bibitem [{\citenamefont {Ostriker}(1999)}]{Ostriker:1998fa}%
  \BibitemOpen
  \bibfield  {author} {\bibinfo {author} {\bibfnamefont {E.~C.}\ \bibnamefont
  {Ostriker}},\ }\bibfield  {title} {\bibinfo {title} {{Dynamical friction in a
  gaseous medium}},\ }\href {https://doi.org/10.1086/306858} {\bibfield
  {journal} {\bibinfo  {journal} {Astrophys. J.}\ }\textbf {\bibinfo {volume}
  {513}},\ \bibinfo {pages} {252} (\bibinfo {year} {1999})},\ \Eprint
  {https://arxiv.org/abs/astro-ph/9810324} {arXiv:astro-ph/9810324}
  \BibitemShut {NoStop}%
\bibitem [{\citenamefont {Barausse}(2007)}]{Barausse:2007ph}%
  \BibitemOpen
  \bibfield  {author} {\bibinfo {author} {\bibfnamefont {E.}~\bibnamefont
  {Barausse}},\ }\bibfield  {title} {\bibinfo {title} {{Relativistic dynamical
  friction in a collisional fluid}},\ }\href
  {https://doi.org/10.1111/j.1365-2966.2007.12408.x} {\bibfield  {journal}
  {\bibinfo  {journal} {Mon. Not. Roy. Astron. Soc.}\ }\textbf {\bibinfo
  {volume} {382}},\ \bibinfo {pages} {826} (\bibinfo {year} {2007})},\ \Eprint
  {https://arxiv.org/abs/0709.0211} {arXiv:0709.0211 [astro-ph]} \BibitemShut
  {NoStop}%
\bibitem [{\citenamefont {{Syer}}\ \emph {et~al.}(1991)\citenamefont {{Syer}},
  \citenamefont {{Clarke}},\ and\ \citenamefont {{Rees}}}]{Syer1991}%
  \BibitemOpen
  \bibfield  {author} {\bibinfo {author} {\bibfnamefont {D.}~\bibnamefont
  {{Syer}}}, \bibinfo {author} {\bibfnamefont {C.~J.}\ \bibnamefont
  {{Clarke}}},\ and\ \bibinfo {author} {\bibfnamefont {M.~J.}\ \bibnamefont
  {{Rees}}},\ }\bibfield  {title} {\bibinfo {title} {{Star-disc interactions
  near a massive black hole}},\ }\href
  {https://doi.org/10.1093/mnras/250.3.505} {\bibfield  {journal} {\bibinfo
  {journal} {\mnras}\ }\textbf {\bibinfo {volume} {250}},\ \bibinfo {pages}
  {505} (\bibinfo {year} {1991})}\BibitemShut {NoStop}%
\bibitem [{\citenamefont {{Fabj}}\ \emph {et~al.}(2020)\citenamefont {{Fabj}},
  \citenamefont {{Nasim}}, \citenamefont {{Caban}}, \citenamefont {{Ford}},
  \citenamefont {{McKernan}},\ and\ \citenamefont {{Bellovary}}}]{Fabj2020}%
  \BibitemOpen
  \bibfield  {author} {\bibinfo {author} {\bibfnamefont {G.}~\bibnamefont
  {{Fabj}}}, \bibinfo {author} {\bibfnamefont {S.~S.}\ \bibnamefont {{Nasim}}},
  \bibinfo {author} {\bibfnamefont {F.}~\bibnamefont {{Caban}}}, \bibinfo
  {author} {\bibfnamefont {K.~E.~S.}\ \bibnamefont {{Ford}}}, \bibinfo {author}
  {\bibfnamefont {B.}~\bibnamefont {{McKernan}}},\ and\ \bibinfo {author}
  {\bibfnamefont {J.~M.}\ \bibnamefont {{Bellovary}}},\ }\bibfield  {title}
  {\bibinfo {title} {{Aligning nuclear cluster orbits with an active galactic
  nucleus accretion disc}},\ }\href {https://doi.org/10.1093/mnras/staa3004}
  {\bibfield  {journal} {\bibinfo  {journal} {\mnras}\ }\textbf {\bibinfo
  {volume} {499}},\ \bibinfo {pages} {2608} (\bibinfo {year} {2020})},\ \Eprint
  {https://arxiv.org/abs/2006.11229} {arXiv:2006.11229 [astro-ph.GA]}
  \BibitemShut {NoStop}%
\bibitem [{\citenamefont {Foreman-Mackey}\ \emph {et~al.}(2013)\citenamefont
  {Foreman-Mackey}, \citenamefont {Hogg}, \citenamefont {Lang},\ and\
  \citenamefont {Goodman}}]{ForemanMackey2013Emcee}%
  \BibitemOpen
  \bibfield  {author} {\bibinfo {author} {\bibfnamefont {D.}~\bibnamefont
  {Foreman-Mackey}}, \bibinfo {author} {\bibfnamefont {D.~W.}\ \bibnamefont
  {Hogg}}, \bibinfo {author} {\bibfnamefont {D.}~\bibnamefont {Lang}},\ and\
  \bibinfo {author} {\bibfnamefont {J.}~\bibnamefont {Goodman}},\ }\bibfield
  {title} {\bibinfo {title} {emcee: The mcmc hammer},\ }\href
  {https://doi.org/10.1086/670067} {\bibfield  {journal} {\bibinfo  {journal}
  {Publications of the Astronomical Society of the Pacific}\ }\textbf {\bibinfo
  {volume} {125}},\ \bibinfo {pages} {306} (\bibinfo {year} {2013})},\ \Eprint
  {https://arxiv.org/abs/1202.3665} {arXiv:1202.3665} \BibitemShut {NoStop}%
\bibitem [{\citenamefont {Vousden}\ \emph {et~al.}(2015)\citenamefont
  {Vousden}, \citenamefont {Farr},\ and\ \citenamefont
  {Mandel}}]{Vousden_2015}%
  \BibitemOpen
  \bibfield  {author} {\bibinfo {author} {\bibfnamefont {W.~D.}\ \bibnamefont
  {Vousden}}, \bibinfo {author} {\bibfnamefont {W.~M.}\ \bibnamefont {Farr}},\
  and\ \bibinfo {author} {\bibfnamefont {I.}~\bibnamefont {Mandel}},\
  }\bibfield  {title} {\bibinfo {title} {Dynamic temperature selection for
  parallel tempering in markov chain monte carlo simulations},\ }\href
  {https://doi.org/10.1093/mnras/stv2422} {\bibfield  {journal} {\bibinfo
  {journal} {Monthly Notices of the Royal Astronomical Society}\ }\textbf
  {\bibinfo {volume} {455}},\ \bibinfo {pages} {1919} (\bibinfo {year}
  {2015})}\BibitemShut {NoStop}%
\bibitem [{\citenamefont {Foreman-Mackey}(2018)}]{autocorrelation}%
  \BibitemOpen
  \bibfield  {author} {\bibinfo {author} {\bibfnamefont {D.}~\bibnamefont
  {Foreman-Mackey}},\ }\href@noop {} {\bibinfo {title} {{Autocorrelation time
  estimation}}},\ \bibinfo {howpublished}
  {\url{https://github.com/dfm/emcee/issues/209}} (\bibinfo {year} {2018}),\
  \bibinfo {note} {accessed 20-July-2022}\BibitemShut {NoStop}%
\bibitem [{\citenamefont {Babak}\ \emph {et~al.}(2021)\citenamefont {Babak},
  \citenamefont {Petiteau},\ and\ \citenamefont {Hewitson}}]{Babak:2021mhe}%
  \BibitemOpen
  \bibfield  {author} {\bibinfo {author} {\bibfnamefont {S.}~\bibnamefont
  {Babak}}, \bibinfo {author} {\bibfnamefont {A.}~\bibnamefont {Petiteau}},\
  and\ \bibinfo {author} {\bibfnamefont {M.}~\bibnamefont {Hewitson}},\
  }\href@noop {} {\bibinfo {title} {{LISA Sensitivity and SNR Calculations}}}
  (\bibinfo {year} {2021}),\ \Eprint {https://arxiv.org/abs/2108.01167}
  {arXiv:2108.01167 [astro-ph.IM]} \BibitemShut {NoStop}%
\bibitem [{\citenamefont {Robson}\ \emph {et~al.}(2019)\citenamefont {Robson},
  \citenamefont {Cornish},\ and\ \citenamefont
  {Liu}}]{robsonConstructionUseLISA2019}%
  \BibitemOpen
  \bibfield  {author} {\bibinfo {author} {\bibfnamefont {T.}~\bibnamefont
  {Robson}}, \bibinfo {author} {\bibfnamefont {N.}~\bibnamefont {Cornish}},\
  and\ \bibinfo {author} {\bibfnamefont {C.}~\bibnamefont {Liu}},\ }\bibfield
  {title} {\bibinfo {title} {The construction and use of {{LISA}} sensitivity
  curves},\ }\href {https://doi.org/10.1088/1361-6382/ab1101} {\bibfield
  {journal} {\bibinfo  {journal} {Class. Quantum Grav.}\ }\textbf {\bibinfo
  {volume} {36}},\ \bibinfo {pages} {105011} (\bibinfo {year} {2019})},\
  \Eprint {https://arxiv.org/abs/1803.01944} {arXiv:1803.01944} \BibitemShut
  {NoStop}%
\bibitem [{\citenamefont {Abbott}\ \emph
  {et~al.}(2021{\natexlab{b}})\citenamefont {Abbott} \emph
  {et~al.}}]{LIGOScientific:2021psn}%
  \BibitemOpen
  \bibfield  {author} {\bibinfo {author} {\bibfnamefont {R.}~\bibnamefont
  {Abbott}} \emph {et~al.} (\bibinfo {collaboration} {LIGO Scientific, VIRGO,
  KAGRA}),\ }\href@noop {} {\bibinfo {title} {{The population of merging
  compact binaries inferred using gravitational waves through GWTC-3}}}
  (\bibinfo {year} {2021}{\natexlab{b}}),\ \Eprint
  {https://arxiv.org/abs/2111.03634} {arXiv:2111.03634 [astro-ph.HE]}
  \BibitemShut {NoStop}%
\bibitem [{\citenamefont {{Tagawa}}\ \emph {et~al.}(2022)\citenamefont
  {{Tagawa}}, \citenamefont {{Kimura}}, \citenamefont {{Haiman}}, \citenamefont
  {{Perna}}, \citenamefont {{Tanaka}},\ and\ \citenamefont
  {{Bartos}}}]{Tagawa2022}%
  \BibitemOpen
  \bibfield  {author} {\bibinfo {author} {\bibfnamefont {H.}~\bibnamefont
  {{Tagawa}}}, \bibinfo {author} {\bibfnamefont {S.~S.}\ \bibnamefont
  {{Kimura}}}, \bibinfo {author} {\bibfnamefont {Z.}~\bibnamefont {{Haiman}}},
  \bibinfo {author} {\bibfnamefont {R.}~\bibnamefont {{Perna}}}, \bibinfo
  {author} {\bibfnamefont {H.}~\bibnamefont {{Tanaka}}},\ and\ \bibinfo
  {author} {\bibfnamefont {I.}~\bibnamefont {{Bartos}}},\ }\bibfield  {title}
  {\bibinfo {title} {{Can Stellar-mass Black Hole Growth Disrupt Disks of
  Active Galactic Nuclei? The Role of Mechanical Feedback}},\ }\href
  {https://doi.org/10.3847/1538-4357/ac45f8} {\bibfield  {journal} {\bibinfo
  {journal} {\apj}\ }\textbf {\bibinfo {volume} {927}},\ \bibinfo {eid} {41}
  (\bibinfo {year} {2022})},\ \Eprint {https://arxiv.org/abs/2112.01544}
  {arXiv:2112.01544 [astro-ph.HE]} \BibitemShut {NoStop}%
\bibitem [{\citenamefont {Levin}(2003)}]{Levin:2003ej}%
  \BibitemOpen
  \bibfield  {author} {\bibinfo {author} {\bibfnamefont {Y.}~\bibnamefont
  {Levin}},\ }\href@noop {} {\bibinfo {title} {{Formation of massive stars and
  black holes in selfgravitating AGN discs, and gravitational waves in LISA
  band}}} (\bibinfo {year} {2003}),\ \Eprint
  {https://arxiv.org/abs/astro-ph/0307084} {arXiv:astro-ph/0307084}
  \BibitemShut {NoStop}%
\bibitem [{\citenamefont {Levin}(2007)}]{Levin:2006uc}%
  \BibitemOpen
  \bibfield  {author} {\bibinfo {author} {\bibfnamefont {Y.}~\bibnamefont
  {Levin}},\ }\bibfield  {title} {\bibinfo {title} {{Starbursts near
  supermassive black holes: young stars in the Galactic Center, and
  gravitational waves in LISA band}},\ }\href
  {https://doi.org/10.1111/j.1365-2966.2006.11155.x} {\bibfield  {journal}
  {\bibinfo  {journal} {Mon. Not. Roy. Astron. Soc.}\ }\textbf {\bibinfo
  {volume} {374}},\ \bibinfo {pages} {515} (\bibinfo {year} {2007})},\ \Eprint
  {https://arxiv.org/abs/astro-ph/0603583} {arXiv:astro-ph/0603583}
  \BibitemShut {NoStop}%
\bibitem [{\citenamefont {{LISA Science Study Team}}(2018)}]{LISASciRD}%
  \BibitemOpen
  \bibfield  {author} {\bibinfo {author} {\bibnamefont {{LISA Science Study
  Team}}},\ }\href@noop {} {\bibinfo {title} {{LISA Science Requirements
  Document}}},\ \bibinfo {howpublished}
  {\url{https://www.cosmos.esa.int/documents/678316/1700384/SciRD.pdf/25831f6b-3c01-e215-5916-4ac6e4b306fb?t=1526479841000}}
  (\bibinfo {year} {14th May 2018}),\ \bibinfo {note} {[Online; accessed
  22-December-2022]}\BibitemShut {NoStop}%
\bibitem [{Note4()}]{Note4}%
  \BibitemOpen
  \bibinfo {note} {This luminosity distance corresponds to the redshift
  $z=0.276$ for a flat $\Lambda $CDM cosmology with Hubble constant $H_0 =
  67.74 \protect \, {\protect \rm km/s/Mpc}$ and matter density $\Omega
  _M=0.3075$.}\BibitemShut {Stop}%
\bibitem [{\citenamefont {Yunes}\ and\ \citenamefont
  {Pretorius}(2009)}]{Yunes:2009ke}%
  \BibitemOpen
  \bibfield  {author} {\bibinfo {author} {\bibfnamefont {N.}~\bibnamefont
  {Yunes}}\ and\ \bibinfo {author} {\bibfnamefont {F.}~\bibnamefont
  {Pretorius}},\ }\bibfield  {title} {\bibinfo {title} {{Fundamental
  Theoretical Bias in Gravitational Wave Astrophysics and the Parameterized
  Post-Einsteinian Framework}},\ }\href
  {https://doi.org/10.1103/PhysRevD.80.122003} {\bibfield  {journal} {\bibinfo
  {journal} {Phys. Rev. D}\ }\textbf {\bibinfo {volume} {80}},\ \bibinfo
  {pages} {122003} (\bibinfo {year} {2009})},\ \Eprint
  {https://arxiv.org/abs/0909.3328} {arXiv:0909.3328 [gr-qc]} \BibitemShut
  {NoStop}%
\bibitem [{\citenamefont {Aird}\ \emph {et~al.}(2018)\citenamefont {Aird},
  \citenamefont {Coil},\ and\ \citenamefont {Georgakakis}}]{Aird:2017cbs}%
  \BibitemOpen
  \bibfield  {author} {\bibinfo {author} {\bibfnamefont {J.}~\bibnamefont
  {Aird}}, \bibinfo {author} {\bibfnamefont {A.~L.}\ \bibnamefont {Coil}},\
  and\ \bibinfo {author} {\bibfnamefont {A.}~\bibnamefont {Georgakakis}},\
  }\bibfield  {title} {\bibinfo {title} {{X-rays across the galaxy population
  \textendash{} II. The distribution of AGN accretion rates as a function of
  stellar mass and redshift}},\ }\href {https://doi.org/10.1093/mnras/stx2700}
  {\bibfield  {journal} {\bibinfo  {journal} {Mon. Not. Roy. Astron. Soc.}\
  }\textbf {\bibinfo {volume} {474}},\ \bibinfo {pages} {1225} (\bibinfo {year}
  {2018})},\ \Eprint {https://arxiv.org/abs/1705.01132} {arXiv:1705.01132
  [astro-ph.HE]} \BibitemShut {NoStop}%
\bibitem [{\citenamefont {Cutler}(1998)}]{Cutler:1997ta}%
  \BibitemOpen
  \bibfield  {author} {\bibinfo {author} {\bibfnamefont {C.}~\bibnamefont
  {Cutler}},\ }\bibfield  {title} {\bibinfo {title} {{Angular resolution of the
  LISA gravitational wave detector}},\ }\href
  {https://doi.org/10.1103/PhysRevD.57.7089} {\bibfield  {journal} {\bibinfo
  {journal} {Phys. Rev. D}\ }\textbf {\bibinfo {volume} {57}},\ \bibinfo
  {pages} {7089} (\bibinfo {year} {1998})},\ \Eprint
  {https://arxiv.org/abs/gr-qc/9703068} {arXiv:gr-qc/9703068} \BibitemShut
  {NoStop}%
\bibitem [{\citenamefont {Lang}\ and\ \citenamefont
  {Hughes}(2006)}]{Lang:2006bsg}%
  \BibitemOpen
  \bibfield  {author} {\bibinfo {author} {\bibfnamefont {R.~N.}\ \bibnamefont
  {Lang}}\ and\ \bibinfo {author} {\bibfnamefont {S.~A.}\ \bibnamefont
  {Hughes}},\ }\bibfield  {title} {\bibinfo {title} {Measuring coalescing
  massive binary black holes with gravitational waves: The impact of
  spin-induced precession},\ }\href
  {https://doi.org/10.1103/PhysRevD.74.122001} {\bibfield  {journal} {\bibinfo
  {journal} {Phys. Rev. D}\ }\textbf {\bibinfo {volume} {74}},\ \bibinfo
  {pages} {122001} (\bibinfo {year} {2006})}\BibitemShut {NoStop}%
\bibitem [{Note5()}]{Note5}%
  \BibitemOpen
  \bibinfo {note} {We ignore errors on the luminosity distance due to lensing
  and peculiar velocity because they are an order of magnitude smaller \protect
  \citep {Speri:2020hwc}.}\BibitemShut {Stop}%
\bibitem [{\citenamefont {Madau}\ \emph {et~al.}(2014)\citenamefont {Madau},
  \citenamefont {Haardt},\ and\ \citenamefont {Dotti}}]{Madau:2014pta}%
  \BibitemOpen
  \bibfield  {author} {\bibinfo {author} {\bibfnamefont {P.}~\bibnamefont
  {Madau}}, \bibinfo {author} {\bibfnamefont {F.}~\bibnamefont {Haardt}},\ and\
  \bibinfo {author} {\bibfnamefont {M.}~\bibnamefont {Dotti}},\ }\bibfield
  {title} {\bibinfo {title} {{Super-Critical Growth of Massive Black Holes from
  Stellar-Mass Seeds}},\ }\href {https://doi.org/10.1088/2041-8205/784/2/L38}
  {\bibfield  {journal} {\bibinfo  {journal} {Astrophys. J. Lett.}\ }\textbf
  {\bibinfo {volume} {784}},\ \bibinfo {pages} {L38} (\bibinfo {year}
  {2014})},\ \Eprint {https://arxiv.org/abs/1402.6995} {arXiv:1402.6995
  [astro-ph.CO]} \BibitemShut {NoStop}%
\bibitem [{Note6()}]{Note6}%
  \BibitemOpen
  \bibinfo {note} {Note that we use different definitions for the accretion
  rate compared to Ref. \cite {Madau:2014pta}.}\BibitemShut {Stop}%
\bibitem [{Note7()}]{Note7}%
  \BibitemOpen
  \bibinfo {note} {At the relatively low accretion rates considered here, the
  bolometric luminosity is proportional to the Eddington ratio $f_{\protect \rm
  Edd}$. The BH mass and spin determined through the GW signal have negligible
  uncertainty, see Fig.~\ref {fig:bias}. Therefore, the relative precision of a
  luminosity measurement translates directly into a relative precision on the
  Eddington ratio.}\BibitemShut {Stop}%
\bibitem [{\citenamefont {{Shankar}}(2013)}]{Shankar2013}%
  \BibitemOpen
  \bibfield  {author} {\bibinfo {author} {\bibfnamefont {F.}~\bibnamefont
  {{Shankar}}},\ }\bibfield  {title} {\bibinfo {title} {{Black hole demography:
  from scaling relations to models}},\ }\href
  {https://doi.org/10.1088/0264-9381/30/24/244001} {\bibfield  {journal}
  {\bibinfo  {journal} {Classical and Quantum Gravity}\ }\textbf {\bibinfo
  {volume} {30}},\ \bibinfo {eid} {244001} (\bibinfo {year} {2013})},\ \Eprint
  {https://arxiv.org/abs/1307.3289} {arXiv:1307.3289 [astro-ph.CO]}
  \BibitemShut {NoStop}%
\bibitem [{\citenamefont {Hopkins}\ \emph {et~al.}(2007)\citenamefont
  {Hopkins}, \citenamefont {Richards},\ and\ \citenamefont
  {Hernquist}}]{Hopkins:2006fq}%
  \BibitemOpen
  \bibfield  {author} {\bibinfo {author} {\bibfnamefont {P.~F.}\ \bibnamefont
  {Hopkins}}, \bibinfo {author} {\bibfnamefont {G.~T.}\ \bibnamefont
  {Richards}},\ and\ \bibinfo {author} {\bibfnamefont {L.}~\bibnamefont
  {Hernquist}},\ }\bibfield  {title} {\bibinfo {title} {{An Observational
  Determination of the Bolometric Quasar Luminosity Function}},\ }\href
  {https://doi.org/10.1086/509629} {\bibfield  {journal} {\bibinfo  {journal}
  {Astrophys. J.}\ }\textbf {\bibinfo {volume} {654}},\ \bibinfo {pages} {731}
  (\bibinfo {year} {2007})},\ \Eprint {https://arxiv.org/abs/astro-ph/0605678}
  {arXiv:astro-ph/0605678} \BibitemShut {NoStop}%
\bibitem [{\citenamefont {{Meidinger}}(2018)}]{Meidinger2018}%
  \BibitemOpen
  \bibfield  {author} {\bibinfo {author} {\bibfnamefont {N.}~\bibnamefont
  {{Meidinger}}},\ }\bibfield  {title} {\bibinfo {title} {{The Wide Field
  Imager instrument for Athena}},\ }\href@noop {} {\bibfield  {journal}
  {\bibinfo  {journal} {Contributions of the Astronomical Observatory Skalnate
  Pleso}\ }\textbf {\bibinfo {volume} {48}},\ \bibinfo {pages} {498} (\bibinfo
  {year} {2018})},\ \Eprint {https://arxiv.org/abs/1702.01079}
  {arXiv:1702.01079 [astro-ph.IM]} \BibitemShut {NoStop}%
\bibitem [{\citenamefont {Gardner}\ \emph {et~al.}(2006)\citenamefont {Gardner}
  \emph {et~al.}}]{Gardner:2006ky}%
  \BibitemOpen
  \bibfield  {author} {\bibinfo {author} {\bibfnamefont {J.~P.}\ \bibnamefont
  {Gardner}} \emph {et~al.},\ }\bibfield  {title} {\bibinfo {title} {{The James
  Webb Space Telescope}},\ }\href {https://doi.org/10.1007/s11214-006-8315-7}
  {\bibfield  {journal} {\bibinfo  {journal} {Space Sci. Rev.}\ }\textbf
  {\bibinfo {volume} {123}},\ \bibinfo {pages} {485} (\bibinfo {year}
  {2006})},\ \Eprint {https://arxiv.org/abs/astro-ph/0606175}
  {arXiv:astro-ph/0606175} \BibitemShut {NoStop}%
\bibitem [{\citenamefont {{J{\"a}rvel{\"a}}}\ \emph {et~al.}(2021)\citenamefont
  {{J{\"a}rvel{\"a}}}, \citenamefont {{Berton}},\ and\ \citenamefont
  {{Crepaldi}}}]{Jarvela2021}%
  \BibitemOpen
  \bibfield  {author} {\bibinfo {author} {\bibfnamefont {E.}~\bibnamefont
  {{J{\"a}rvel{\"a}}}}, \bibinfo {author} {\bibfnamefont {M.}~\bibnamefont
  {{Berton}}},\ and\ \bibinfo {author} {\bibfnamefont {L.}~\bibnamefont
  {{Crepaldi}}},\ }\bibfield  {title} {\bibinfo {title} {{Narrow-line Seyfert 1
  galaxies with absorbed jets -insights from radio spectral index maps}},\
  }\href {https://doi.org/10.3389/fspas.2021.735310} {\bibfield  {journal}
  {\bibinfo  {journal} {Frontiers in Astronomy and Space Sciences}\ }\textbf
  {\bibinfo {volume} {8}},\ \bibinfo {eid} {147} (\bibinfo {year} {2021})},\
  \Eprint {https://arxiv.org/abs/2108.08521} {arXiv:2108.08521 [astro-ph.GA]}
  \BibitemShut {NoStop}%
\bibitem [{Note8()}]{Note8}%
  \BibitemOpen
  \bibinfo {note} {Since this procedure is computationally expensive, we run
  this analysis only over the intrinsic EMRI parameters. We do not expect this
  choice to affect the conclusions. { For more information on EMRI searches we
  refer the reader to \cite {Babak:2017tow}}}\BibitemShut {NoStop}%
\bibitem [{Note9()}]{Note9}%
  \BibitemOpen
  \bibinfo {note} {Even though we use a naive, brute-force search method and
  explore only a portion of the parameter space, our results qualitatively
  suggest how challenging EMRI search and inference studies could be when
  ignoring environmental effects.}\BibitemShut {Stop}%
\bibitem [{\citenamefont {Yunes}\ \emph {et~al.}(2010)\citenamefont {Yunes},
  \citenamefont {Pretorius},\ and\ \citenamefont {Spergel}}]{Yunes:2009bv}%
  \BibitemOpen
  \bibfield  {author} {\bibinfo {author} {\bibfnamefont {N.}~\bibnamefont
  {Yunes}}, \bibinfo {author} {\bibfnamefont {F.}~\bibnamefont {Pretorius}},\
  and\ \bibinfo {author} {\bibfnamefont {D.}~\bibnamefont {Spergel}},\
  }\bibfield  {title} {\bibinfo {title} {{Constraining the evolutionary history
  of Newton's constant with gravitational wave observations}},\ }\href
  {https://doi.org/10.1103/PhysRevD.81.064018} {\bibfield  {journal} {\bibinfo
  {journal} {Phys. Rev. D}\ }\textbf {\bibinfo {volume} {81}},\ \bibinfo
  {pages} {064018} (\bibinfo {year} {2010})},\ \Eprint
  {https://arxiv.org/abs/0912.2724} {arXiv:0912.2724 [gr-qc]} \BibitemShut
  {NoStop}%
\bibitem [{\citenamefont {Chamberlain}\ and\ \citenamefont
  {Yunes}(2017)}]{Chamberlain:2017fjl}%
  \BibitemOpen
  \bibfield  {author} {\bibinfo {author} {\bibfnamefont {K.}~\bibnamefont
  {Chamberlain}}\ and\ \bibinfo {author} {\bibfnamefont {N.}~\bibnamefont
  {Yunes}},\ }\bibfield  {title} {\bibinfo {title} {{Theoretical Physics
  Implications of Gravitational Wave Observation with Future Detectors}},\
  }\href {https://doi.org/10.1103/PhysRevD.96.084039} {\bibfield  {journal}
  {\bibinfo  {journal} {Phys. Rev. D}\ }\textbf {\bibinfo {volume} {96}},\
  \bibinfo {pages} {084039} (\bibinfo {year} {2017})},\ \Eprint
  {https://arxiv.org/abs/1704.08268} {arXiv:1704.08268 [gr-qc]} \BibitemShut
  {NoStop}%
\bibitem [{\citenamefont {Barbieri}\ \emph {et~al.}(2022)\citenamefont
  {Barbieri}, \citenamefont {Savastano}, \citenamefont {Speri}, \citenamefont
  {Antonelli}, \citenamefont {Sberna}, \citenamefont {Burke}, \citenamefont
  {Gair},\ and\ \citenamefont {Tamanini}}]{Barbieri:2022zge}%
  \BibitemOpen
  \bibfield  {author} {\bibinfo {author} {\bibfnamefont {R.}~\bibnamefont
  {Barbieri}}, \bibinfo {author} {\bibfnamefont {S.}~\bibnamefont {Savastano}},
  \bibinfo {author} {\bibfnamefont {L.}~\bibnamefont {Speri}}, \bibinfo
  {author} {\bibfnamefont {A.}~\bibnamefont {Antonelli}}, \bibinfo {author}
  {\bibfnamefont {L.}~\bibnamefont {Sberna}}, \bibinfo {author} {\bibfnamefont
  {O.}~\bibnamefont {Burke}}, \bibinfo {author} {\bibfnamefont
  {J.}~\bibnamefont {Gair}},\ and\ \bibinfo {author} {\bibfnamefont
  {N.}~\bibnamefont {Tamanini}},\ }\href@noop {} {\bibinfo {title}
  {{Constraining the evolution of Newton's constant with slow inspirals
  observed from spaceborne gravitational-wave detectors}}} (\bibinfo {year}
  {2022}),\ \Eprint {https://arxiv.org/abs/2207.10674} {arXiv:2207.10674
  [gr-qc]} \BibitemShut {NoStop}%
\bibitem [{\citenamefont {Wang}\ \emph {et~al.}(2022)\citenamefont {Wang},
  \citenamefont {Zhao}, \citenamefont {An}, \citenamefont {Shao},\ and\
  \citenamefont {Cao}}]{Wang:2022yxb}%
  \BibitemOpen
  \bibfield  {author} {\bibinfo {author} {\bibfnamefont {Z.}~\bibnamefont
  {Wang}}, \bibinfo {author} {\bibfnamefont {J.}~\bibnamefont {Zhao}}, \bibinfo
  {author} {\bibfnamefont {Z.}~\bibnamefont {An}}, \bibinfo {author}
  {\bibfnamefont {L.}~\bibnamefont {Shao}},\ and\ \bibinfo {author}
  {\bibfnamefont {Z.}~\bibnamefont {Cao}},\ }\href
  {https://doi.org/10.1016/j.physletb.2022.137416} {\bibinfo {title}
  {{Simultaneous bounds on the gravitational dipole radiation and varying
  gravitational constant from compact binary inspirals}}} (\bibinfo {year}
  {2022}),\ \Eprint {https://arxiv.org/abs/2208.11913} {arXiv:2208.11913
  [gr-qc]} \BibitemShut {NoStop}%
\bibitem [{\citenamefont {Maselli}\ \emph {et~al.}(2022)\citenamefont
  {Maselli}, \citenamefont {Franchini}, \citenamefont {Gualtieri},
  \citenamefont {Sotiriou}, \citenamefont {Barsanti},\ and\ \citenamefont
  {Pani}}]{Maselli:2021men}%
  \BibitemOpen
  \bibfield  {author} {\bibinfo {author} {\bibfnamefont {A.}~\bibnamefont
  {Maselli}}, \bibinfo {author} {\bibfnamefont {N.}~\bibnamefont {Franchini}},
  \bibinfo {author} {\bibfnamefont {L.}~\bibnamefont {Gualtieri}}, \bibinfo
  {author} {\bibfnamefont {T.~P.}\ \bibnamefont {Sotiriou}}, \bibinfo {author}
  {\bibfnamefont {S.}~\bibnamefont {Barsanti}},\ and\ \bibinfo {author}
  {\bibfnamefont {P.}~\bibnamefont {Pani}},\ }\bibfield  {title} {\bibinfo
  {title} {{Detecting fundamental fields with LISA observations of
  gravitational waves from extreme mass-ratio inspirals}},\ }\href
  {https://doi.org/10.1038/s41550-021-01589-5} {\bibfield  {journal} {\bibinfo
  {journal} {Nature Astron.}\ }\textbf {\bibinfo {volume} {6}},\ \bibinfo
  {pages} {464} (\bibinfo {year} {2022})},\ \Eprint
  {https://arxiv.org/abs/2106.11325} {arXiv:2106.11325 [gr-qc]} \BibitemShut
  {NoStop}%
\bibitem [{\citenamefont {Barsanti}\ \emph
  {et~al.}(2022{\natexlab{a}})\citenamefont {Barsanti}, \citenamefont
  {De~Luca}, \citenamefont {Maselli},\ and\ \citenamefont
  {Pani}}]{Barsanti:2021ydd}%
  \BibitemOpen
  \bibfield  {author} {\bibinfo {author} {\bibfnamefont {S.}~\bibnamefont
  {Barsanti}}, \bibinfo {author} {\bibfnamefont {V.}~\bibnamefont {De~Luca}},
  \bibinfo {author} {\bibfnamefont {A.}~\bibnamefont {Maselli}},\ and\ \bibinfo
  {author} {\bibfnamefont {P.}~\bibnamefont {Pani}},\ }\bibfield  {title}
  {\bibinfo {title} {{Detecting Subsolar-Mass Primordial Black Holes in Extreme
  Mass-Ratio Inspirals with LISA and Einstein Telescope}},\ }\href
  {https://doi.org/10.1103/PhysRevLett.128.111104} {\bibfield  {journal}
  {\bibinfo  {journal} {Phys. Rev. Lett.}\ }\textbf {\bibinfo {volume} {128}},\
  \bibinfo {pages} {111104} (\bibinfo {year} {2022}{\natexlab{a}})},\ \Eprint
  {https://arxiv.org/abs/2109.02170} {arXiv:2109.02170 [gr-qc]} \BibitemShut
  {NoStop}%
\bibitem [{\citenamefont {Antonelli}\ \emph {et~al.}(2021)\citenamefont
  {Antonelli}, \citenamefont {Burke},\ and\ \citenamefont
  {Gair}}]{Antonelli:2021vwg}%
  \BibitemOpen
  \bibfield  {author} {\bibinfo {author} {\bibfnamefont {A.}~\bibnamefont
  {Antonelli}}, \bibinfo {author} {\bibfnamefont {O.}~\bibnamefont {Burke}},\
  and\ \bibinfo {author} {\bibfnamefont {J.~R.}\ \bibnamefont {Gair}},\
  }\bibfield  {title} {\bibinfo {title} {{Noisy neighbours: inference biases
  from overlapping gravitational-wave signals}},\ }\href
  {https://doi.org/10.1093/mnras/stab2358} {\bibfield  {journal} {\bibinfo
  {journal} {Mon. Not. Roy. Astron. Soc.}\ }\textbf {\bibinfo {volume} {507}},\
  \bibinfo {pages} {5069} (\bibinfo {year} {2021})},\ \Eprint
  {https://arxiv.org/abs/2104.01897} {arXiv:2104.01897 [gr-qc]} \BibitemShut
  {NoStop}%
\bibitem [{\citenamefont {{Tsang}}(2011)}]{Tsang2011}%
  \BibitemOpen
  \bibfield  {author} {\bibinfo {author} {\bibfnamefont {D.}~\bibnamefont
  {{Tsang}}},\ }\bibfield  {title} {\bibinfo {title} {{Protoplanetary Disk
  Resonances and Type I Migration}},\ }\href
  {https://doi.org/10.1088/0004-637X/741/2/109} {\bibfield  {journal} {\bibinfo
   {journal} {\apj}\ }\textbf {\bibinfo {volume} {741}},\ \bibinfo {eid} {109}
  (\bibinfo {year} {2011})},\ \Eprint {https://arxiv.org/abs/1107.4069}
  {arXiv:1107.4069 [astro-ph.EP]} \BibitemShut {NoStop}%
\bibitem [{\citenamefont {{Miranda}}\ and\ \citenamefont
  {{Lai}}(2018)}]{Miranda2018}%
  \BibitemOpen
  \bibfield  {author} {\bibinfo {author} {\bibfnamefont {R.}~\bibnamefont
  {{Miranda}}}\ and\ \bibinfo {author} {\bibfnamefont {D.}~\bibnamefont
  {{Lai}}},\ }\bibfield  {title} {\bibinfo {title} {{Trapping of low-mass
  planets outside the truncated inner edges of protoplanetary discs}},\ }\href
  {https://doi.org/10.1093/mnras/stx2706} {\bibfield  {journal} {\bibinfo
  {journal} {\mnras}\ }\textbf {\bibinfo {volume} {473}},\ \bibinfo {pages}
  {5267} (\bibinfo {year} {2018})},\ \Eprint {https://arxiv.org/abs/1708.07872}
  {arXiv:1708.07872 [astro-ph.EP]} \BibitemShut {NoStop}%
\bibitem [{\citenamefont {{Bellovary}}\ \emph {et~al.}(2016)\citenamefont
  {{Bellovary}}, \citenamefont {{Mac Low}}, \citenamefont {{McKernan}},\ and\
  \citenamefont {{Ford}}}]{Bellovary2016}%
  \BibitemOpen
  \bibfield  {author} {\bibinfo {author} {\bibfnamefont {J.~M.}\ \bibnamefont
  {{Bellovary}}}, \bibinfo {author} {\bibfnamefont {M.-M.}\ \bibnamefont {{Mac
  Low}}}, \bibinfo {author} {\bibfnamefont {B.}~\bibnamefont {{McKernan}}},\
  and\ \bibinfo {author} {\bibfnamefont {K.~E.~S.}\ \bibnamefont {{Ford}}},\
  }\bibfield  {title} {\bibinfo {title} {{Migration Traps in Disks around
  Supermassive Black Holes}},\ }\href
  {https://doi.org/10.3847/2041-8205/819/2/L17} {\bibfield  {journal} {\bibinfo
   {journal} {\apjl}\ }\textbf {\bibinfo {volume} {819}},\ \bibinfo {eid} {L17}
  (\bibinfo {year} {2016})},\ \Eprint {https://arxiv.org/abs/1511.00005}
  {arXiv:1511.00005 [astro-ph.GA]} \BibitemShut {NoStop}%
\bibitem [{\citenamefont {Zrake}\ \emph {et~al.}(2021)\citenamefont {Zrake},
  \citenamefont {Tiede}, \citenamefont {MacFadyen},\ and\ \citenamefont
  {Haiman}}]{Zrake:2020zkw}%
  \BibitemOpen
  \bibfield  {author} {\bibinfo {author} {\bibfnamefont {J.}~\bibnamefont
  {Zrake}}, \bibinfo {author} {\bibfnamefont {C.}~\bibnamefont {Tiede}},
  \bibinfo {author} {\bibfnamefont {A.}~\bibnamefont {MacFadyen}},\ and\
  \bibinfo {author} {\bibfnamefont {Z.}~\bibnamefont {Haiman}},\ }\bibfield
  {title} {\bibinfo {title} {{Equilibrium Eccentricity of Accreting
  Binaries}},\ }\href {https://doi.org/10.3847/2041-8213/abdd1c} {\bibfield
  {journal} {\bibinfo  {journal} {Astrophys. J. Lett.}\ }\textbf {\bibinfo
  {volume} {909}},\ \bibinfo {pages} {L13} (\bibinfo {year} {2021})},\ \Eprint
  {https://arxiv.org/abs/2010.09707} {arXiv:2010.09707 [astro-ph.HE]}
  \BibitemShut {NoStop}%
\bibitem [{\citenamefont {Zwick}\ \emph {et~al.}(2023)\citenamefont {Zwick},
  \citenamefont {Capelo},\ and\ \citenamefont {Mayer}}]{Zwick:2022dih}%
  \BibitemOpen
  \bibfield  {author} {\bibinfo {author} {\bibfnamefont {L.}~\bibnamefont
  {Zwick}}, \bibinfo {author} {\bibfnamefont {P.~R.}\ \bibnamefont {Capelo}},\
  and\ \bibinfo {author} {\bibfnamefont {L.}~\bibnamefont {Mayer}},\ }\bibfield
   {title} {\bibinfo {title} {{Priorities in gravitational waveforms for future
  space-borne detectors: vacuum accuracy or environment?}},\ }\href
  {https://doi.org/10.1093/mnras/stad707} {\bibfield  {journal} {\bibinfo
  {journal} {Monthly Notices of the Royal Astronomical Society}\ }\textbf
  {\bibinfo {volume} {521}},\ \bibinfo {pages} {4645} (\bibinfo {year}
  {2023})},\ \Eprint
  {https://arxiv.org/abs/https://academic.oup.com/mnras/article-pdf/521/3/4645/49714307/stad707.pdf}
  {https://academic.oup.com/mnras/article-pdf/521/3/4645/49714307/stad707.pdf}
  \BibitemShut {NoStop}%
\bibitem [{\citenamefont {Hannuksela}\ \emph {et~al.}(2020)\citenamefont
  {Hannuksela}, \citenamefont {Ng},\ and\ \citenamefont
  {Li}}]{Hannuksela:2019vip}%
  \BibitemOpen
  \bibfield  {author} {\bibinfo {author} {\bibfnamefont {O.~A.}\ \bibnamefont
  {Hannuksela}}, \bibinfo {author} {\bibfnamefont {K.~C.~Y.}\ \bibnamefont
  {Ng}},\ and\ \bibinfo {author} {\bibfnamefont {T.~G.~F.}\ \bibnamefont
  {Li}},\ }\bibfield  {title} {\bibinfo {title} {{Extreme dark matter tests
  with extreme mass ratio inspirals}},\ }\href
  {https://doi.org/10.1103/PhysRevD.102.103022} {\bibfield  {journal} {\bibinfo
   {journal} {Phys. Rev. D}\ }\textbf {\bibinfo {volume} {102}},\ \bibinfo
  {pages} {103022} (\bibinfo {year} {2020})},\ \Eprint
  {https://arxiv.org/abs/1906.11845} {arXiv:1906.11845 [astro-ph.CO]}
  \BibitemShut {NoStop}%
\bibitem [{\citenamefont {Annulli}\ \emph {et~al.}(2020)\citenamefont
  {Annulli}, \citenamefont {Cardoso},\ and\ \citenamefont
  {Vicente}}]{Annulli:2020ilw}%
  \BibitemOpen
  \bibfield  {author} {\bibinfo {author} {\bibfnamefont {L.}~\bibnamefont
  {Annulli}}, \bibinfo {author} {\bibfnamefont {V.}~\bibnamefont {Cardoso}},\
  and\ \bibinfo {author} {\bibfnamefont {R.}~\bibnamefont {Vicente}},\
  }\bibfield  {title} {\bibinfo {title} {{Stirred and shaken: Dynamical
  behavior of boson stars and dark matter cores}},\ }\href
  {https://doi.org/10.1016/j.physletb.2020.135944} {\bibfield  {journal}
  {\bibinfo  {journal} {Phys. Lett. B}\ }\textbf {\bibinfo {volume} {811}},\
  \bibinfo {pages} {135944} (\bibinfo {year} {2020})},\ \Eprint
  {https://arxiv.org/abs/2007.03700} {arXiv:2007.03700 [astro-ph.HE]}
  \BibitemShut {NoStop}%
\bibitem [{\citenamefont {Kavanagh}\ \emph {et~al.}(2020)\citenamefont
  {Kavanagh}, \citenamefont {Nichols}, \citenamefont {Bertone},\ and\
  \citenamefont {Gaggero}}]{Kavanagh:2020cfn}%
  \BibitemOpen
  \bibfield  {author} {\bibinfo {author} {\bibfnamefont {B.~J.}\ \bibnamefont
  {Kavanagh}}, \bibinfo {author} {\bibfnamefont {D.~A.}\ \bibnamefont
  {Nichols}}, \bibinfo {author} {\bibfnamefont {G.}~\bibnamefont {Bertone}},\
  and\ \bibinfo {author} {\bibfnamefont {D.}~\bibnamefont {Gaggero}},\
  }\bibfield  {title} {\bibinfo {title} {{Detecting dark matter around black
  holes with gravitational waves: Effects of dark-matter dynamics on the
  gravitational waveform}},\ }\href
  {https://doi.org/10.1103/PhysRevD.102.083006} {\bibfield  {journal} {\bibinfo
   {journal} {Phys. Rev. D}\ }\textbf {\bibinfo {volume} {102}},\ \bibinfo
  {pages} {083006} (\bibinfo {year} {2020})},\ \Eprint
  {https://arxiv.org/abs/2002.12811} {arXiv:2002.12811 [gr-qc]} \BibitemShut
  {NoStop}%
\bibitem [{\citenamefont {Coogan}\ \emph {et~al.}(2022)\citenamefont {Coogan},
  \citenamefont {Bertone}, \citenamefont {Gaggero}, \citenamefont {Kavanagh},\
  and\ \citenamefont {Nichols}}]{Coogan:2021uqv}%
  \BibitemOpen
  \bibfield  {author} {\bibinfo {author} {\bibfnamefont {A.}~\bibnamefont
  {Coogan}}, \bibinfo {author} {\bibfnamefont {G.}~\bibnamefont {Bertone}},
  \bibinfo {author} {\bibfnamefont {D.}~\bibnamefont {Gaggero}}, \bibinfo
  {author} {\bibfnamefont {B.~J.}\ \bibnamefont {Kavanagh}},\ and\ \bibinfo
  {author} {\bibfnamefont {D.~A.}\ \bibnamefont {Nichols}},\ }\bibfield
  {title} {\bibinfo {title} {{Measuring the dark matter environments of black
  hole binaries with gravitational waves}},\ }\href
  {https://doi.org/10.1103/PhysRevD.105.043009} {\bibfield  {journal} {\bibinfo
   {journal} {Phys. Rev. D}\ }\textbf {\bibinfo {volume} {105}},\ \bibinfo
  {pages} {043009} (\bibinfo {year} {2022})},\ \Eprint
  {https://arxiv.org/abs/2108.04154} {arXiv:2108.04154 [gr-qc]} \BibitemShut
  {NoStop}%
\bibitem [{\citenamefont {Cardoso}\ \emph {et~al.}(2022)\citenamefont
  {Cardoso}, \citenamefont {Destounis}, \citenamefont {Duque}, \citenamefont
  {Macedo},\ and\ \citenamefont {Maselli}}]{Cardoso:2021wlq}%
  \BibitemOpen
  \bibfield  {author} {\bibinfo {author} {\bibfnamefont {V.}~\bibnamefont
  {Cardoso}}, \bibinfo {author} {\bibfnamefont {K.}~\bibnamefont {Destounis}},
  \bibinfo {author} {\bibfnamefont {F.}~\bibnamefont {Duque}}, \bibinfo
  {author} {\bibfnamefont {R.~P.}\ \bibnamefont {Macedo}},\ and\ \bibinfo
  {author} {\bibfnamefont {A.}~\bibnamefont {Maselli}},\ }\bibfield  {title}
  {\bibinfo {title} {{Black holes in galaxies: Environmental impact on
  gravitational-wave generation and propagation}},\ }\href
  {https://doi.org/10.1103/PhysRevD.105.L061501} {\bibfield  {journal}
  {\bibinfo  {journal} {Phys. Rev. D}\ }\textbf {\bibinfo {volume} {105}},\
  \bibinfo {pages} {L061501} (\bibinfo {year} {2022})},\ \Eprint
  {https://arxiv.org/abs/2109.00005} {arXiv:2109.00005 [gr-qc]} \BibitemShut
  {NoStop}%
\bibitem [{\citenamefont {Vicente}\ and\ \citenamefont
  {Cardoso}(2022)}]{Vicente:2022ivh}%
  \BibitemOpen
  \bibfield  {author} {\bibinfo {author} {\bibfnamefont {R.}~\bibnamefont
  {Vicente}}\ and\ \bibinfo {author} {\bibfnamefont {V.}~\bibnamefont
  {Cardoso}},\ }\bibfield  {title} {\bibinfo {title} {{Dynamical friction of
  black holes in ultralight dark matter}},\ }\href
  {https://doi.org/10.1103/PhysRevD.105.083008} {\bibfield  {journal} {\bibinfo
   {journal} {Phys. Rev. D}\ }\textbf {\bibinfo {volume} {105}},\ \bibinfo
  {pages} {083008} (\bibinfo {year} {2022})},\ \Eprint
  {https://arxiv.org/abs/2201.08854} {arXiv:2201.08854 [gr-qc]} \BibitemShut
  {NoStop}%
\bibitem [{\citenamefont {Speeney}\ \emph {et~al.}(2022)\citenamefont
  {Speeney}, \citenamefont {Antonelli}, \citenamefont {Baibhav},\ and\
  \citenamefont {Berti}}]{Speeney:2022ryg}%
  \BibitemOpen
  \bibfield  {author} {\bibinfo {author} {\bibfnamefont {N.}~\bibnamefont
  {Speeney}}, \bibinfo {author} {\bibfnamefont {A.}~\bibnamefont {Antonelli}},
  \bibinfo {author} {\bibfnamefont {V.}~\bibnamefont {Baibhav}},\ and\ \bibinfo
  {author} {\bibfnamefont {E.}~\bibnamefont {Berti}},\ }\href@noop {} {\bibinfo
  {title} {{The impact of relativistic corrections on the detectability of
  dark-matter spikes with gravitational waves}}} (\bibinfo {year} {2022}),\
  \Eprint {https://arxiv.org/abs/2204.12508} {arXiv:2204.12508 [gr-qc]}
  \BibitemShut {NoStop}%
\bibitem [{\citenamefont {Cole}\ \emph {et~al.}(2022)\citenamefont {Cole},
  \citenamefont {Coogan}, \citenamefont {Kavanagh},\ and\ \citenamefont
  {Bertone}}]{Cole:2022ucw}%
  \BibitemOpen
  \bibfield  {author} {\bibinfo {author} {\bibfnamefont {P.~S.}\ \bibnamefont
  {Cole}}, \bibinfo {author} {\bibfnamefont {A.}~\bibnamefont {Coogan}},
  \bibinfo {author} {\bibfnamefont {B.~J.}\ \bibnamefont {Kavanagh}},\ and\
  \bibinfo {author} {\bibfnamefont {G.}~\bibnamefont {Bertone}},\ }\href@noop
  {} {\bibinfo {title} {{Measuring dark matter spikes around primordial black
  holes with Einstein Telescope and Cosmic Explorer}}} (\bibinfo {year}
  {2022}),\ \Eprint {https://arxiv.org/abs/2207.07576} {arXiv:2207.07576
  [astro-ph.CO]} \BibitemShut {NoStop}%
\bibitem [{\citenamefont {Cardoso}\ \emph {et~al.}(2021)\citenamefont
  {Cardoso}, \citenamefont {Duque},\ and\ \citenamefont
  {Khanna}}]{Cardoso:2021vjq}%
  \BibitemOpen
  \bibfield  {author} {\bibinfo {author} {\bibfnamefont {V.}~\bibnamefont
  {Cardoso}}, \bibinfo {author} {\bibfnamefont {F.}~\bibnamefont {Duque}},\
  and\ \bibinfo {author} {\bibfnamefont {G.}~\bibnamefont {Khanna}},\
  }\bibfield  {title} {\bibinfo {title} {{Gravitational tuning forks and
  hierarchical triple systems}},\ }\href
  {https://doi.org/10.1103/PhysRevD.103.L081501} {\bibfield  {journal}
  {\bibinfo  {journal} {Phys. Rev. D}\ }\textbf {\bibinfo {volume} {103}},\
  \bibinfo {pages} {L081501} (\bibinfo {year} {2021})},\ \Eprint
  {https://arxiv.org/abs/2101.01186} {arXiv:2101.01186 [gr-qc]} \BibitemShut
  {NoStop}%
\bibitem [{\citenamefont {Maselli}\ \emph {et~al.}(2020)\citenamefont
  {Maselli}, \citenamefont {Franchini}, \citenamefont {Gualtieri},\ and\
  \citenamefont {Sotiriou}}]{Maselli:2020zgv}%
  \BibitemOpen
  \bibfield  {author} {\bibinfo {author} {\bibfnamefont {A.}~\bibnamefont
  {Maselli}}, \bibinfo {author} {\bibfnamefont {N.}~\bibnamefont {Franchini}},
  \bibinfo {author} {\bibfnamefont {L.}~\bibnamefont {Gualtieri}},\ and\
  \bibinfo {author} {\bibfnamefont {T.~P.}\ \bibnamefont {Sotiriou}},\
  }\bibfield  {title} {\bibinfo {title} {{Detecting scalar fields with Extreme
  Mass Ratio Inspirals}},\ }\href
  {https://doi.org/10.1103/PhysRevLett.125.141101} {\bibfield  {journal}
  {\bibinfo  {journal} {Phys. Rev. Lett.}\ }\textbf {\bibinfo {volume} {125}},\
  \bibinfo {pages} {141101} (\bibinfo {year} {2020})},\ \Eprint
  {https://arxiv.org/abs/2004.11895} {arXiv:2004.11895 [gr-qc]} \BibitemShut
  {NoStop}%
\bibitem [{\citenamefont {Barsanti}\ \emph
  {et~al.}(2022{\natexlab{b}})\citenamefont {Barsanti}, \citenamefont
  {Franchini}, \citenamefont {Gualtieri}, \citenamefont {Maselli},\ and\
  \citenamefont {Sotiriou}}]{Barsanti:2022ana}%
  \BibitemOpen
  \bibfield  {author} {\bibinfo {author} {\bibfnamefont {S.}~\bibnamefont
  {Barsanti}}, \bibinfo {author} {\bibfnamefont {N.}~\bibnamefont {Franchini}},
  \bibinfo {author} {\bibfnamefont {L.}~\bibnamefont {Gualtieri}}, \bibinfo
  {author} {\bibfnamefont {A.}~\bibnamefont {Maselli}},\ and\ \bibinfo {author}
  {\bibfnamefont {T.~P.}\ \bibnamefont {Sotiriou}},\ }\href@noop {} {\bibinfo
  {title} {{Extreme mass-ratio inspirals as probes of scalar fields: eccentric
  equatorial orbits around Kerr black holes}}} (\bibinfo {year}
  {2022}{\natexlab{b}}),\ \Eprint {https://arxiv.org/abs/2203.05003}
  {arXiv:2203.05003 [gr-qc]} \BibitemShut {NoStop}%
\bibitem [{\citenamefont {Harris}\ \emph {et~al.}(2020)\citenamefont {Harris},
  \citenamefont {Millman}, \citenamefont {van~der Walt}, \citenamefont
  {Gommers}, \citenamefont {Virtanen}, \citenamefont {Cournapeau},
  \citenamefont {Wieser}, \citenamefont {Taylor}, \citenamefont {Berg},
  \citenamefont {Smith}, \citenamefont {Kern}, \citenamefont {Picus},
  \citenamefont {Hoyer}, \citenamefont {van Kerkwijk}, \citenamefont {Brett},
  \citenamefont {Haldane}, \citenamefont {del R{\'{i}}o}, \citenamefont
  {Wiebe}, \citenamefont {Peterson}, \citenamefont {G{\'{e}}rard-Marchant},
  \citenamefont {Sheppard}, \citenamefont {Reddy}, \citenamefont {Weckesser},
  \citenamefont {Abbasi}, \citenamefont {Gohlke},\ and\ \citenamefont
  {Oliphant}}]{harris2020array}%
  \BibitemOpen
  \bibfield  {author} {\bibinfo {author} {\bibfnamefont {C.~R.}\ \bibnamefont
  {Harris}}, \bibinfo {author} {\bibfnamefont {K.~J.}\ \bibnamefont {Millman}},
  \bibinfo {author} {\bibfnamefont {S.~J.}\ \bibnamefont {van~der Walt}},
  \bibinfo {author} {\bibfnamefont {R.}~\bibnamefont {Gommers}}, \bibinfo
  {author} {\bibfnamefont {P.}~\bibnamefont {Virtanen}}, \bibinfo {author}
  {\bibfnamefont {D.}~\bibnamefont {Cournapeau}}, \bibinfo {author}
  {\bibfnamefont {E.}~\bibnamefont {Wieser}}, \bibinfo {author} {\bibfnamefont
  {J.}~\bibnamefont {Taylor}}, \bibinfo {author} {\bibfnamefont
  {S.}~\bibnamefont {Berg}}, \bibinfo {author} {\bibfnamefont {N.~J.}\
  \bibnamefont {Smith}}, \bibinfo {author} {\bibfnamefont {R.}~\bibnamefont
  {Kern}}, \bibinfo {author} {\bibfnamefont {M.}~\bibnamefont {Picus}},
  \bibinfo {author} {\bibfnamefont {S.}~\bibnamefont {Hoyer}}, \bibinfo
  {author} {\bibfnamefont {M.~H.}\ \bibnamefont {van Kerkwijk}}, \bibinfo
  {author} {\bibfnamefont {M.}~\bibnamefont {Brett}}, \bibinfo {author}
  {\bibfnamefont {A.}~\bibnamefont {Haldane}}, \bibinfo {author} {\bibfnamefont
  {J.~F.}\ \bibnamefont {del R{\'{i}}o}}, \bibinfo {author} {\bibfnamefont
  {M.}~\bibnamefont {Wiebe}}, \bibinfo {author} {\bibfnamefont
  {P.}~\bibnamefont {Peterson}}, \bibinfo {author} {\bibfnamefont
  {P.}~\bibnamefont {G{\'{e}}rard-Marchant}}, \bibinfo {author} {\bibfnamefont
  {K.}~\bibnamefont {Sheppard}}, \bibinfo {author} {\bibfnamefont
  {T.}~\bibnamefont {Reddy}}, \bibinfo {author} {\bibfnamefont
  {W.}~\bibnamefont {Weckesser}}, \bibinfo {author} {\bibfnamefont
  {H.}~\bibnamefont {Abbasi}}, \bibinfo {author} {\bibfnamefont
  {C.}~\bibnamefont {Gohlke}},\ and\ \bibinfo {author} {\bibfnamefont {T.~E.}\
  \bibnamefont {Oliphant}},\ }\bibfield  {title} {\bibinfo {title} {Array
  programming with {NumPy}},\ }\href
  {https://doi.org/10.1038/s41586-020-2649-2} {\bibfield  {journal} {\bibinfo
  {journal} {Nature}\ }\textbf {\bibinfo {volume} {585}},\ \bibinfo {pages}
  {357} (\bibinfo {year} {2020})}\BibitemShut {NoStop}%
\bibitem [{\citenamefont {Hunter}(2007)}]{Hunter:2007}%
  \BibitemOpen
  \bibfield  {author} {\bibinfo {author} {\bibfnamefont {J.~D.}\ \bibnamefont
  {Hunter}},\ }\bibfield  {title} {\bibinfo {title} {Matplotlib: A 2d graphics
  environment},\ }\href {https://doi.org/10.1109/MCSE.2007.55} {\bibfield
  {journal} {\bibinfo  {journal} {Computing in Science \& Engineering}\
  }\textbf {\bibinfo {volume} {9}},\ \bibinfo {pages} {90} (\bibinfo {year}
  {2007})}\BibitemShut {NoStop}%
\bibitem [{\citenamefont {Virtanen}\ \emph {et~al.}(2020)\citenamefont
  {Virtanen}, \citenamefont {Gommers}, \citenamefont {Oliphant}, \citenamefont
  {Haberland}, \citenamefont {Reddy}, \citenamefont {Cournapeau}, \citenamefont
  {Burovski}, \citenamefont {Peterson}, \citenamefont {Weckesser},
  \citenamefont {Bright}, \citenamefont {{van der Walt}}, \citenamefont
  {Brett}, \citenamefont {Wilson}, \citenamefont {Millman}, \citenamefont
  {Mayorov}, \citenamefont {Nelson}, \citenamefont {Jones}, \citenamefont
  {Kern}, \citenamefont {Larson}, \citenamefont {Carey}, \citenamefont {Polat},
  \citenamefont {Feng}, \citenamefont {Moore}, \citenamefont {{VanderPlas}},
  \citenamefont {Laxalde}, \citenamefont {Perktold}, \citenamefont {Cimrman},
  \citenamefont {Henriksen}, \citenamefont {Quintero}, \citenamefont {Harris},
  \citenamefont {Archibald}, \citenamefont {Ribeiro}, \citenamefont
  {Pedregosa}, \citenamefont {{van Mulbregt}},\ and\ \citenamefont {{SciPy 1.0
  Contributors}}}]{2020SciPy-NMeth}%
  \BibitemOpen
  \bibfield  {author} {\bibinfo {author} {\bibfnamefont {P.}~\bibnamefont
  {Virtanen}}, \bibinfo {author} {\bibfnamefont {R.}~\bibnamefont {Gommers}},
  \bibinfo {author} {\bibfnamefont {T.~E.}\ \bibnamefont {Oliphant}}, \bibinfo
  {author} {\bibfnamefont {M.}~\bibnamefont {Haberland}}, \bibinfo {author}
  {\bibfnamefont {T.}~\bibnamefont {Reddy}}, \bibinfo {author} {\bibfnamefont
  {D.}~\bibnamefont {Cournapeau}}, \bibinfo {author} {\bibfnamefont
  {E.}~\bibnamefont {Burovski}}, \bibinfo {author} {\bibfnamefont
  {P.}~\bibnamefont {Peterson}}, \bibinfo {author} {\bibfnamefont
  {W.}~\bibnamefont {Weckesser}}, \bibinfo {author} {\bibfnamefont
  {J.}~\bibnamefont {Bright}}, \bibinfo {author} {\bibfnamefont {S.~J.}\
  \bibnamefont {{van der Walt}}}, \bibinfo {author} {\bibfnamefont
  {M.}~\bibnamefont {Brett}}, \bibinfo {author} {\bibfnamefont
  {J.}~\bibnamefont {Wilson}}, \bibinfo {author} {\bibfnamefont {K.~J.}\
  \bibnamefont {Millman}}, \bibinfo {author} {\bibfnamefont {N.}~\bibnamefont
  {Mayorov}}, \bibinfo {author} {\bibfnamefont {A.~R.~J.}\ \bibnamefont
  {Nelson}}, \bibinfo {author} {\bibfnamefont {E.}~\bibnamefont {Jones}},
  \bibinfo {author} {\bibfnamefont {R.}~\bibnamefont {Kern}}, \bibinfo {author}
  {\bibfnamefont {E.}~\bibnamefont {Larson}}, \bibinfo {author} {\bibfnamefont
  {C.~J.}\ \bibnamefont {Carey}}, \bibinfo {author} {\bibfnamefont
  {{\.I}.}~\bibnamefont {Polat}}, \bibinfo {author} {\bibfnamefont
  {Y.}~\bibnamefont {Feng}}, \bibinfo {author} {\bibfnamefont {E.~W.}\
  \bibnamefont {Moore}}, \bibinfo {author} {\bibfnamefont {J.}~\bibnamefont
  {{VanderPlas}}}, \bibinfo {author} {\bibfnamefont {D.}~\bibnamefont
  {Laxalde}}, \bibinfo {author} {\bibfnamefont {J.}~\bibnamefont {Perktold}},
  \bibinfo {author} {\bibfnamefont {R.}~\bibnamefont {Cimrman}}, \bibinfo
  {author} {\bibfnamefont {I.}~\bibnamefont {Henriksen}}, \bibinfo {author}
  {\bibfnamefont {E.~A.}\ \bibnamefont {Quintero}}, \bibinfo {author}
  {\bibfnamefont {C.~R.}\ \bibnamefont {Harris}}, \bibinfo {author}
  {\bibfnamefont {A.~M.}\ \bibnamefont {Archibald}}, \bibinfo {author}
  {\bibfnamefont {A.~H.}\ \bibnamefont {Ribeiro}}, \bibinfo {author}
  {\bibfnamefont {F.}~\bibnamefont {Pedregosa}}, \bibinfo {author}
  {\bibfnamefont {P.}~\bibnamefont {{van Mulbregt}}},\ and\ \bibinfo {author}
  {\bibnamefont {{SciPy 1.0 Contributors}}},\ }\bibfield  {title} {\bibinfo
  {title} {{{SciPy} 1.0: Fundamental Algorithms for Scientific Computing in
  Python}},\ }\href {https://doi.org/10.1038/s41592-019-0686-2} {\bibfield
  {journal} {\bibinfo  {journal} {Nature Methods}\ }\textbf {\bibinfo {volume}
  {17}},\ \bibinfo {pages} {261} (\bibinfo {year} {2020})}\BibitemShut
  {NoStop}%
\bibitem [{\citenamefont {Abramowicz}\ \emph {et~al.}(1988)\citenamefont
  {Abramowicz}, \citenamefont {Czerny}, \citenamefont {Lasota},\ and\
  \citenamefont {Szuszkiewicz}}]{Abramowicz:1988sp}%
  \BibitemOpen
  \bibfield  {author} {\bibinfo {author} {\bibfnamefont {M.~A.}\ \bibnamefont
  {Abramowicz}}, \bibinfo {author} {\bibfnamefont {B.}~\bibnamefont {Czerny}},
  \bibinfo {author} {\bibfnamefont {J.~P.}\ \bibnamefont {Lasota}},\ and\
  \bibinfo {author} {\bibfnamefont {E.}~\bibnamefont {Szuszkiewicz}},\
  }\bibfield  {title} {\bibinfo {title} {{Slim accretion disks}},\ }\href
  {https://doi.org/10.1086/166683} {\bibfield  {journal} {\bibinfo  {journal}
  {Astrophys. J.}\ }\textbf {\bibinfo {volume} {332}},\ \bibinfo {pages} {646}
  (\bibinfo {year} {1988})}\BibitemShut {NoStop}%
\bibitem [{\citenamefont {Speri}\ \emph {et~al.}(2021)\citenamefont {Speri},
  \citenamefont {Tamanini}, \citenamefont {Caldwell}, \citenamefont {Gair},\
  and\ \citenamefont {Wang}}]{Speri:2020hwc}%
  \BibitemOpen
  \bibfield  {author} {\bibinfo {author} {\bibfnamefont {L.}~\bibnamefont
  {Speri}}, \bibinfo {author} {\bibfnamefont {N.}~\bibnamefont {Tamanini}},
  \bibinfo {author} {\bibfnamefont {R.~R.}\ \bibnamefont {Caldwell}}, \bibinfo
  {author} {\bibfnamefont {J.~R.}\ \bibnamefont {Gair}},\ and\ \bibinfo
  {author} {\bibfnamefont {B.}~\bibnamefont {Wang}},\ }\bibfield  {title}
  {\bibinfo {title} {{Testing the Quasar Hubble Diagram with LISA Standard
  Sirens}},\ }\href {https://doi.org/10.1103/PhysRevD.103.083526} {\bibfield
  {journal} {\bibinfo  {journal} {Phys. Rev. D}\ }\textbf {\bibinfo {volume}
  {103}},\ \bibinfo {pages} {083526} (\bibinfo {year} {2021})},\ \Eprint
  {https://arxiv.org/abs/2010.09049} {arXiv:2010.09049 [astro-ph.CO]}
  \BibitemShut {NoStop}%
\bibitem [{\citenamefont {Perkins}\ \emph {et~al.}(2021)\citenamefont
  {Perkins}, \citenamefont {Yunes},\ and\ \citenamefont
  {Berti}}]{Perkins:2020tra}%
  \BibitemOpen
  \bibfield  {author} {\bibinfo {author} {\bibfnamefont {S.~E.}\ \bibnamefont
  {Perkins}}, \bibinfo {author} {\bibfnamefont {N.}~\bibnamefont {Yunes}},\
  and\ \bibinfo {author} {\bibfnamefont {E.}~\bibnamefont {Berti}},\ }\bibfield
   {title} {\bibinfo {title} {{Probing Fundamental Physics with Gravitational
  Waves: The Next Generation}},\ }\href
  {https://doi.org/10.1103/PhysRevD.103.044024} {\bibfield  {journal} {\bibinfo
   {journal} {Phys. Rev. D}\ }\textbf {\bibinfo {volume} {103}},\ \bibinfo
  {pages} {044024} (\bibinfo {year} {2021})},\ \Eprint
  {https://arxiv.org/abs/2010.09010} {arXiv:2010.09010 [gr-qc]} \BibitemShut
  {NoStop}%
\bibitem [{\citenamefont {Babak}\ \emph {et~al.}(2009)\citenamefont {Babak},
  \citenamefont {Gair},\ and\ \citenamefont {Porter}}]{Babak:2009ua}%
  \BibitemOpen
  \bibfield  {author} {\bibinfo {author} {\bibfnamefont {S.}~\bibnamefont
  {Babak}}, \bibinfo {author} {\bibfnamefont {J.~R.}\ \bibnamefont {Gair}},\
  and\ \bibinfo {author} {\bibfnamefont {E.~K.}\ \bibnamefont {Porter}},\
  }\bibfield  {title} {\bibinfo {title} {{An Algorithm for detection of extreme
  mass ratio inspirals in LISA data}},\ }\href
  {https://doi.org/10.1088/0264-9381/26/13/135004} {\bibfield  {journal}
  {\bibinfo  {journal} {Class. Quant. Grav.}\ }\textbf {\bibinfo {volume}
  {26}},\ \bibinfo {pages} {135004} (\bibinfo {year} {2009})},\ \Eprint
  {https://arxiv.org/abs/0902.4133} {arXiv:0902.4133 [gr-qc]} \BibitemShut
  {NoStop}%
\end{thebibliography}%

\clearpage

\onecolumngrid
\appendix

\section{Additional dependence on the radius in analytic models}
\label{app:analytical_models}

Our analysis relies on analytical torque models that are simple powers in the radius. However analytical models could be more complicated than this simple prescription. For instance, accretion-induced torques often carry over the factor $ F \equiv ( 1- \sqrt{r_{\rm in}/r})^{1/4}$, originating from the solution for the disk temperature and surface density \citep{Frank92}. In this work we have not included it both to maintain a more agnostic approach (potentially across environmental effects of different origins than accretion disks) and to avoid introducing too many parameters in the Monte Carlo analyses of Sec.~\ref{sec:measuring_accretion_properties}. Our omission is a conservative choice, since $F$ always increases the disk temperature and density in the inner region of the disk (perhaps unrealistically, that migration torques will also be affected by the lack gas in the innermost region). Here we partly amend this omission by presenting how the analytical prescriptions of~\citep{Yunes:2011ws,Kocsis:2011dr} would change in the presence of this factor.

To reintroduce the F-factor in the expressions of the main body of the paper, the relevant quantities are the temperature in the central disk plane and surface density \citep{Frank92},
\begin{equation}\label{eq:TempSigma}
    T^4= \frac{3 \Sigma k_R}{4}T_\text{eff}^4, \qquad \Sigma = \frac{\dot{M}_1}{3 \pi \nu} F^4,
\end{equation}
\new{where $\nu = \alpha \beta^b c_s^2 H c_s $ is the kinematic viscosity coefficient in the disk, with $ b=0 $ ($ b=1 $) for $\alpha$ ($\beta$) disks}, $k_R=0.348 \text{ cm}^{2}/g$ is the electron-scattering opacity for a gas of hydrogen and helium, and $\beta$ is a parameter that we will define shortly. We have also introduced an effective temperature $T_\text{eff}$ and sound speed $c_s$,
\begin{equation}
    T_\text{eff} = \left(\frac{3 M_1 \dot M_1}{8\pi r^3 \sigma}\right)^{1/4} F, \qquad c_s^2 = \frac{(p_\text{gas}+p_\text{rad})}{\rho},\quad p_{\rm gas} =\frac{\rho k_B T}{\mu_0 m_p},\quad \text{and }\quad p_{\rm rad} = \frac{4\sigma}{3} T^4,
\end{equation}
which carry further dependencies on the  Boltzmann constant $k_B$, proton mass $m_p$, mean molecular weight $\mu_0=0.615$, and Stefan-Boltzmann constant $\sigma$. We introduce $H(r) = c_s M_1^{-1/2} r^{3/2}$ \citep{Frank92}. 

Notice finally that the $\beta$ parameter is implicitly defined by
\begin{equation}\label{eq:betaimpl}
\frac{\beta}{1-\beta} =\frac{p_{\rm gas}}{p_{\rm rad}} = \frac{3 c k_B^{1/2}}{8 \sigma m_p^{1/2}} \beta^{1/2} \frac{\Sigma}{M_1T^{7/2}r^{3/2}}.
\end{equation}
Solving this for $\beta\ll 1$ and $\alpha$ disks (b=0) yields $\beta_\alpha \approx 1.14 r^{21/8}F^{-8}$, while solving it for $\beta$ disks (b=1) gives $\beta_\beta \approx 1.11 r^{21/10}F^{-32/5}$. Inserting these values in Eq.~\eqref{eq:TempSigma} and the definition of $H$ then leads to the following modifications of the surface densities in Eq.~\eqref{eq:Sigma_H},
\begin{align}\label{eq:Sigma_H_app}
 \Sigma_{\alpha}\bigg[\frac{\rm kg}{\rm m^2}\bigg] \approx& 
5.4 \times 10^{3}\alphanorm^{-1} \feddnorm^{-1}\rnorm^{3/2} F^{-4},\\
\Sigma_{\beta}\bigg[\frac{\rm kg}{\rm m^2}\bigg] \approx&
2.1\times 10^{{7}}\alphanorm^{-4/5} \feddnorm^{3/5}\,
\Monenorm^{1/5}\rnorm^{-3/5}\, F^{12/5}, \\
H[\rm m]\approx&  2.3 \times 10^9 \feddnorm\, M_1 \, F^4,
\end{align}
which in turn gives the following densities
\begin{align}\label{eq:rho_def_app}
\rho_{\alpha}\bigg[\frac{\rm kg}{\rm m^3}\bigg] \approx &
1.3 \times 10^{-6} \alphanorm^{-1} \feddnorm^{-2}\Monenorm^{-1} \rnorm^{3/2} F^{-8},\nonumber\\
\rho_{\beta}\bigg[\frac{\rm kg}{\rm m^3}\bigg] \approx &
4.7 \times 10^{-3} \alphanorm^{-4/5} \feddnorm^{-2/5}\Monenorm^{-4/5} \rnorm^{-3/5} F^{-8/5}.
\end{align}

Carrying over the F factors through $H$, $\Sigma$ and $\rho$ in the expression of the torque \eqref{eq:Ldot_mig} leads to the addition of a factor in the parametrization of Eq. \eqref{eq:L_GW},
\begin{equation*}
    \dot L_{\rm disk} =  A\, \, \rnorm^{ n_{\small r}} F^{n_{\small F}}\, \dot L_{ \rm GW}^{(0)},
\end{equation*}
where $n_{\small F}=\{-12, -28/5\}$ for migration in $\alpha$ and $\beta$ disks. The other parameters in Table \ref{tab:torque_params} remain unchanged.

\section{Upper limit on the amplitude of effects with power law-like radial dependence}
\label{app:upper_limit}
\begin{figure*}
\centering
\includegraphics[width=0.55\linewidth]{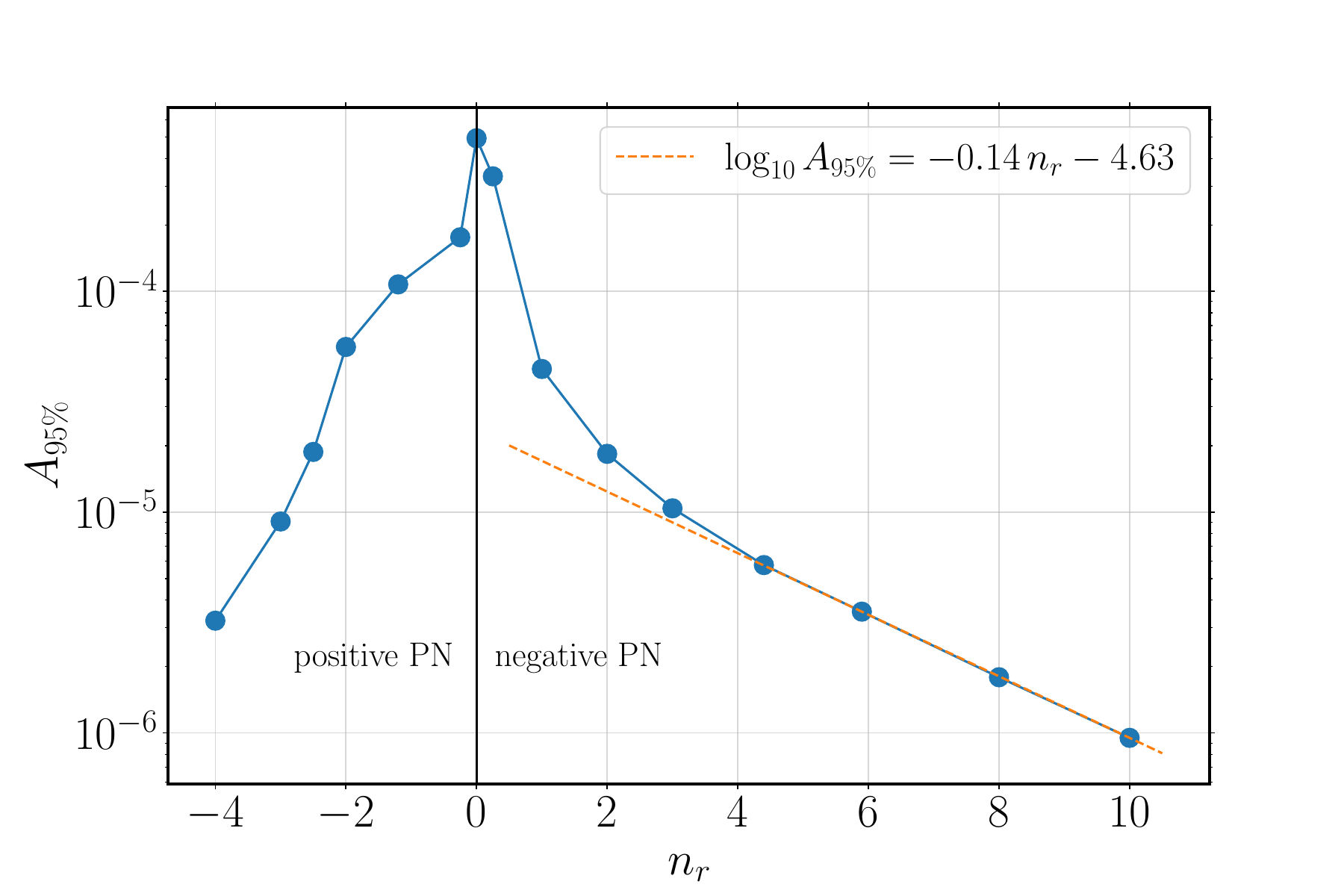}
\caption{Upper limit on the amplitude at $95\%$ (symmetric) for different slopes $n_r$ (blue dots). A fit of the last four points is shown as a dashed orange line.
}
\label{fig:bounds_A_nr}
\end{figure*}

In our analysis we derived the constraints LISA could put on the amplitude of two disk-induced effects, which predict different torque powers $n_r$. Other beyond-vacuum effects might also manifest with a specific power law-like dependence on the orbital separation. Here, we explore how the constraints change as a function of $n_r$ for our reference EMRI source. We show the results in Fig.~\ref{fig:bounds_A_nr} in terms of the symmetric $95\%$ bounds on the amplitude $A_{95\%}$. We find that for $n_r\gtrsim 3$ the bound can be fitted with a straight line in log-space as follows,
\begin{equation}
\log_{10} A_{95\%} = -4.63 -0.14 \, n_{r} \, .
\end{equation}
Similar results are found in Fig.~8 and 9 of Ref.~\cite{Perkins:2020tra} and in Fig.~2 of Ref.~\cite{Chamberlain:2017fjl}, where the bounds are set on a different amplitude parameter. In future work, we plan to investigate how to map our parametrization to the parametrized post-Einsteinian expansions~\citep{Yunes:2009ke}.

\section{Full posterior probability with detectable accretion effect}
\label{app:full_posterior}
\begin{figure*}
\centering
\includegraphics[width=\linewidth]{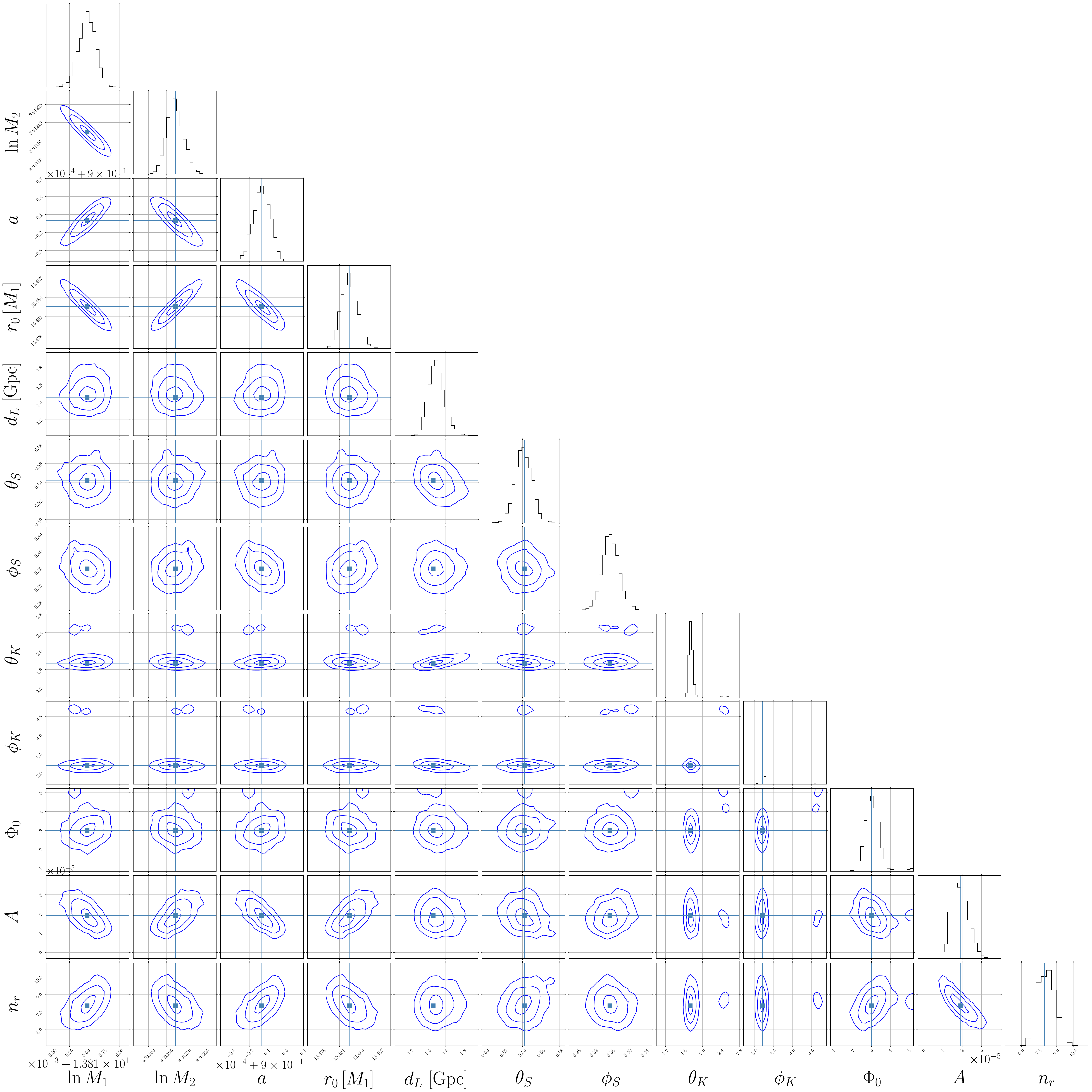}
\caption{Full posterior for the reference EMRI injection described in Sec.\ref{sec:measuring_accretion_properties}.}
\label{fig:fullcorner_plot_alpha}
\end{figure*}

In Sec.~\ref{sec:detection}, we consider the case in which the effect of the environment is detectable in the GW signal. We present in Fig.~\ref{fig:fullcorner_plot_alpha} the full posterior probability distribution for our reference EMRI. 
The posterior is multimodal, although with secondary peaks much suppressed compared to the primary. This is due to the physical degeneracy in the spin orientation parameters $(\theta_K,\phi_K)$ (studied in detail in \citep{Babak:2009ua}) and the initial phase $\Phi_0$.
We note that this the first appearance of the posterior of a circular-equatorial EMRI in the literature. We present in Table.~\ref{tab:inferred_params} the inferred parameters in terms of the median and 95\% credible interval.

\begin{table}[htbp]
    \centering
    \caption{Inferred parameters of the circular-equatorial EMRI affected by migration in $\alpha$ disk. The median and the credible intervals have been estimated using the posterior distribution of Fig.~\ref{fig:fullcorner_plot_alpha} }
    \label{tab:inferred_params}
    \begin{tabular}{cccc}
      \toprule
      \textbf{Parameter} & \textbf{Injected Value} & \textbf{Median and 95\% Credible Interval} \\
      \midrule
      $\ln M_1$ & 13.815510 & $13.815521^{+2.3\times 10^{-4}}_{-2.6 \times 10^{-4}}$ \\
      $\ln M_2$ & 3.912023 & $3.912017^{+1.4\times 10^{-4}}_{-1.2 \times 10^{-4}}$ \\
      $a$ & 0.9 & $0.900001^{+2.5\times 10^{-5}}_{-2.8 \times 10^{-5}}$ \\
      $r_0 \, [M_1]$ & 15.482608 & $15.482507^{+2.7\times 10^{-3}}_{-2.3 \times 10^{-3}}$ \\
      $d_L$ [Gpc] & 1.456479 & $1.492336^{+2.2 \times 10^{-1}}_{-1.6 \times 10^{-1}}$ \\
      $\theta_S$ & 0.542088 & $0.541022^{+1.8 \times 10^{-2}}_{-1.7 \times 10^{-2}}$ \\
      $\phi_S$ & 5.357656 & $5.359016^{+4.1 \times 10^{-2}}_{-3.7 \times 10^{-2}}$ \\
      $\theta_K$ & 1.734812 & $1.747338^{+1.4 \times 10^{-1}}_{-7.1 \times 10^{-2}}$ \\
      $\phi_K$ & 3.200417 & $3.195550^{+8.8 \times 10^{-2}}_{-7.9 \times 10^{-2}}$ \\
      $\Phi_0$ & 3.0 & $3.024194^{+9.1 \times 10^{-1}}_{-6.9 \times 10^{-1}}$ \\
      $A$ & $1.92 \times 10^{-5}$ & $1.84^{+0.9}_{-0.7} \times 10^{-5}$ \\
      $n_r$ & 8.0 & $8.15^{+1.4}_{-1.3}$ \\
      \bottomrule
    \end{tabular}
  \end{table}


\end{document}